\documentclass{JHEP3}

\usepackage{latexsym,bm,amsmath,amssymb,amsfonts, dsfont}
\usepackage{epsfig,graphics,graphicx,mathrsfs}
\usepackage{harpoon}
\usepackage{slashed}
\usepackage{cite}

  \newcommand{\lan}{\left\langle}
  \newcommand{\ran}{\right\rangle}
  \newcommand{\mcal}{\mathcal}
  
  \newcommand{\mcN}{\mathcal{N}}
  \newcommand{\beq}{\begin{eqnarray}}
  \newcommand{\eeq}{\end{eqnarray}}
\vspace{1.5cm}

\title{\rm \LARGE TMD factorization and the gluon distribution in high energy QCD }

\author{Emil Avsar\\
\!\!$^{a}$104 Davey Lab, Penn State University, University Park, 16802 PA,  USA\\  
E-mail:  \email{eavsar@phys.psu.edu}}

\abstract{ This paper is a part of a series of works where we in detail examine the concept of 
Transverse Momentum Dependent (TMD), or $k_\perp$, factorization, 
which is frequently encountered in the literature and is widely used in the phenomenological applications 
of QCD at very high energies.  
We address the question of what exactly factorization is, as it is 
meant in different contexts and formalisms, and we compare the formalisms to each other. We clarify 
some basic concepts regarding factorization and how it exactly is applied in high energy QCD, and we 
make important notes on some key and fundamental points that are often overlooked.  We offer 
an extensive analysis of single inclusive particle production, and we analyze the TMD gluon distribution 
that plays a pivotal role in high energy QCD. 
 }

\begin{document}

\section{Introduction}

Parton distributions, supplemented by factorization theorems, play a crucial role in the understanding 
and exploration of QCD \cite{qcdbook}.  In formulating factorization theorems it is desirable to make as little approximations 
in the kinematics as possible, so as to capture more of the underlying dynamics. Frequently then one 
encounters the concept of transverse-momentum-dependent (TMD), or $k_\perp$-dependent, parton distributions
which follow from TMD factorization ($k_\perp$-factorization). The TMD distributions are 
important because they capture more of the parton kinematics than do the canonical integrated parton distributions, 
the PDFs, and they therefore play an important role in the study of less inclusive 
hadronic observables which are sensitive to the details of the parton kinematics  \cite{Collins:2005uv}. 

In the high energy, small-$x$, limit of QCD even inclusive 
cross sections are sensitive to the TMD distributions, as the so-called Regge kinematics is dominated by 
the transverse components of the momenta. 
 Large contributions arise from large rapidity separations, and the typical 
contributing momenta are slightly off-shell, the off-shellness determined by the transverse momentum. 
Much of the intuition about the TMD distributions is based on concepts directly 
borrowed from the  parton 
model, and it is for example very frequent to find in the literature the assertion that the TMD parton distributions are field 
theoretical number densities, and for example that the underlying mechanism of the phenomenon of saturation is related to the saturation of the phase space occupation number of gluons in 
a hadron, thus implying that there is a upper limit for the number of partons per phase space in the hadron wave 
function. 

While intuitive notions may be helpful in interpreting the dynamics, what is important 
is the exact formulation of TMD factorization that is a must for any proper definition of the 
relevant parton distribution, and the resulting distribution may or may not have the number density interpretation. 
In the small-$x$ literature we find many statements regarding factorization, yet looking 
closely at these statements, we find that the necessary proofs are not always provided. We have moreover found different 
meanings attached to the word ``factorization'', and we therefore take the task of illuminating what exactly 
is being meant in different formalisms. We will do this in section \ref{sec:factorization} where we compare 
different formalisms with each other. 

We should here mention that when we do speak of factorization we shall sometimes use different names 
to distinguish different formalisms. For example, we frequently use the words ``hard scattering factorization"
with which we are referring to the basic factorization of QCD processes where a hard scattering is present \cite{Collins:1981uw, Collins:1981uk, Collins:1984kg, Meng:1995yn, Collins:2003fm, qcdbook}.  The hard scale sets the relevant 
momentum scale by which contributions can be classified according to their power as being leading or suppressed. 
The latter classification is achieved using the power 
counting arguments of \cite{Sterman:1978bi, Libby:1978bx}. We will go through this factorization approach in section 
\ref{sec:hardscatfact}. 
We note that usually the hard scattering 
factorization is referred to as the ``collinear factorization" while the small-$x$ Regge type formalisms go under
the name of ``$k_\perp$-factorization". This is rather misleading, however, since $k_\perp$-factorization (TMD factorization)
is also a central part of the hard scattering factorization approach so that it is important to realize that TMD factorization 
is not only relevant for small-$x$ physics. Depending on the exact final state studied, TMD factorization is 
a necessary tool for QCD studies even when $x$ is not small. We will in section \ref{sec:factorization} 
also go through the Color Glass Condensate  \cite{JalilianMarian:1997jx, JalilianMarian:1997gr, JalilianMarian:1997dw, Iancu:2000hn, 
Iancu:2001ad, Ferreiro:2001qy} formalism which is based on a physical picture of classical color fields. 
One of our main objectives will be to compare the picture of factorization that emerges from the CGC with the 
hard scattering factorization approach. This is important and relevant for understanding much of the phenomenology
based on these formalisms that is currently being used.

In section \ref{sec:gluonprod} we give a detailed analysis on the validity of factorization in single inclusive 
particle production at small-$x$. The main small-$x$ formula, equation \eqref{GLRfact}, or some 
variation of it, has been
widely used in the applications of particle production in proton-proton ($pp$), proton-nucleus ($pA$)
and nucleus-nucleus ($AA$) collisions (see e.g. \cite{Kharzeev:2003wz, Gelis:2003vh, Blaizot:2004wu, Marquet:2004xa, Kharzeev:2004if, Fujii:2006ab, Gelis:2006yv, Gelis:2006dv, Gelis:2008rw, Gelis:2008sz, Levin:2010dw, Levin:2010zy, Albacete:2010bs,  Levin:2011hr, ALbacete:2010ad, Levin:2010br, Tribedy:2011aa} and references therein). 
We shall examine the foundations of the formula, the arguments given 
for its validity, and we shall clarify the exact pre-factor involved in the formula (as there are variations in the literature
regarding the pre-factor). Additionally we shall examine 
what exactly the definition of the corresponding TMD gluon distribution is.

The standard arguments for the validity of the $k_\perp$-factorization formula are usually based on the use of the light-cone gauge. Here, simplifications occur because the leading gluon contributions are suppressed, and Faddeev-Popov
ghosts are absent. 
However, there appear severe technical difficulties 
by the introduction of the unphysical singularities in the light-cone gauge propagators. One issue is that these can potentially
obstruct the contour deformations that are needed for the complete proof of factorization. Additionally, 
for the TMD distributions, the singularities of the gauge propagator imply rapidity divergences starting from 
one loop order, and one must then  consistently regularize those divergences. 

While in the moderate-$x$ region the important gluon momenta are collinear to the hadron momentum, in the 
small-$x$ region one enters the Regge kinematics where actually the transverse momentum components 
are dominating. If $k$ is the gluon momentum then $k^+k^- \ll k_\perp^2$. 
In this case the gluons are also said to be in the Glauber region. 
In light-cone gauge then, transversely
polarized gluons are no longer power-suppressed. This complicates the general treatment 
because one can then have arbitrarily many transversely polarized gluons exchanged 
without power-suppression. 
 To remove the extra gluon contributions and 
establish factorization, one must then be able to perform contour deformations
on the loop momenta out of Glauber region. 
It is then important that the unphysical singularities 
in the gauge propagators do not block the necessary contour deformations. 

In reference \cite{Kovchegov:1998bi} it is shown at least in the deep inelastic scattering of a color-singlet gauge 
invariant gluon current on a hadron that the contour deformations are possible  in low order graphs. 
However, in  \cite{Kovchegov:1998bi} specific assumptions 
are made on the target state that make the application of the Ward identities simpler, at least for the low order graphs. 
Going to higher order graphs, however, complications can easily arise, and a systematic treatment is therefore needed. 
We will examine the applications of  
axial gauge on the particle production process in sections \ref{sec:lcgauge}, \ref{sec:axialgauge}
and \ref{sec:singlehadron}, addressing in particular the ability of making the necessary contour deformations.


Apart from the technical details of the proof of factorization, another issue we address here concerns the 
exact definition of the TMD gluon distribution that is associated with the factorization formula, 
equation  \eqref{GLRfact}. 
The definitions found in the literature  all center 
around the so-called ``dipole gluon distribution" that is related to a (slightly modified) Fourier transform of the coordinate space dipole scattering 
amplitude, see equations \eqref{dipgluedistrb} and \eqref{KTgluon}. In the arguments leading to the factorization formula, however,  one makes use 
of the axial gauge. In the axial gauge, one necessarily obtains 
a definition for the gluon distribution that is an expectation value over the transverse gluon fields, $\langle A^iA^i\rangle$.
This is 
canonically identified, not with the dipole distribution, but with the so-called small-$x$ Weizsacker-Williams (WW) distribution 
which is meant to represent a number density of gluons 
\cite{McLerran:1993ni, McLerran:1993ka, Kovchegov:1996ty, Kovchegov:1997pc, McLerran:1998nk, Iancu:2002xk}. 
The WW distribution naturally appears also in the calculation of certain classical quantities, such as the energy density 
of the so-called Glasma, see for example \cite{Lappi:2006hq}.
There is therefore a potential confusion 
as to what exactly the gluon distribution is, this is for example apparent in reference \cite{Kovchegov:2001sc}. 
We discuss further the form of the gluon distribution in section  \ref{sec:tmdgluon}. 

We should also mention here that this work is part of a larger project initiated in order to understand the connections and 
differences between the various TMD factorization formalisms and the TMD gluon distributions which they give rise 
to. Related points that are not covered here will therefore be discussed and addressed  in two separate papers 
\cite{ourpaper, mypaper2}. 

This paper is somewhat long, the reason being that we cover a variety of topics
which are important for the questions regarding factorization and the correct definitions of the TMD gluon distribution,
and we do not wish to skip important and subtle points  but rather try to explain and illuminate them, as 
this is the goal of our project. We have also aimed at providing a coherent exposition of the various topics 
that appear in different formalisms and different set of works but nevertheless all are centered around the concepts 
of TMD factorization and TMD parton distributions. We have therefore  decided to present all the material 
in a single paper.  We believe that it will be of interest for both 
experimentalists and theoreticians working on related topics. 


The paper is organized as follows. In section \ref{sec:unint} we analyze and explain some fundamental aspects 
of unintegrated parton distributions, starting from the elementary parton model definition. We concentrate on the two type of distributions commonly found 
in the small-$x$ literature.  Section 
\ref{sec:factorization} contains our main discussion on factorization. In section \ref{sec:hardscatfact}
we provide an analysis of the hard scattering factorization approach which leads to both collinear and 
TMD factorization. Then in sections \ref{sec:bfklfact} and \ref{sec:cgcfact} we analyze the formulation of $k_\perp$-factorization 
in the small-$x$ region and we compare these to the hard scattering TMD factorization. 
Section \ref{sec:hybrid} gives an account 
of the formalisms that combine collinear factorization with the small-$x$ formulas. Section \ref{sec:gluonprod}
gives the detailed analysis of the single inclusive particle production in the small-$x$ region as already explained 
above. We have divided this section into several subsections according to the different points we cover, as
was summarized above. Finally, section \ref{sec:summary} contains a brief summary.

\section{Unintegrated parton distributions}
\label{sec:unint}

Our aim in this section is to first  recall the basic 
idea of parton densities. We will outline the basic definition as given by the parton model, and then shortly 
discuss some of the modifications induced by the dynamics of QCD. We also examine the validity of  the intuitive ideas 
borrowed from the parton model in the formulation of small-$x$ QCD. We will therefore here  go through the commonly used ``number density" and ``dipole" distributions from the small-$x$ literature.

The concept of parton distributions dates back to the introduction of the parton model itself by Feynman 
\cite[p.\ 135]{Feynman:1972r}. In there, partons of a particular flavor are considered to have a number density in the target hadron. While for the parton model calculation in DIS it is sufficient to consider number densities in the longitudinal 
momentum component $x$,  the concept also naturally extends to a number density in both $x$ and $k_\perp$. 
 The intuitive concept of a number density of partons can be 
formalized using light-front quantization and writing 
\begin{equation}
\label{eq:pdf.lf.def}
  f_{j/h}(x,k_\perp) = \sum_\alpha 
         \frac{1}{ 2x (2\pi)^3 }
         \frac{\langle P,h |  a_{k,\alpha,j}^{\dag} a_{k,\alpha,j} |P,h\rangle}
              {\langle P,h | P,h\rangle}.
\end{equation}
Here $j$ and $h$ label parton and hadron flavor, $\alpha$ is a parton
helicity index, $|P,h\rangle$ is the target state of momentum $P$, and
$a^\dagger$ and $a$ are parton creation and annihilation operators respectively.

While intuitively clear, definition \eqref{eq:pdf.lf.def} above is not really correct in full QCD, and it cannot be used 
in the exact form just given \cite{qcdbook}. 
In the above formula for example,  the kinematic variables $x$ and $k_\perp$ are literally the momentum 
fraction and transverse momentum 
of the parton probed by the electromagnetic current in DIS. Therefore the unintegrated distribution above is indeed
a simultaneous distribution of the partons in both $x$ and $k_\perp$. In QCD, however, several modifications 
do occur. The variables $x$ and $k_\perp$ no longer correspond to the literal momentum fractions 
of any single parton in the hadron state, and additional variables must be introduced which are connected to the 
divergences that occur in loop calculations (see section \ref{sec:rapidityvariable}
below and in addition the discussions in section \ref{sec:hardscatfact}). 

\subsection{The gluon ``number density"}

It is in the small-$x$ literature often implied that the TMD gluon distribution 
indeed has the meaning of a phase space number density 
as in the above formula. Thus we often find the statement of a certain ``number of gluons per unit phase space". 
In the Color Glass Condensate (CGC) model at least, this statement is meant in the sense of the Weizsacker-Williams 
method of virtual quanta. We recall that in electrodynamics this method replaces the energy density of the classical electromagnetic field created by fast moving charged particles  by the equivalent field of pulse radiation. 
The latter is interpreted semi-classically  as 
 consisting of  a distribution of energy quanta, that is, photons. From the average energy density of the classical field, 
 $\langle |E|^2 \rangle$, one can 
 then calculate the equivalent number of photons. This is the reason why the gluon distribution appearing in the 
 CGC formalism is referred to as the Weizsacker-Williams (WW) gluon distribution.  In the CGC then, one solves the 
 classical Yang-Mills equations for the non-Abelian color field. The energy density of
 the classical field then relates to the  equivalent number density of energy quanta, in this case identified with the gluons. 
 In the light-cone gauge $A^+=0$ one defines (for a hadron with large $P^+$)
\beq
f_{WW}(x,k_\perp) &=& \sum_{i,a} \frac{1}{2(2\pi)^3}  \lan a_a^{\dagger i}(x^+\!\!, k) \, a_a^{i}(x^+\!\!, k) \ran \nonumber \\
&=& \sum_{i,a} \frac{2(k^+)^2}{(2\pi)^3} \lan
A^i_a(x^+\!\!, k) \, A^i_a(x^+\!\!, -k) \ran \nonumber \\
&=&  \sum_{i,a} \frac{2}{(2\pi)^3} \lan
F^{i+}_a(x^+\!\!, k) \, F^{i+}_a(x^+\!\!, -k) \ran 
\label{numberdens}
\eeq
where $k=(k^+,k_\perp)$, and  $a$ and $a^\dagger$,  as in  \eqref{eq:pdf.lf.def}, denote the parton (in this case gluon) 
annihilation and creation operators in the sense of light front quantization where $x^+$ plays 
the role of time. Notice that $x=k^+/P^+$ should not be confused with the time variable $x^+$. 
The last identity, $\langle F^{+i}F^{+i} \rangle$, can be calculated in a classical approximation, for example using 
the McLerran-Venugopalan model \cite{McLerran:1993ni, McLerran:1993ka}, from which an explicit expression can be obtained for $f_{WW}$.  

The definition of the WW distribution is thus essentially identical to the parton model definition \eqref{eq:pdf.lf.def}. 
One trivial difference is that, by convention, the $1/x$ term in \eqref{eq:pdf.lf.def} is not included in \eqref{numberdens}.  
As a less trivial difference we also note that while in \eqref{eq:pdf.lf.def} the quantum mechanical averaging is taken over the 
momentum eigenstates of the target, $|P\rangle$, in the CGC definition \eqref{numberdens} one rather specifies 
a classical charge density $\rho(x^-,x_\perp)$ in the transverse and longitudinal planes, and the classical averaging
is then performed with respect to the specified profile, using a classical weight functional\footnote{This functional should not be confused with our generic notation for Wilson lines which is also $W$. We therefore always explicitly indicate the $\rho$ dependence of the classical CGC functional and write $W[\rho]$.}  $W[\rho]$. One is then 
clearly not averaging over momentum eigenstates. 
The brackets are defined such that  
any function, $\mcal{O}$, of the classical source $\rho$ has the average
\beq
\langle \mcal{O}\rangle = \int D\rho \,\mcal{O}[\rho] W[\rho].
\label{cgcaverage}
\eeq
This averaging is normalized to unity, so that $\langle \mathds{1}  \rangle = 1$, 
\emph{i.e.} the classical weight functional $W[\rho]$ is such that
\beq
\int \mcal{D}\rho \,W[\rho] = 1.
\label{cgcunity}
\eeq

 A gauge invariant version of \eqref{numberdens} can be written as
 (where we now expand the $F^{+i}(x^+\!\!,k)$ in terms of $F^{+i}(x^+\!\!,x^-\!\!,x_\perp)$)
\beq
f_{WW}(x,k_\perp) =  \frac{2}{(2\pi)^3}
\int dx^-dy^-\!\!\int d^2x_\perp && \!\!\! d^2y_\perp e^{ixP^+ (x^- -y^-) - ik_\perp (x_\perp-y_\perp)} \nonumber \\
&&\lan  F_a^{+i}(x) W_{ab}(x,y)F_b^{+i}(y) \ran.
\label{WWdistrbadj}
\eeq
Here $W$ denotes a Wilson line in the adjoint representation needed to make the operators
within the expectation value gauge invariant. We write down the explicit definitions of the Wilson lines 
in the following sections. 
  
 \subsection{The dipole gluon distribution}
 \label{sec:dipoledistrb}
 
 The most commonly encountered ``unintegrated gluon distribution" in the small-$x$ formalism 
 is actually different than the above distribution and is related to the so-called dipole scattering amplitude 
 which itself is specified in coordinate space. The dipole scattering amplitude, and the associated ``gluon 
 distribution" appears as a result of the use of the dipole formalism \cite{Nikolaev:1990ja, Nikolaev:1991et, Mueller:1993rr, Mueller:1994gb, Mueller:2001fv} which canonically is applied 
 to DIS at small-$x$. 
  
 The basic object that enters any definition of the dipole ``gluon distribution" is the coordinate space 
 dipole ``scattering amplitude", $\mcal{N}$. The standard definition of this object in 
 DIS, or in $\gamma^*\gamma^*$ scattering is given by (see for example \cite{Balitsky:1995ub, Balitsky:2001gj, 
 Kovner:2001vi, Iancu:2002xk})
 \beq
\mcal{N}(x_\perp,y_\perp;y) \equiv 1 - \frac{1}{N_c}
\lan \mathrm{Tr} \{ W^\dagger (x_\perp)W (y_\perp)
\}\ran_y, 
\label{dipN}
\eeq
 where we shall freely switch between the coordinates $x_\perp$ and $y_\perp$, and 
 \beq
 r_\perp=x_\perp-y_\perp,  \\ 
  b_\perp=(x_\perp+y_\perp)/2,
 \eeq
 which are respectively the dipole ``size"
and  ``impact parameter" in transverse coordinate space.  In \eqref{dipN},  $W$ denotes the eikonal 
 Wilson line given by 
 \beq
 W(x_\perp) = P \exp \left( -i g_s \int_{-\infty}^\infty d \lambda \, n \!\cdot \!A^a(x_\perp \!\!+ \!\!\lambda n )\, t_F^a
\right ).
\label{Wilsonfund}
\eeq
 Here $P$ denotes path ordering with respect to $\lambda$, and $t_F^a$ is the SU(3) color matrix in the 
 fundamental representation.  The vector $n$ is taken along the light-like direction, and the trace in \eqref{dipN}
 is meant with respect to the color matrices $t_F$. The assertion of the dipole model is that this quantity is 
 relevant for DIS \cite{Buchmuller:1996xw, Buchmuller:1998jv, Mueller:2001fv}, $\gamma^*\gamma^*$ scattering \cite{Balitsky:1995ub, Balitsky:2001gj}, and also for  quark, or prompt photon production 
 in hadron-hadron collisions (see for example \cite{Gelis:2002ki, Gelis:2002nn, Gelis:2006hy}).

As for the momentum distribution referred to as the ``dipole gluon distribution"  \cite{Kharzeev:2003wz, Kharzeev:2004if, Albacete:2010bs, Dominguez:2011wm}, or also very commonly 
as simply the ``unintegrated gluon density" \cite{Blaizot:2004wu, Marquet:2004xa, Fujii:2006ab, Gelis:2006yv, Gelis:2006dv, Gelis:2008rw, Levin:2010dw, Levin:2010zy}, it is given by a modified Fourier transform of the dipole scattering 
amplitude. Most commonly we do in the literature find the definition
\beq
f_{dip}(k_\perp; y) = 
\mcal{C} \int d^2r_\perp d^2b_\perp e^{-i k_\perp \cdot r_\perp} \nabla_{r}^2 \,
\mcal{N}(r_\perp,b_\perp; y),
\label{dipgluedistrb}
\eeq 
where now we have used the variables $r_\perp$ and $b_\perp$ instead of $x_\perp$ and $y_\perp$. 
We write the pre-factor simply  as $\mcal{C}$ 
since there does not seem to be any universally accepted value for it, and different papers use different 
pre-factors.  Note also that a fully gauge invariant definition of \eqref{dipN}, and therefore also of 
\eqref{dipgluedistrb}, requires that one also insert transverse gauge links at $\pm \infty$. 

Formula \eqref{dipgluedistrb} is not exactly linked to the parton model definition of the unintegrated gluon distribution
in \eqref{eq:pdf.lf.def}. It is therefore also distinct from the Weizsacker-Williams distribution, and also 
from the gluon distributions obtained in the TMD factorization approach that we go through in section 
\ref{sec:TMD}. 
 We examine the derivation of the Wilson lines in the definition \eqref{dipgluedistrb} in  \cite{ourpaper}.

A version of the dipole gluon distribution in the adjoint representation appears also in single inclusive
gluon production, equation \eqref{KTgluon}, which we shall examine in detail in section \ref{sec:gluonprod}. 

\subsection{On the rapidity variable in the gluon distribution}
\label{sec:rapidityvariable}

It is also common to denote the rapidity dependence of the dipole distribution \eqref{dipgluedistrb} 
by $x$, using $y=\ln 1/x$. 
We emphasize, however, that the rapidity variable in \eqref{dipgluedistrb} is conceptually different 
than the variable $x$ which appears in  \eqref{eq:pdf.lf.def} and \eqref{numberdens}. In the dipole 
distribution, $y=\ln 1/x$ enters as a rapidity cut-off, either as the scale  in the CGC formalism where the functional $W_y[\rho]$
is evaluated, or as 
the non-zero slope of the Wilson lines  in the formalism by Balitsky \cite{Balitsky:1995ub, Balitsky:2001gj}. On the other hand, in \eqref{eq:pdf.lf.def},
$x =k^+/P^+$, where $k^+$ is the momentum of the parton entering the hard scattering.
Similarly in the light-cone gauge definition of the WW distribution \eqref{numberdens} it again has the meaning 
of the momentum fraction of the gluon entering the hard scattering. Of course, to avoid rapidity 
divergences in \eqref{numberdens} a cut-off must be inserted just as in \eqref{dipgluedistrb}. There must therefore
be present an additional variable, $\zeta$, which plays the same role as $y=\ln 1/x$ in \eqref{dipgluedistrb}.
Thus we have 
\beq
f_{WW}=f_{WW}(x,k_\perp;\zeta),
\eeq
 and we must generally distinguish $x$ and $\zeta$.  It is customary to 
 choose $\zeta=x$ where for example in DIS $x$ is taken to be the Bjorken variable. 
 
 One may then naturally ask why only $y$ and $k_\perp$ appear in the definition of the dipole 
 distribution. The answer is that $k^+$ is actually set to 0 (this is why the Wilson line \eqref{Wilsonfund} is integrated 
 in $x^-$ from $-\infty$ to $+\infty$). Thus the variable  $x$ which appears in $f_{WW}$
is instead set to 0 in $f_{dip}$. If therefore for example the brackets in $f_{dip}$ are evaluated 
fully in the classical approximation without any effects of quantum corrections, say in the MV model, 
then there is no $x$ dependence, unlike $f_{WW}$ which has a $x$ dependence even 
in the classical computation. 
 
 
   


\section{Factorization}
\label{sec:factorization}

As the word ``factorization" is often used in the literature, and as there are many formalisms which 
go under the name of ``$k_\perp$-factorization", we want to examine these formalisms, to explain the similarities 
and the differences among them. We believe this to be a relevant task since it is important especially for the 
experimental community to have clear understanding on what exactly is meant in the different formalisms. This 
is also of interest for theorists, however, and especially in the case of small-$x$ physics where many statements are 
put forward, particularly regarding $k_\perp$-factorization and unintegrated parton distributions. We must 
then once for all analyze these statements and the assertions made. 

The original concept of factorization is to be found in the hard scattering factorization 
approach \cite{Collins:1981uw, Collins:1981uk, Collins:1984kg, Meng:1995yn, Collins:2003fm} 
where for a given process the contributing Feynman graphs are shown to be factorizable into 
different components each of which is associated with a particular type of momentum region. The leading
momentum regions are determined by a power counting analysis that we go through in section \ref{sec:powercount}. 
There is a hard part  specified by the large momentum scale $Q$, and dominated by short distance, $d\sim 1/Q$, contributions.
The hard scattering factorization does not directly deal with the small-$x$ region where $\sqrt{s}$ is 
asymptotically large, and where there may or may not be present in addition the hard scale $Q$. 
For an up-to-date and comprehensive overview of factorization in QCD, see \cite{qcdbook}.
We go through the hard scattering factorization in section \ref{sec:hardscatfact}.

After going through the hard scattering factorization,  we shall in section \ref{sec:bfklfact} 
examine the basic aspects of the BFKL formalism \cite{Fadin:1975cb, Kuraev:1977fs, Balitsky:1978ic}. Here the emphasis is put on the so-called Multi-Regge-Kinematics (MRK), and ideas 
borrowed from the pre-QCD Regge theory \cite{Weis:1972ir, Bartels:1974tj, Bartels:1974tk} play an important role. 
Even though the methods are rather different than the hard scattering factorization,
one can actually identify a structure where different factors are associated with different momentum regions  as in the hard scattering factorization (see \cite{ourpaper} for further discussions).

There is also the CCH approach \cite{Catani:1990xk, Catani:1990eg, Catani:1994sq}  which is based on BFKL but is meant to build on a structure 
that is closely related to the hard scattering factorization since again emphasis is put on a hard 
scattering coefficient. 
We will here not go through CCH since we give a detailed analysis 
in \cite{ourpaper}. In \cite{ourpaper} we also go through in more detail the CCFM formalism \cite{Ciafaloni:1987ur, Catani:1989yc, Catani:1989sg}    that is also 
based on the CCH approach and is meant to interpolate between the small-$x$ BFKL formalism and 
the collinear limit at high $Q$ encoded in the DGLAP evolution. 

There is then the CGC approach \cite{JalilianMarian:1997jx, JalilianMarian:1997gr, JalilianMarian:1997dw, Iancu:2000hn, Iancu:2001ad, Ferreiro:2001qy, Iancu:2002xk, Blaizot:2004wu, Gelis:2006yv, Gelis:2006dv, Gelis:2008rw, Gelis:2008sz} which uses a very different language in terms of classical fields, $A_{cl}$, and their corresponding sources, $\rho$. 
In this case emphasis is put on a power counting in $g_s \rho$ where the strong coupling $g_s$ 
is taken as a fixed variable which can be made as small as possible. A difference between 
``dilute" and ``dense" systems is emphasized, where for dilute systems $g_s \rho \ll 1$ while 
for dense systems $g_s \rho \sim 1$.  The structure of the factorization formula is therefore rather different than the hard scattering factorization. We analyze factorization within the CGC formalism in section \ref{sec:cgcfact}. 
We shall then in section \ref{sec:hybrid} analyze some formalisms where the ideas of collinear factorization and 
the CGC are mixed. 

We may also mention the dipole approach encountered above where the scattering process of parton impinging upon 
a target hadron is modeled via the insertion of Wilson lines as in \eqref{Wilsonfund}, where for a quark the Wilson 
line is taken in the fundamental representation while for a gluon the color matrices in \eqref{Wilsonfund} are instead 
taken in the adjoint representation. The dipole formalism is easily embedded into the CGC picture because 
the CGC formalism, or the MV formalism, gives an explicit way of calculating the averages of the Wilson lines 
that are present in the dipole formalism. Actually factorization is more or less asserted in the dipole formalism. 
In  \cite{ourpaper, mypaper2} we analyze the underlying structure in more detail. 
 
\subsection{Hard scattering factorization}
\label{sec:hardscatfact}

We now review and explain the factorization which is applied to processes where a 
hard scale is present. As we shall see, however, there is a structure which does not depend
on the existence of the hard factor. It will then be important to understand the overall structure here,
since it can also be applied to the Regge 
region. 
We will start with the most simple case of the parton model, and then 
move on to the more complicated cases in QCD, and eventually to TMD factorization which is 
the main interest of this paper. 

\subsubsection{Basic parton model}
\label{sec:partonmod}

\begin{figure}[t]
\begin{center}
\includegraphics[angle=0, scale=0.6]{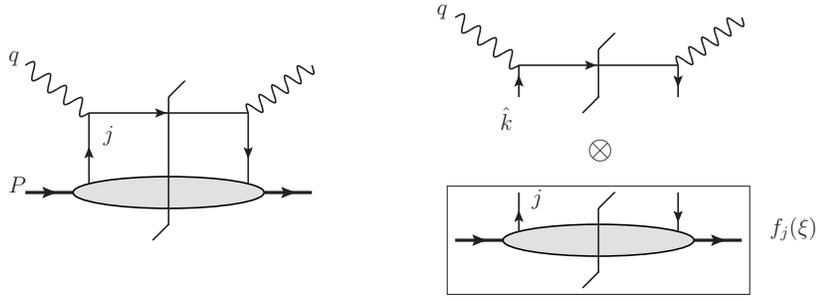}
\end{center}
\caption{\label{partonDIS} DIS in the simple parton model. Right: Factorized structure in the parton model.} 
\end{figure}

In order to understand the basic idea of the hard scattering factorization, it is useful to first look at the 
interpretation of DIS within the parton model. The advantage of the simple parton model is that 
the intuitive ideas  about the scattering and the structure of hadrons can be quantified in a mathematical 
manner which then paves the way for an understanding of the more complicated case of full QCD. 
The quantitive analysis of the model is simplified by the understanding of the kinematics
involved, and in DIS it is convenient to consider the frame where the target hadron 
has momentum  $P=(P^+\!,m^2/2P^+\!,0_\perp)$, 
while the virtual photon has momentum $q=(q^+,q^-,0_\perp)$ where of course $-2q^+q^-=Q^2$.
The scattering in the parton model approximation proceeds as shown in figure \ref{partonDIS} (left graph). 
The parton which is struck by the virtual photon has momentum $k$. 
In the rest frame of the target all the components of $k$  are of the order of the typical hadronic
scale $m$. A large boost in the plus direction then brings the momentum of $P$ into the above form, and implies
that $k^+$ is the largest component, being of order $Q$, while $k^-$ and $k_\perp$ are of order $m^2/Q$ and $m$ 
respectively. This corresponds to the region where the longitudinal momentum fraction $\xi=k^+/P^+$  is not much smaller
than 1. 

According to the parton model one can neglect the effects of the strong interaction during the 
time of the interaction with the photon, and all the effects of the long distance strong interactions is put into 
the parton distribution functions. This structure is shown in figure \ref{partonDIS} (right graph).  In the 
upper part which contains the hard scattering, one can set 
$k=\xi P$. In particular since the minus component of $P$ is power suppressed with respect to the plus component,
one can make the collinear approximation whereby only $k^+$ is kept in the calculation of the hard scattering
coefficient. We denote by $\hat{k}=(k^+\!,0^-\!,0_\perp)$  the approximated momentum.

We define the DIS hadronic tensor $W^{\mu\nu}$ as 
\beq
W^{\mu\nu}(q,P) &=& \frac{1}{4\pi}\int d^4z \, e^{i q\cdot z} \langle P| J^{\mu}(z)J^\nu(0) |P\rangle \nonumber \\
&=& 4\pi^3 \sum_X \delta(p_X-P-q) \langle P| J^{\mu}(0)J^\nu(0) |P\rangle.
\eeq
A factorization formula using the basic assumptions of the parton model can then be easily obtained 
for $W^{\mu\nu}$. Using the general structure of the contributing graphs shown in  figure \ref{partonDIS}, we can write the hadronic tensor as
\beq
W^{\mu\nu} = \sum_j \frac{e_j^2}{4\pi} \int \frac{d^4k}{(2\pi)^4} \mathrm{Tr}\, \gamma^\mu U_j(k+q)
\gamma^\nu L_j(k,P)
\eeq
where $U$ refers to the upper part of the diagram while $L$ refers to the lower blob. The trace refers to the 
Dirac trace. In the upper part only 
$k^+$ is important so we replace $k$ by $\hat{k}$.  Then in the lower part one can replace
$k^+ \to xP^+$ since $\xi=x(1+\mcal{O}(m^2/Q^2))$. Thus we get 
\beq
W^{\mu\nu} \!=\! \sum_j \frac{e_j^2}{4\pi} \mathrm{Tr} \,\gamma^\mu \! \left [ \int \! dk^+
 U_j(k^+\!\!,q^-\!\!,0_\perp)\right ] \! \gamma^\nu \! \left [ \int \! \frac{dk^-d^2k_\perp}{(2\pi)^4}   L_j(xP^+\!\!,k^-\!\!,k_\perp,P)\right ] 
\! + \mathrm{p.s.c.}
 \label{partonmod1}
\eeq
where ``p.s.c." stands for ``power suppressed corrections". 
To finally obtain a fully factorized structure we notice that the leading contribution from the lower part 
comes from the component which is enhanced by the factor $Q$ in the boost along the plus direction from the hadron rest frame. 
Using Lorentz invariance, this leading component can be written as $L_{leading}= \gamma^-\tilde{L}^+= (1/4)\mathrm{Tr} \,
\gamma^+ L$. Thus the factorized structure is given by 
\beq
W^{\mu\nu} = \sum_j \frac{e_j^2}{4\pi} \!\! && \!\!\!\!\!\! \mathrm{Tr}  \left [  \gamma^\mu\int \frac{d\xi}{\xi}
 U_j(\xi P^+\!\!,q^-\!\!,0_\perp)\gamma^\nu \frac{\hat{\slashed{k}}}{2}\right ] \nonumber \\ 
 &\times& \!\! \mathrm{Tr}\left [ \int \frac{dk^-d^2k_\perp}{(2\pi)^4}  \frac{1}{2} \gamma^+ L_j(xP^+\!\!,k^-\!\!,k_\perp,P)\right ]
 + \mathrm{p.s.c.}
 \label{pmfact}
\eeq
The factor in the second row defines the unpolarized integrated quark distribution in the parton model and it can be 
shown to be equivalent to  \eqref{eq:pdf.lf.def}. 
The unintegrated density is obtained simply by undoing the $k_\perp$ integral. Thus 
\beq
f_j(\xi) = \int d^2k_\perp f_j(\xi, k_\perp)
\label{intvsunintpm}
\eeq
in the parton model.
Note that the integral is over \emph{all} $k_\perp$. 
Actually as we review in detail in \cite{ourpaper}, much of the 
literature on the TMD gluon distribution in small-$x$ physics uses very much the same ideas as above. 
We shall also see in section \ref{sec:gluonprod} that very similar arguments  are used in the 
treatment of single inclusive gluon production in small-$x$ QCD. 

\subsubsection{On the leading momentum regions in field theory}
\label{sec:powercount}

In trying to simplify generic graphs in a field theory, so as to extract a factorized form, it is important 
to systematically classify the structure of the leading contributions. In each graph at any given order in 
perturbation theory there may be many loop momenta that give rise to a rather complicated manifold of 
momentum regions. It turns out, however, that there is a correspondence between divergences in massless 
theories and the leading configurations in high-energy processes \cite{Sterman:1978bi, Libby:1978bx}. 

These leading regions are non-UV regions that are important when the hard scale $Q$ gets large. 
The UV region for momenta above $Q$ of course gives divergent contributions but these contributions are handled 
by renormalization which effectively cuts off the integrals above the renormalization scale $\mu$ that 
conveniently may be taken as $Q$.  

If one considers the complex momentum plane, then as $Q\to \infty$, many of the momentum integrations 
can be deformed away from the propagator singularities, and those give therefore vanishing contributions 
at asymptotic $Q$. There may, however, be contributions which cannot be deformed away from the propagator poles.
These contributions arise from surfaces in loop momentum space which are called ``pinch-singular surfaces" (PSSs). 
The PSSs therefore give important contributions which must be taken into account. To determine the strengths
of the different PSSs a power counting analysis is employed. Via the power counting one also can see the 
appropriate approximations to be made in the different momentum regions, and this is highly relevant for 
factorization. 

The interesting regions where there might be large contributions to the graphs for any given process
are thus regions where a given loop momentum $k$ has small virtuality, $|k^2| \ll Q^2$.  Consider semi-inclusive 
DIS where a hadron of momentum $p_B$ is produced  away from the target, \emph{i.e} the large 
component of $p_B$ is its minus component. The target hadron has momentum $p_A$ which is large in the 
plus direction.  

\begin{figure}[t]
\begin{center}
\includegraphics[angle=0, scale=0.6]{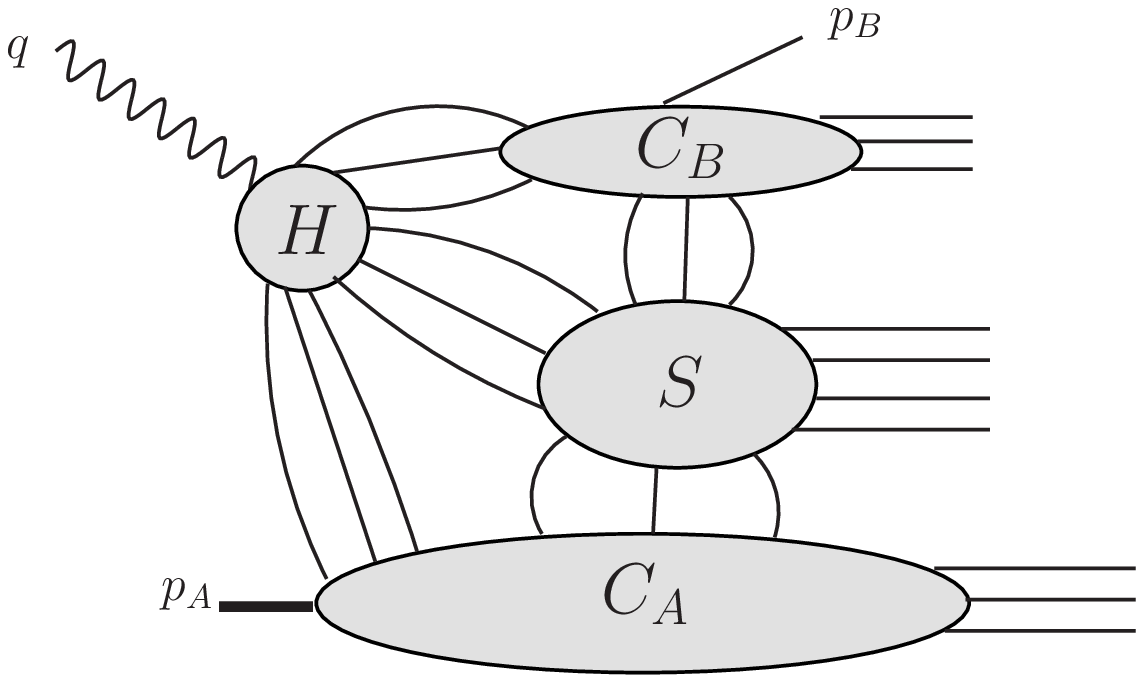}
\includegraphics[angle=0, scale=0.6]{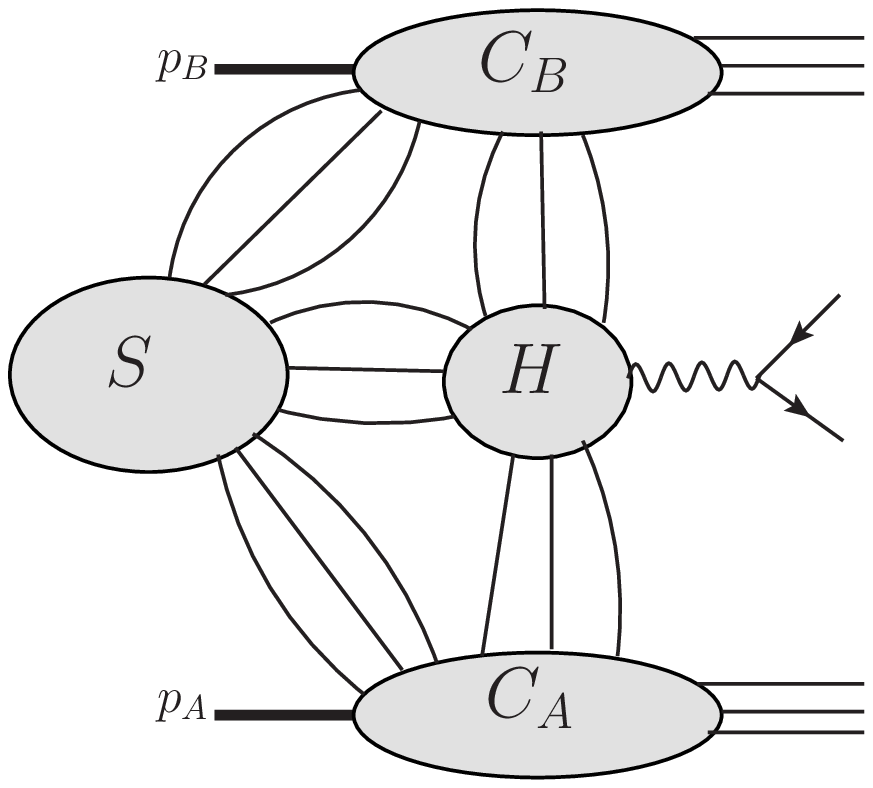}
\end{center}
\caption{\label{SIDISreduced} Left: Reduced graphs for SIDIS where a hadron with momentum 
$p_B$ is detected. Right: Reduced graphs for the Drell-Yan process of lepton pair production in hadron-hadron 
scattering.} 
\end{figure}

We show in figure (left graph) \ref{SIDISreduced} a so-called ``reduced graph" for the important PSSs. 
In obtaining a reduced graph from the full Feynman graph one contracts to points all the lines whose denominators 
are not pinched. This follows from the observation that those lines in the limit $Q^2\to \infty$ carry much larger 
momentum than the pinched lines and therefore in a space-time picture they would reduce to points. The regions 
$H, C_A, C_B, S$ denote the different momentum regions where the momenta are large and of order $Q$ (for $H$), 
collinear to $p_A$ (for $C_A$), collinear to $p_B$ (for $C_B$), and small of order $m$ (for $S$). In the asymptotic 
limit, $p_A$ and $p_B$ become exactly light-like, and the exact PSSs correspond to these limits where the virtuality 
vanishes. Of course in the realistic (non-asymptotic) case the momenta are not exactly light-like so the exact PSSs
form a sort of skeleton of the corresponding region (for example the PSS for $C_A$ is the skeleton where the 
given momentum $k$ is exactly parallel to the light-like limit of $p_A$, while the whole region of $C_A$ also contains 
momenta which are approximately collinear to $p_A$).  The soft  PSS corresponds to the exact 
limit of $S$ where all momentum components of $k$ are 0. Thus in general, momenta belonging to $S$ have 
all their component small (no component is enhanced by any factor of $Q$, and they stay fixed as $Q\to \infty$).
The soft lines can therefore connect to any other region. If $k_S$ is a soft line and is added to say $k_A$ which 
is in $C_A$, then $k_S+k_A$ still belongs to $C_A$. We notice, however, that lines in $C_B$ and $C_A$ cannot 
be directly added to each other because adding two light-like momenta in opposite directions gives a non-light-like
momentum far off shell, and such a line does not belong to any of the two regions (it actually belongs to 
the hard region $H$). 
The collinear lines can, however, be added to the hard part since the result is again a hard 
momentum. Thus one finds the connections between the regions as in figure \ref{SIDISreduced}. 
We also show in figure  \ref{SIDISreduced} (right graph)  the Drell-Yan lepton pair production where 
again there are two collinear regions associated with the incoming momenta $p_A$ and $p_B$, and in 
addition there is the hard part where all momenta are of order $Q$, and there is again  the  soft graph connecting possibly 
to any of the other regions. 

In a collinear pinch, say collinear to the $+$ direction, the typical scales for the momenta are $k^+ \sim Q$, 
$k^- \sim m^2/Q$ and $k_\perp \sim m$. In the soft pinch on the other hand all components satisfy 
$k^\mu \sim m$, while in the hard region  the virtuality is large $|k^2| \sim Q^2$. 
There can also be several collinear regions $C_i$ in a given process. For example in DIS we can have 
several jets emerging from the hard scattering, each defining its own collinear region.  Notice also that a \emph{single} Feynman graph can have 
multiple leading PSSs. This is so because for any given momentum line $k$ in the original graph, we have 
the possibility that $k$ is in any of the allowed regions for that graph. 

\begin{figure}[t]
\begin{center}
\includegraphics[angle=0, scale=0.55]{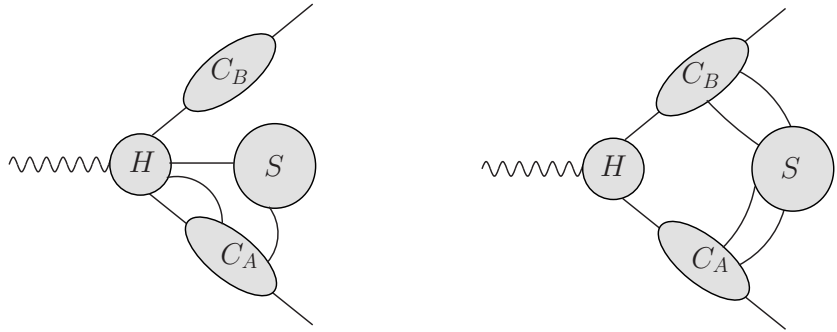}
\end{center}
\caption{\label{sudreduced} Examples of reduced graphs for the two loop Sudakov form factor.} 
\end{figure}

Consider now in QCD gluons exchanged between the different regions. Let us assume we have a collinear-to-$A$ 
gluon $k$ exchanged between the hard part $H$ and $C_A$.  We then have a contribution of the type
\beq
H^\mu N_{\mu\nu}(k)C_A^\nu.
\eeq
Since $C_A$ contains momenta which are large in the $+$ direction, the contribution proportional to $C_A^+$ is boosted by a factor $Q$, and we see that the leading contribution 
satisfies
\beq
H^\mu N_{\mu\nu}(k)C_A^\nu \approx H^-N^{+-}(k)C_A^+.
\label{HtoAapprox}
\eeq
Similar relations hold for gluons exchanged 
between $H$ and $C_B$. If, however, a gluon is exchanged between $H$ and the soft region $S$, 
there is no large boost factor associated with $S$. In fact the $H$-to-$S$ couplings give power suppressed corrections and therefore
the leading power contribution does not contain any lines attaching $H$ to $S$ (see below). As a simple example consider 
figure \ref{sudexample} where a time-like photon $q$ 
produces an exclusive pair of an anti-quark with large
minus momentum $p_B$, and a quark with large plus momentum $p_A$ (this is a two-loop contribution to the Sudakov form factor). In the Feynman graph shown in figure 
\ref{sudexample}, one possibility is that the gluon $k_1$ is collinear to $p_A$, while $k_2$ is  soft. It is then easily seen that 
$p_A-k_1-k_2$ and $p_A-k_2$ are collinear to $p_A$, while $p_B+k_1$ and $p_B+k_1+k_2$ are hard lines (since their virtualities are of order $Q^2$). The reduced graph for this Feynman graph is shown in 
figure \ref{sudreduced} (left graph). 
The contribution  is proportional to 
\beq
g_s^4\, \bar{u}(p_A) \gamma^{\mu_2}\frac{\slashed{p}_A-\slashed{k}_2}{(p_A-k_2)^2+i\epsilon}\gamma^{\mu_1}
\frac{\slashed{p}_A-\slashed{k}_1-\slashed{k}_2}{(p_A-k_1-k_2)^2+i\epsilon} \frac{N_{\mu_1\nu_1}(k_1)}
{k_1^2+i\epsilon}\gamma^\mu
\nonumber \\
\frac{\slashed{p}_B+\slashed{k}_1+\slashed{k}_2}{(p_B+k_1+k_2)^2+i\epsilon}\gamma^{\nu_2}
\frac{\slashed{p}_B+\slashed{k}_1}{(p_B+k_1)^2+i\epsilon}\gamma^{\nu_1}v(p_B)\frac{N_{\mu_2 \nu_2}(k_2)}{k_2^2+i\epsilon}.
\label{sudtwoloop}
\eeq 
To pick up the leading contributions we project out the $+$ component inside the $C_A$ part (which 
consists of the factors to the left of $\gamma^\mu$). This  part can then be written as 
\beq
\gamma^{+}\frac{2p_A^+}{-2p_A^+k_2^-+i\epsilon}
\frac{\slashed{p}_A-\slashed{k}_1}{-2(p_A^+-k_1^+)k_2^-+i\epsilon} \frac{N^{-+}(k_1)}
{k_1^2+i\epsilon} \sim \frac{Q}{Q\, \lambda_s}\frac{Q}{Q\, \lambda_s}\frac{1}{\lambda_A^2}.
\label{twoloopgraph}
\eeq
\begin{figure}[t]
\begin{center}
\includegraphics[angle=0, scale=0.55]{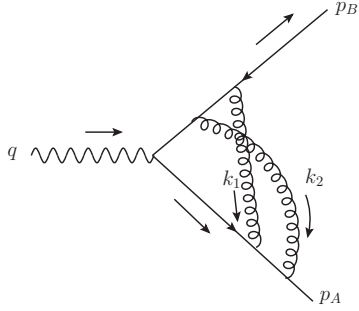}
\end{center}
\caption{\label{sudexample} A two loop contribution to the Sudakov form factor. } 
\end{figure}
Here we have introduced typical momentum scales for the collinear and soft regions, $\lambda_A$
and $\lambda_s$ respectively, such that for any collinear-to-$A$ ($C_A$) momentum, $k_A$, we have $k_A^2 \sim \lambda^2_A$,
while for the soft momentum, $k_s$, we have $k_s^2 \sim \lambda_s^2$.  Notice that since $k_A^+\sim Q$, this 
means that $k_A^- \sim \lambda_A^2/Q$.  The soft region in \eqref{sudtwoloop} simply consists 
of the soft propagator $1/k_2^2 \sim 1/\lambda_s^2$, and the momentum integral $\int d^4k_s \sim \int d\lambda_s 
\lambda_s^3$. 
The collinear-to-$B$ region, $C_B$, is elementary while the hard region power counts as 
\beq
\frac{\slashed{p}_B}{2p_B^-k_1^++i\epsilon}\gamma^{-}\frac{\slashed{p}_B}{2p_B^-k_1^++i\epsilon}\gamma^{-}
\sim \frac{Q}{Q^2} \frac{Q}{Q^2}.
\eeq
The PSSs then give 
\beq
\int^{\sim Q}\!\! d\lambda_A \lambda_A^3 \int^{\sim Q}\!\! d\lambda_s \lambda_s^3 \frac{1}{Q^2}  \frac{1}{\lambda_A^2}\frac{1}{\lambda_s^2}
\frac{Q}{Q\, \lambda_s}\frac{Q}{Q\, \lambda_s} = \int^{\sim Q}\frac{d\lambda_A}{\lambda_A} \left ( \frac{\lambda_A}{Q} \right )^2
\int^{\sim Q} \frac{d \lambda_s}{\lambda_s}.
\label{2loopsuppressed}
\eeq
The complete result is given by multiplying \eqref{2loopsuppressed} with the LO graph. 

In figure \ref{sudreduced} (right graph) we show the case where both $k_1$ and $k_2$ are soft gluons. 
Here, the hard part is elementary while the soft part now 
contains both gluon propagators. It is easy to see that we get in this case
\beq
\int^{\sim Q} \!\!d\lambda_{s,1} \lambda_{s,1}^3 \int^{\sim Q}\!\! d\lambda_{s,2} \lambda_{s,2}^3
\frac{Q^2}{(Q\lambda_{s,1})^2} \frac{Q^2}{(Q\lambda_{s,2})^2} \frac{1}{\lambda_{s,1}^2} \frac{1}{\lambda_{s,2}^2} 
= \int^{\sim Q} \frac{d\lambda_{s,1}}{\lambda_{s,1}}\int^{\sim Q} \frac{d\lambda_{s,2}}{\lambda_{s,2}}. 
\label{twoloopleading}
\eeq
The contribution from the  PSS \eqref{twoloopleading} as we see has no suppression compared to the LO graph, while 
\eqref{2loopsuppressed} has a power suppression. The power suppression comes from the coupling 
of the soft part to the hard part.\footnote{It may seem in \eqref{2loopsuppressed} that performing the $\lambda_A$
integral gives a contribution of order unity since we integrate all the way up to $Q$. However, the integral is 
completely dominated by the upper limit where the momentum is no longer collinear-to-$A$ but is instead is 
hard. In the definition of the hard region there will be a subtraction of the smaller PSSs, for example $C_A$. 
That subtraction will cancel the dominant contribution of the integral and ensure that \eqref{2loopsuppressed} is truly power-suppressed.}

For a given amplitude, cross section or structure function to be analyzed we denote the leading 
power obtained by dimensional analysis as $Q^p$, where $p=4-E_L$ with $E_L$ counting the 
number of external lines. For the Sudakov form factor in figure \ref{sudexample}, $E_L=3$, so 
the lowest order contribution grows as $Q$.  For DIS, $E_L=4$ and the leading power is $Q^0$.
 For a given PSS, we then generally have integrals of the form 
\beq
Q^{p_1}\int^{\sim Q} \frac{d\lambda}{\lambda}\lambda^{p_2},
\label{lambdaint}
\eeq
where $p_1$ and $p_2$ are different powers. 

Making use of dimensional analysis and Lorentz invariance, one then finds in 
QCD the following results \cite{qcdbook}:  For a collinear 
region $C$, every line joining $C$ to $H$ gives a power $\lambda/Q$ \emph{except} for longitudinally 
polarized gluons, carrying  polarization $N^{+-}$, 
for which there is no suppression. For the soft region, every gluon coupling $S$ to $H$
gives a factor $\lambda/Q$ (as in the example of \eqref{2loopsuppressed}) while every fermion gives $(\lambda/Q)^{3/2}$. Every fermion coupling $S$ 
to $C$ gives a factor $(\lambda/Q)^{1/2}$. Thus all couplings between $S$ and other regions are suppressed, 
\emph{except} for longitudinally polarized gluons between $S$ and $C$ for which there is no suppression. There is 
thus no penalty for coupling $C$ and $H$, and $S$ and $C$ via longitudinally polarized gluons. For more details, 
see \cite{Sterman:1978bi, Libby:1978bx, qcdbook}.  In the cases 
where there is no suppression, the integrals \eqref{lambdaint} usually produce logarithms $\ln Q^2/m^2$ 
that accompany the leading power (this is due to the renormalizable nature of QCD in which the coupling 
is dimensionless), as for example in \eqref{twoloopleading}. 
 
\subsubsection{Factorization in simple theory}
\label{sec:simplefact}

\begin{figure}[t]
\begin{center}
\includegraphics[angle=0, scale=0.65]{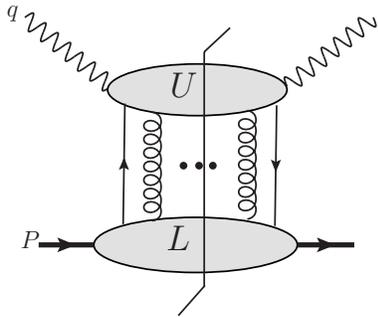}
\end{center}
\caption{\label{gaugeDIS} Generic contribution to inclusive DIS in simplified case.} 
\end{figure}

\begin{figure}[t]
\begin{center}
\includegraphics[angle=0, scale=0.6]{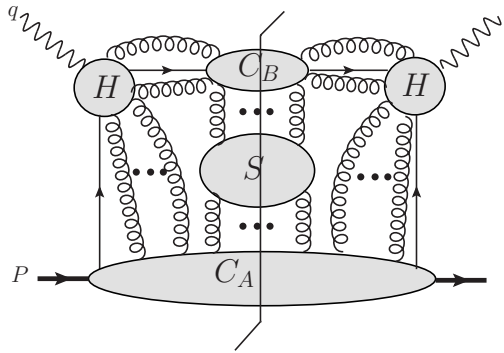}
\end{center}
\caption{\label{gaugeDIS3} Generic contribution to inclusive DIS. } 
\end{figure}

The results above show that in QCD one has to take into account arbitrarily many gluon exchanges, of longitudinal 
polarizations, between the different regions (except for $S$-to-$H$ couplings which are always power suppressed 
regardless of polarization). The proof for factorization is then more complicated compared to the simple parton 
model in figure \ref{partonDIS} where gauge bosons are not present. 
Let us first, however, study a simplified situation by using the results from the power counting. 
This example will be illustrative for understanding the small-$x$ calculations in section \ref{sec:gluonprod}.

 In figure \ref{gaugeDIS} we show an example of inclusive DIS where arbitrarily many gluons are exchanged between the lower part $L$, which is collinear to the target 
hadron $P$, and the upper part $U$, which contains the hard scattering. Of course where the final state cut goes through
$U$, the cut lines are necessarily on-shell, but the bubble will still contain internal lines that are far off-shell. In a more 
complete picture one must consider instead the class of graphs shown figure \ref{gaugeDIS3}. 
It can, however, be shown in the inclusive case by a sum-over-cuts argument that the momenta
in the collinear region can be deformed out to the region where it is far off-shell, effectively reducing the 
leading graphs to that shown in figure \ref{gaugeDIS}. We thus treat the upper part of the diagram 
as the hard region. According to the analysis in the previous section, we then see that soft gluon couplings
do not arise in the leading contributions.

\begin{figure}[t]
\begin{center}
\includegraphics[angle=0, scale=0.65]{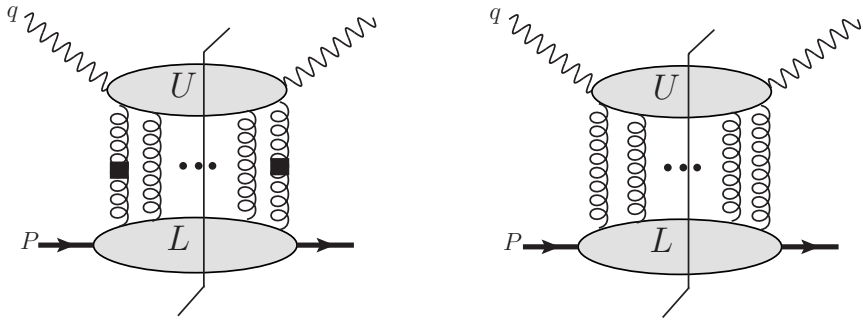}
\end{center}
\caption{\label{gaugeDIS2} Pure gluonic contributions to DIS. Left: The black squares indicate transversely polarized 
gluons while all other gluons are longitudinally polarized.  Right: Longitudinally polarized gluons only give 
a super-leading contribution in the hard scattering region.} 
\end{figure}

We notice that one may also consider pure gluon exchanges between the upper and lower parts. If all gluons are longitudinally polarized, \emph{i.e} contributing via $N^{-+}$, then a super-leading contribution 
arises which has power $Q^2/m^2$ relative to the leading case. However, Ward identities apply for these contributions,
and a careful treatment shows that the super-leading piece actually cancels, leaving behind a remainder term that is leading 
only \cite{Collins:2008sg}. A leading contribution is also obtained 
when one of the gluons at each side of the cut is transversely polarized, we show this in figure 
\ref{gaugeDIS2} (left graph) where we  denote the transversely polarized gluons using the black squares.  
Pure gluon exchange terms 
are important for the analysis in the small-$x$ region which we come back to later. 

\begin{figure}[t]
\begin{center}
\includegraphics[angle=0, scale=0.65]{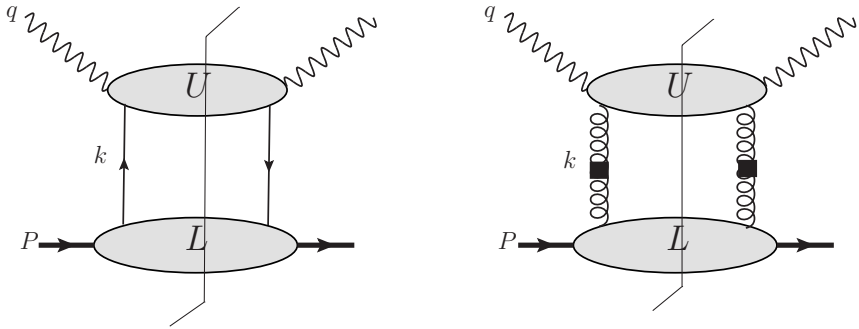}
\end{center}
\caption{\label{nongaugeDIS} Leading contribution in the simplified case in non-gauge theory (only left graph) or in light-cone 
gauge QCD (both graphs).} 
\end{figure}

The parton model result reviewed above can be exactly reproduced in a model field theory which is non-gauge 
(this removes all gauge boson attachments between $L$ and $U$) and super-renormalizable (this implies that 
the hard part $U$ is trivial as in figure \ref{partonDIS}).  As a simplified case we instead imagine a theory which is still non-gauge but is renormalizable.  This means that 
the higher order corrections to the hard part are not power suppressed anymore.  Moreover it means that one has 
to also take into account the UV renormalization. At the same time it implies that 
the gauge boson exchanges shown in figure \ref{gaugeDIS} are absent, and one obtains instead figure \ref{nongaugeDIS}. Now, another way to think of this case is to actually consider full QCD in light-cone gauge $A^+=0$. In this case the leading gluon coupling vanishes since 
\beq
N^{-+}(k)=g^{-+} - \frac{n^-k^++n^+k^-}{k^+n^-} = 1 - 1 = 0.
\eeq
Therefore in figure \ref{gaugeDIS}, all gluon couplings again vanish to leading power. In figure \ref{gaugeDIS2} 
it means on the other hand that only the two transversely polarized gluons remain, as shown in 
figure \ref{nongaugeDIS}. 

A factorization formula for figure \ref{nongaugeDIS} can now be obtained rather easily by assuming that there is a 
clear separation in momenta for the exchanged line $k$, namely that it can either be hard or collinear 
to $P$.  We can then write the hadronic tensor as (neglecting photon indices)
\beq
W = \int \frac{d^{4-2\epsilon} k}{(2\pi)^{4-2\epsilon}} U^{\{\alpha\}}(k, q) L_{\{\alpha\}}(k,P),
\eeq
where the index $\{\alpha\}$ collectively denotes all relevant labels such as flavor, color, polarization\footnote{Of course 
in a non-gauge theory we need not consider the color indices but as the analysis is also relevant for light-cone QCD 
we include all quantum labels.}.  We again
make the  approximation of replacing $k$ in $U$ by $\hat{k}=(k^+\!,0,0_\perp)$. Thus one gets 
\beq
W \sim \int \frac{d k^+}{k^+} U^{\{\alpha\}}(\hat{k}, q) \,\,k^+\!\!\int \frac{dk^- d^{2-2\epsilon} k_\perp}{(2\pi)^{4-2\epsilon}} 
L_{\{\alpha\}}(k,P).
\eeq
This formula is not yet in a fully factorized form, however, since there is still the sum over the labels $\{\alpha\}$.
We note that $U$ must be diagonal in the color indices since the photon is color singlet. Consider first the 
quark contribution shown in figure \ref{nongaugeDIS} (left graph). To fully factorize $F$ we can then apply 
exactly the same argument as in the parton model case in going from equation \eqref{partonmod1} to \eqref{pmfact}. 
We then get just as in \eqref{pmfact}
\beq
W \sim   \int \frac{d\xi}{\xi} \left [ \mathrm{Tr} \,
 U_j(\xi P^+\!\!,q^-\!\!,0_\perp) \frac{\hat{\slashed{k}}}{2}\right ] \left [  \mathrm{Tr}
  \int \frac{dk^-d^{2-2\epsilon}k_\perp}{(2\pi)^{4-2\epsilon}}  \frac{1}{2} \gamma^+ L_j(k,P)\right ].
\eeq
Summation over the color indices in $L$ is kept implicit.  Corrections to the factorization formula 
are power suppressed by the analysis in section \ref{sec:powercount}. 

For the gluon contribution shown in the right graph of figure \ref{nongaugeDIS} we instead find
\beq
W \sim \int \frac{d k^+}{k^+} U^{ij}(\hat{k}, q)\,\, k^+\!\!\int \frac{dk^- d^{2-2\epsilon} k_\perp}{(2\pi)^{4-2\epsilon}} 
L^{ij}_{aa}(k,P).
\label{simplequarkfact}
\eeq
We then notice that the upper part $U$ is diagonal in the transverse and color indices which gives the 
factorized form 
\beq
W \sim \int \frac{d \xi}{\xi} \left [ \frac{1}{2}U^{jj}(\hat{k}, q) \right ] \left [\xi P^+\!\!\int \frac{dk^- d^{2-2\epsilon} k_\perp}{(2\pi)^{4-2\epsilon}} 
L^{ii}_{aa}((\xi P^+\!\!,k^-\!,k_\perp),P)\right ].
\label{simplegluonfact}
\eeq
The second factor here defines, preliminarily, the integrated gluon distribution.  We shall see in section \ref{sec:gluonprod}
that the elementary definition 
of the TMD gluon distribution in axial gauge in the small-$x$ limit is given by the very same 
set of approximations. 

This simple derivation of factorization cannot be strictly true, however. 
Namely, the main assumption that a clear separation of scales is possible 
is not generally true in a renormalizable theory like QCD.  For example in the above calculation we assume that 
$k_\perp \sim m$, while the case $k_\perp \sim Q$ would have instead contributed to the next-to-leading order correction 
to the hard part $H$. There is, however, also an intermediate region, where 
$m \lesssim k_\perp \lesssim Q$, and $k$ is neither exactly target collinear nor exactly hard, and as a consequence it is not 
clear in the above formalism how to exactly handle $k$ in this case. For the assumptions above 
to thus hold, it must be true that this intermediate region can be safely omitted.  This is, however, not the case. In 
fact, the renormalizability of QCD implies that there are in general logarithmic contributions, 
\beq
\int_{\sim m^2}^{\sim Q^2} \frac{dk_\perp^2}{k_\perp^2} \sim \ln Q^2/m^2.
\eeq

\begin{figure}[t]
\begin{center}
\includegraphics[angle=0, scale=0.5]{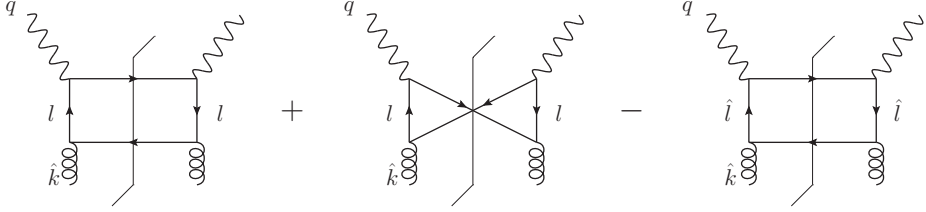}
\end{center}
\caption{\label{dissubtracted} Example of subtraction in the NLO gluon coefficient. The subtraction removes the 
contribution where the loop momentum $l$ is target-collinear, indicated by $\hat{l}$ in the last graph. } 
\end{figure}

\begin{figure}[t]
\begin{center}
\includegraphics[angle=0, scale=0.5]{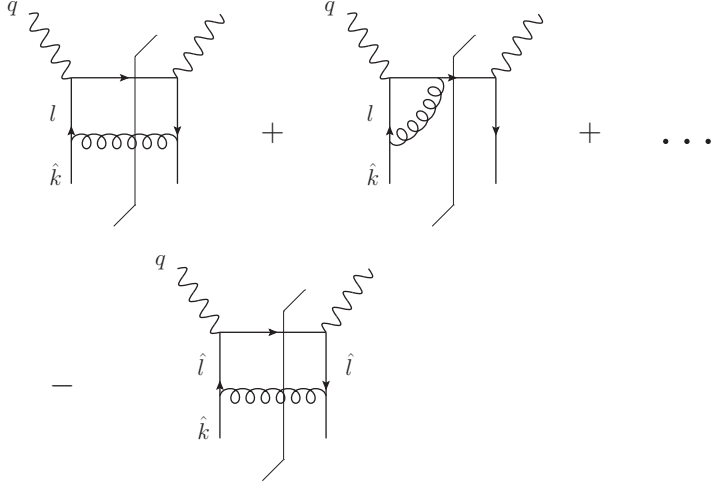}
\end{center}
\caption{\label{dissubtracted2} Example of subtraction in the NLO quark coefficient. The subtraction removes the 
contribution where the loop momentum $l$ is target-collinear, indicated by $\hat{l}$ in the last graph.} 
\end{figure}

There is therefore no power suppression of the intermediate region, and in fact it is even enhanced by a 
logarithm.  A full treatment must therefore treat such regions correctly, and this can in general be done by a 
subtractive formalism \cite{qcdbook}. This means that each PSS is defined with subtractions 
of the smaller PSSs that it contains, to prevent double counting and ensure that it indeed is dominated 
by the momenta associated with it. For the hard part $U$ in figure \ref{nongaugeDIS}, one should therefore
include a subtraction of the target-collinear PSS.  We show examples of these subtractions 
in DIS for the gluon and quark contributions in figures \ref{dissubtracted} and \ref{dissubtracted2} 
respectively. If we denote by $d\Pi$ the phase 
space measure for the momenta contained in $U$ then a more correct version of \eqref{simplegluonfact} reads
\beq
W \sim \int \frac{d \xi}{\xi} \!\!\!\!&&\left [ \frac{1}{2}\int d\Pi \left [ U^{jj}(\hat{k}, q)
-\mathrm{subtractions} \right ] \right ] \nonumber \\
&&\times  \left [\xi P^+\!\!\int \frac{dk^- d^{2-2\epsilon} k_\perp}{(2\pi)^{4-2\epsilon}} 
L^{ii}_{aa}((\xi P^+\!\!,k^-\!,k_\perp),P)\right ].
\label{subtractedfact}
\eeq
The integrated (bare) gluon distribution is thus given by 
\beq
f_g^{(0)}(\xi) &=& \xi P^+\!\!\int \frac{dk^- d^{2-2\epsilon} k_\perp}{(2\pi)^{4-2\epsilon}} 
L^{ii}_{aa}((\xi P^+\!,k^-\!,k_\perp),P) \nonumber \\
&=& \int \frac{dx^-}{2\pi \, \xi P^+} e^{i\xi P^+ x^-} \langle P | F_{(0) \, a}^{+i}(0^+\!\!,x^-\!\!,0_\perp)F_{(0) \, a}^{+i}(0)
|P\rangle
\label{intgluonpdf}
\eeq
where the last result holds in $A^+=0$ gauge in QCD, apart from some technical problems associated with this 
gauge that we are neglecting.

As indicated in \eqref{intgluonpdf}, the basic operator definitions of the parton distributions are for the bare fields of the Lagrangian. Note that 
it is these fields which have the canonical gauge transformation properties, and thus in discussing the gauge 
transformation properties of the parton distributions one necessarily refers to the operator definitions constructed 
out of the bare fields. 
The renormalization of the bare parton distributions
is then an issue of the renormalization of non-local operators. While in the case of local 
field operators, the renormalization factor can be taken as a multiplicative constant which is independent of 
momenta and masses, for the non-local operators appearing in the definitions of the bare parton distributions
one instead finds that there is a convolution with a renormalization factor. Basically if we denote the bare parton 
distribution for a parton of flavor $j$ as obtained from either \eqref{simplequarkfact} or \eqref{simplegluonfact} 
by $f^{(0)}_j(\xi)$, and the renormalized distribution by $f_j(\xi)$, we find 
\beq
f_j(x; \mu) = \lim_{\epsilon \to 0} Z_{jj'}(\xi, \mu, \epsilon) \otimes_{\xi} f_{j'}^{(0)}(x/\xi; \mu, \epsilon),
\label{pdfrenorm}
\eeq 
where the convolution is an integral in $\xi$ as in \eqref{simplequarkfact} and \eqref{simplegluonfact}. 
The evolution of $f_j(x; \mu)$ with respect to $\mu$ is given by the DGLAP equations. 

\subsubsection{Including the gluons, and the Glauber region}

For a fully satisfactory treatment of factorization in full QCD one needs, however, to deal with the gluon 
emissions. As we recall from section \ref{sec:powercount}, in QCD we can without any power suppression exchange arbitrarily many 
longitudinally polarized gluons between the hard and collinear, and the soft and collinear regions respectively.
We indicated this possibility already in figures \ref{gaugeDIS3} and \ref{gaugeDIS2}.  In the previous section 
we argued that in the collinear factorization of inclusive DIS at least, the structure of the leading graphs can 
be simplified by choosing the light-cone gauge $A^+=0$ which eliminates the leading longitudinally polarized gluons.

There is, however, a good reason to try to avoid the light-cone gauge in the generic treatment (see also 
sections \ref{sec:lcgauge}, \ref{sec:axialgauge} and \ref{sec:singlehadron} below). Note from the 
arguments in the previous sections that the treatment of factorization is based on first analyzing  the analytic structure 
of the Feynman graphs, identifying the PSSs, and then using power counting to extract the leading PSSs. To 
guarantee that the power counting arguments work properly, contour deformations must be performed when necessary. 
In particular, if $k$ is a momentum in the soft region, then there is the possibility that the components of $k$ 
do not all scale with the same power $\lambda_s$, but that the longitudinal components $k^+$ and $k^-$ might 
be parametrically much smaller than $k_\perp$. This happens  if $k^+$ or $k^-$ is pinched by the collinear lines it attaches to. For example, if $k$ couples to a collinear line $p_A$ then a propagator,
\beq
(p_A + k)^2 + i\epsilon,
\eeq
arises. The pole for $k^-$ is then 
\beq
k^- \sim \frac{m^2}{Q} - i\epsilon.
\label{kminuspole}
\eeq
Thus $k^-$ is parametrically much smaller than $\lambda_s \sim m$.  When this happens, we say the
momentum is in the Glauber region, $k^+k^- \ll k_\perp^2$. Now, if no other such pole is present, 
or if all such poles lie in the same part of the imaginary plane (all below or above the real axis), then we can deform the contour away from 
this pole to keep $k^- \sim \lambda_s$.  If, however, another pole exists simultaneously, such that 
\beq
k^- \sim \frac{m^2}{Q} + i\epsilon
\label{kminuspole2}
\eeq
then the $k^-$ contour is pinched, and cannot be deformed. It might still be possible to deform on $k^+$ but if not, then the standard power counting fails. The longitudinal polarizations 
then no longer dominate and one cannot use the eikonal approximations needed to obtain factorization. 

The use of the light-cone gauge implies that the analytic structure of the individual Feynman graphs is altered,
since now an additional pole $1/k^+$ is introduced with each propagator. This has obvious implications for the 
factorization proofs.  These poles might for example introduce pinch points that are not present in a covariant gauge. 
Moreover, the gauge poles $1/k^+$ commonly give rise to integrals of the form 
\beq
\int_0^\infty dk^+ \frac{1}{k^+}I(k^+,k_\perp),
\label{rapdiv}
\eeq
and these diverge as $k^+\to 0$. Notice that the divergences arise from end point singularities and can 
therefore not be treated by any $i\epsilon$ prescription or principal value. In fact there exists no generalized
function which is a ``canonical regularization", in the sense described in \cite{gelfand}, of this integral. 

These divergences are in fact the rapidity divergences we mentioned 
in sections \ref{sec:dipoledistrb} and \ref{sec:rapidityvariable}. They also arise when the eikonal approximation 
is used in a covariant gauge. In the integrated distribution, there is actually a cancellation between real and virtual 
terms, which means that in \eqref{rapdiv} 
\beq
\int d^2k_\perp I(k^+=0,k_\perp) = 0.
\eeq
This leads to the well-known ``plus prescription", $\left ( \frac{1}{1-z} \right )_+$. In TMD distributions, however, no cancellation occurs, since $I(0,k_\perp) \neq 0$,
and the light-cone gauge therefore introduces problems. 
The light-cone gauge is moreover not useful when several different collinear directions are relevant.

The general method for factorizing the arbitrary order  gluon couplings between the different regions 
is based on exploiting the gauge symmetries of the leading terms, and to use Ward identities (Slavnov-Taylor-Ward identities). 
The basic technique can be understood as follows. In Feynman gauge, let $k$ 
be a soft gluon coupling the regions $S$ and $A$. We then have a contribution of the type
\beq
A^\mu(k, p_A) \, g_{\mu\nu}\, S^\nu(k).
\eeq
Generally of course there will be many other couplings, and $A$ and $S$ will depend on additional 
momenta but that does not matter for the approximation we are explaining. The leading contribution is then 
\beq
A^\mu(k, p_A) \, g_{\mu\nu} \, S^\nu(k) \sim A^+(\hat{k}_B, p_A)S^-(k) \nonumber \\
= A^\mu(\hat{k}_B,p_A)\frac{\hat{k}_{B,\mu} \, n_{A, \nu}}{k\cdot n_A} S^\nu(k),
\label{AtoSapprox}
\eeq
where 
\beq
\hat{k}_B=(k \cdot n_A)\, n_B=(0^+\!\!, k^-\!\!, 0_\perp).
\eeq
Here $n_A$ is a light-like vector in the direction of $p_A$, 
with $n_A\cdot V = V^-$ for any $V$. Thus $\hat{k}_B\cdot n_A = k \cdot n_A$.  Since now the 
polarization of the gluon $k$ is multiplied by its momentum in the coupling to $A$, Ward identities 
can be applied. The eikonal denominator in \eqref{AtoSapprox} gives a contribution in $S$ from a 
Wilson line. The all-order gluon couplings between $A$ and $S$ can then be successively factorized
into a Wilson line contribution in $S$. 

\begin{figure}[t]
\begin{center}
\includegraphics[angle=0, scale=0.65]{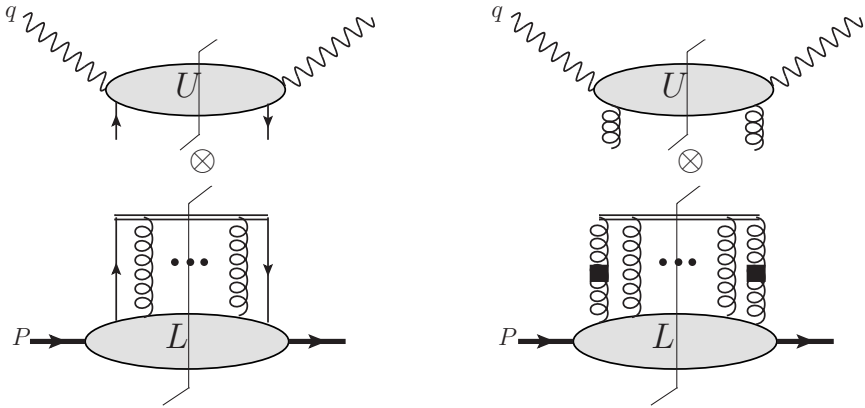}
\end{center}
\caption{\label{DISwilsonlines} Factorized structure in inclusive DIS in covariant gauge. The longitudinal 
gluon emissions are factorized into eikonal Wilson lines (double lines) to provide gauge invariant definitions 
of the parton distributions. Left: Quark distribution. Right: Gluon distribution where the gluons with black 
squares are transversely polarized gluons. } 
\end{figure}

One can similarly make approximations for the $H$-to-$A$ couplings. The eikonal terms that arise are then 
absorbed into $A$ to provide gauge invariant definitions of the basic parton distributions (or fragmentation 
functions).  An example in the case of inclusive DIS is shown in 
figure \ref{DISwilsonlines} where the Wilson lines are indicated by double lines. 
The procedure of using the Ward identities for extracting the gluon exchanges between the different 
regions proceeds very much the same whether one is formulating collinear factorization or TMD factorization.

As we have seen, Wilson lines appear in the small-$x$ formalisms as well, both in the Weizsacker-Williams 
distribution \eqref{WWdistrbadj} and the dipole distribution \eqref{dipgluedistrb}.  It is then rather important 
to understand the exact structure and derivation of these lines, in particular since differences appear between 
the dipole definition and the TMD distributions. We analyze these points in detail in \cite{ourpaper}.

\subsubsection{TMD factorization}
\label{sec:TMD}

\begin{figure}[t]
\begin{center}
\includegraphics[angle=0, scale=0.55]{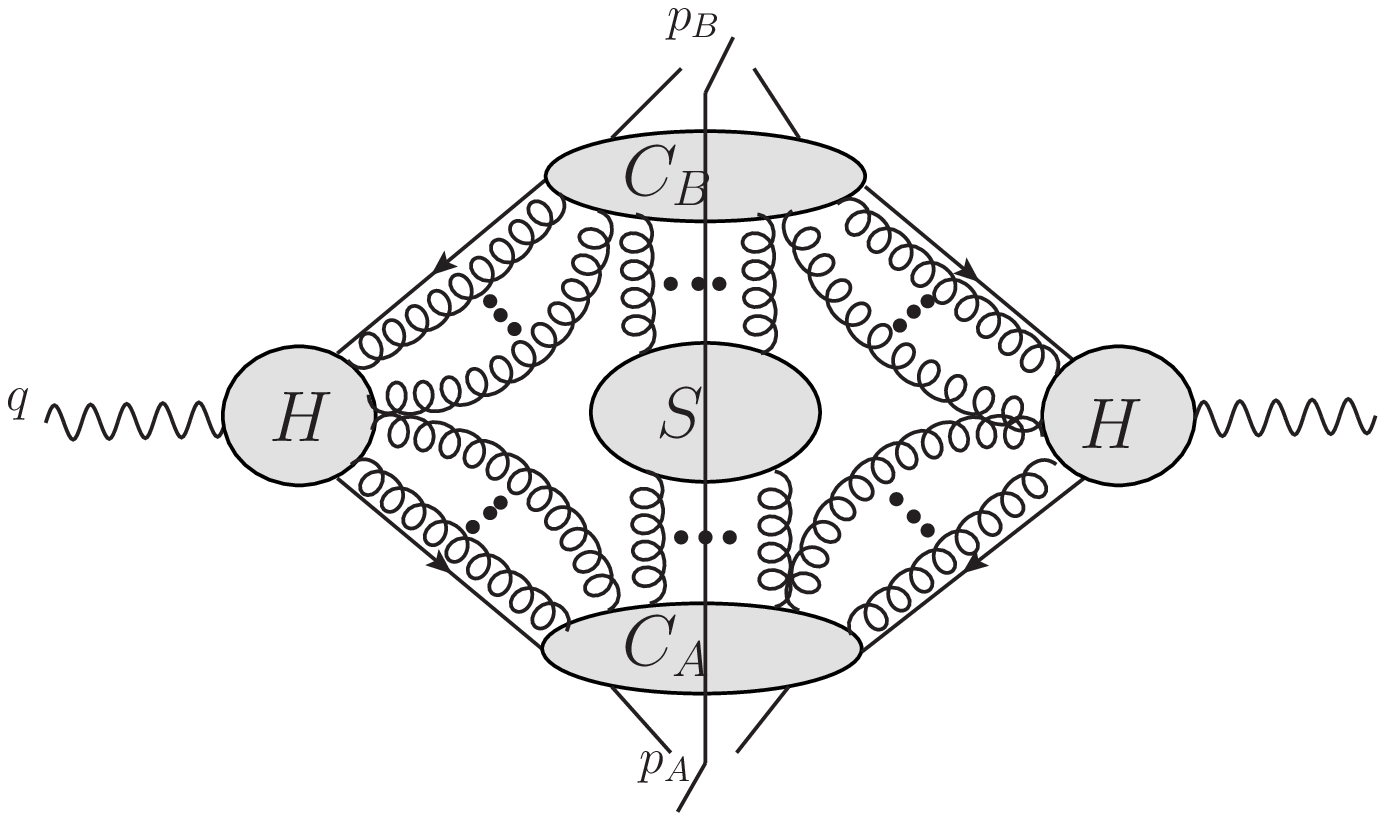}
\includegraphics[angle=0, scale=0.55]{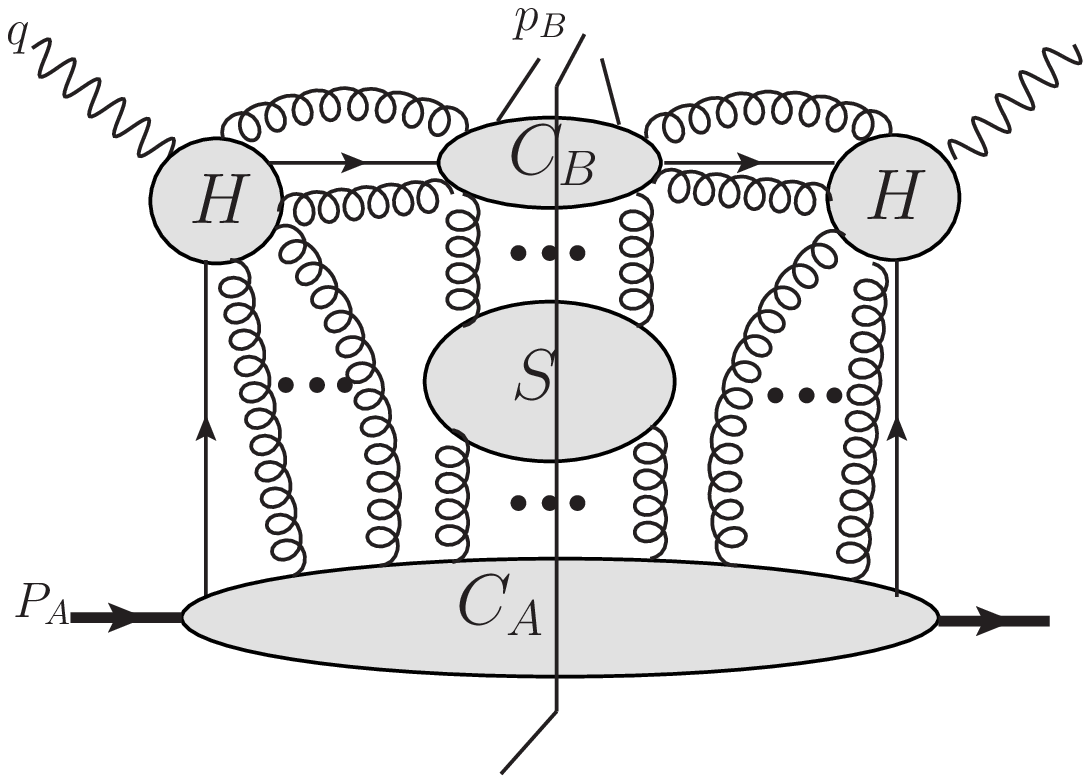}
\end{center}
\caption{\label{TMDgraphs1} Processes where TMD factorization is relevant. Left: Di-hadron production in $e^+e^-$. 
Right: Hadron production in SIDIS. } 
\end{figure}

\begin{figure}[t]
\begin{center}
\includegraphics[angle=0, scale=0.55]{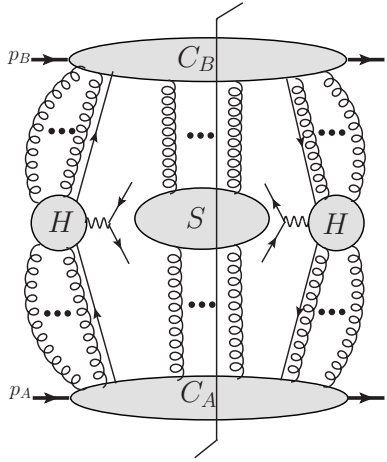}
\end{center}
\caption{\label{TMDgraphs2} Leading regions for TMD factorization in Drell-Yan lepton pair production. } 
\end{figure}

In the hard scattering formalism, the need for TMD factorization becomes obvious when one considers observables 
which are more sensitive to the exact kinematics of the final state.  
A typical example concerns the almost back-to-back production of 
hadrons \cite{Collins:1981uk} in $e^+e^-$ annihilation shown in figure \ref{TMDgraphs1}.  Other relevant processes 
where one needs to consider TMD factorization are 
single-inclusive hadron production at low $p_\perp$ 
in DIS (SIDIS) also shown in figure \ref{TMDgraphs1}, and Drell-Yan lepton pair production shown in figure 
\ref{TMDgraphs2}   where the total transverse momentum of the lepton pair
is much smaller than the hard scale. In all these cases the kinematics is sensitive to low values of the observable
transverse momentum $q_\perp$, 
and one cannot therefore neglect any of the transverse momenta flowing through  the regions $C_A$, $C_B$ and $S$,  
as doing so would significantly change the kinematics of the observable final state products. 
If on the other hand the relevant transverse momentum observables are large, of the order of the hard scale $Q$, 
then the effects of the transverse momentum flowing out from the collinear regions via the soft region 
is power suppressed and can be neglected. In that case one obtains the standard integrated (collinear) factorization. 

Note, however, that the transverse momentum flowing directly into the hard part $H$ from the collinear regions 
$C_A$ and $C_B$ can still 
be neglected, since the error involved in this approximation is of order $q_\perp/Q$ 
which is small in the validity region of TMD factorization. As $q_\perp \to Q$ the TMD formula loses its accuracy
but then one enters the region where ordinary integrated factorization is valid.  When $q_\perp \sim Q$, the 
transverse momentum must be a part of the hard region, physically it corresponds to the case where several 
high $q_\perp$ partons emerge from $H$.  Thus 
what determines the need for TMD parton distributions and fragmentation functions is the kinematics of the 
final state. The momenta entering $H$ from the collinear region $C_A$ or $C_B$ can still be approximated to be on-shell, 
even in the case of TMD factorization. This is somewhat different than the small-$x$ formulation 
where the gluon momentum entering the hard scattering (if there is any) is off-shell, its virtuality being determined 
by the transverse momentum. 

\begin{figure}[t]
\begin{center}
\includegraphics[angle=0, scale=0.65]{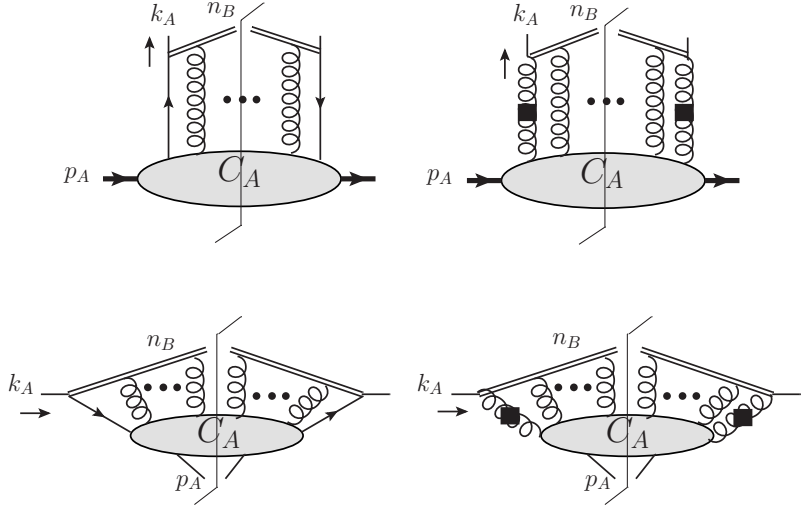}
\end{center}
\caption{\label{unsubtracted} Graphical representation of the unsubtracted collinear part 
  after the 
gluon couplings to the hard part have been factorized into Wilson lines in the direction $n_B$. Left: Quark 
distribution. The black squares indicate transversely polarized gluons. Top: Collinear part in a parton distribution. Bottom: Collinear part in a fragmentation 
function. 
 } 
\end{figure}

The factorization formula in case of hadron pair production in $e^+e^-$ annihilation involves the transverse momentum convolution of 
two fragmentation functions (since there is no hadronic initial state in this process). The factorized formula for the relevant hadronic tensor is obtained by applying the appropriate Ward identities for the longitudinally polarized gluons exchanged between leading 
regions shown in figure \ref{TMDgraphs1} (left graph). If the momentum entering regions $C_A$, $C_B$ and $S$ is denoted 
respectively by $k_A$, $k_B$ and $k_S$, then the factorized formula is given by (we denote $C_{A}$ by 
$A$,  and $C_B$ by $B$ for clarity)  
\beq
W^{\mu\nu} = \int d^4k_A \, d^4k_B \, d^4k_S\,  A(k_A) \, B(k_B)\, S(k_S) \, H^{\mu\nu}(q) 
\delta^{(4)}\!(q\!-k_A\!\!-k_B\!\!-k_S).
\eeq
The delta function can be used to fix $k_{S,\perp}$, $k^+_A$ and $k_B^-$. One furthermore makes the 
approximation of ignoring $k_A^-$ ($k_B^+$) everywhere but in $A$ ($B$), and ignoring $k_S^{\pm}$ 
everywhere but in $S$. These approximations are allowed since the corrections are power-suppressed
at least as $m^2/Q^2$.
The integrals over these variables can  then all be short circuited and one gets
\beq
W^{\mu\nu}\!\! &=& \!\! \int d^2k_{A,\perp} d^2k_{B,\perp} \! \left ( \int dk_A^-A(k_A)\right)\!\! \left ( \int dk_B^- B(k_B)\right ) \!\!\left ( \int d k_S^+ dk_S^- S(k_S) \right )\!H^{\mu\nu}(q) \nonumber \\
&=&  \!\!\int d^2k_{A,\perp} d^2k_{B,\perp} \,A(z_A,k_{A,\perp})\, B(z_B,k_{B,\perp})\, S(q_\perp\!\! -  k_{A,\perp}\!\! -k_{B,\perp})
H^{\mu\nu}(q).
\label{pretmdepem}
\eeq 
Each respective factor in the parentheses gives the basic operator definition of the fragmentation 
functions and the soft factor. 
We mentioned in sections \ref{sec:powercount} and \ref{sec:simplefact} that each given PSS contains subtractions of the smaller PSSs.  Thus the collinear 
factors $A$ and $B$ in \eqref{pretmdepem} contain subtractions of the soft region.  
Now, the unsubtracted collinear 
parts contain Wilson lines which arise from the factorized gluon couplings to the hard part $H$. 
This is done by using the approximation in \eqref{HtoAapprox}, 
rewriting this as in \eqref{AtoSapprox} and applying the Ward identities.  For the $A$-to-$H$ couplings, the approximated momenta 
from \eqref{HtoAapprox} are $\hat{k}_A = (k^+,0^-,0_\perp)=(k\cdot n_B)\, n_A$ and therefore we get a Wilson line 
in the direction $n_B$:
\beq
W(x;n_B) = P \exp \left (-ig_s \int_0^\infty d \lambda \, A(x+n_B \lambda)\cdot n_B \right ).
\label{wilsonnb}
\eeq
For the $B$ part we instead get a Wilson line in the direction $n_A$.  In figure \ref{unsubtracted} we 
graphically represent the unsubtracted collinear part, including the Wilson line \eqref{wilsonnb} 
shown by double lines, for both a parton distribution (top two graphs) and a fragmentation function (bottom two graphs). 
The color representation of the Wilson line \eqref{wilsonnb} is determined by the particle at the end of 
the double lines in figure \ref{unsubtracted}: Fundamental for a quark (top and bottom left), adjoint 
for a gluon (top and bottom right). 

\begin{figure}[t]
\begin{center}
\includegraphics[angle=0, scale=0.65]{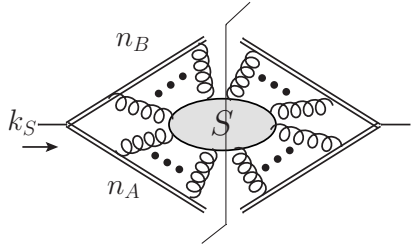}
\end{center}
\caption{\label{softfigure}  The factorized soft part. On each side of the cut, the gluons that couple 
to regions $A$ and $B$ are factorized into Wilson lines in the directions $n_A$ and $n_B$ respectively. 
}
\end{figure}

The soft gluons are similarly summed into Wilson lines using \eqref{AtoSapprox}.  From the $A$ side we 
see we get a line in the direction of $n_A$ while from the $B$ side we instead get a line in the direction of 
$n_B$. The definition of the collinear part involves always the hadron state $|P\rangle$, either as incoming 
(for a parton distribution) or as outgoing (for a fragmentation function). The soft factor on the other hand does 
not contain such a hadron so it is defined as a vacuum expectation value which we represent in figure 
\ref{softfigure}.

 As seen from \eqref{pretmdepem}, it is convenient to make a  Fourier transform into transverse coordinate $b_\perp$ to obtain 
\beq
W^{\mu\nu} = \int d^2b_\perp e^{-iq_\perp \cdot b_\perp} A(z_A,b_{\perp}) B(z_B,b_{\perp}) S(b_\perp)
H^{\mu\nu}(q)
\eeq
which is simpler than the momentum convolution written above. 

\begin{figure}[t]
\begin{center}
\includegraphics[angle=0, scale=0.37]{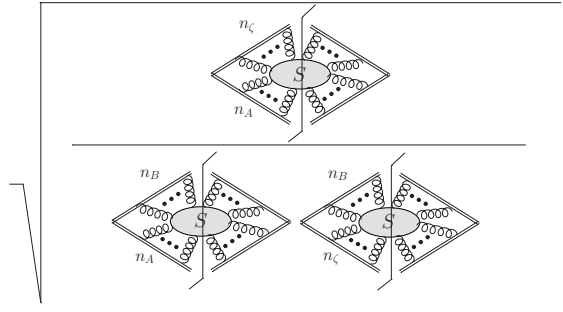}
\end{center}
\caption{\label{subtractedpdf}  The soft factor absorbed into the unsubtracted parton 
distributions and fragmentation functions. In the final result, $n_A$ and $n_B$ can be taken 
exactly light-like since the rapidity divergences cancel those in the unsubtracted collinear factor. 
The vector $n_\zeta$ cannot be taken light-like, however. 
}
\end{figure}

In the final definition, the soft factor is absorbed completely into the collinear factors to define 
the final subtracted fragmentation functions given by \cite{qcdbook}
\beq
D_{H_A/f}(z_A,b_{\perp}; \zeta, \mu) = D^{\mathrm{unsub}}_{H_A/f}(z_A,b_\perp;n_B)
 \! \times \! \sqrt{\frac{S(b_\perp;n_A,n_\zeta)}{S(b_\perp;n_A,n_B)S(b_\perp;n_\zeta,n_B)}}
 \times Z
 \label{subtractedtmd}
\eeq
Here $n_A$ and $n_B$ are taken light-like, and $Z$ is the UV renormalization factor. 
The somewhat strange looking factor in the square root 
is shown in figure \ref{subtractedpdf}. The precise motivation for it is described in detail in 
\cite[Ch.\ 13]{qcdbook}. The final definition is free from divergences associated with Wilson line self-energy 
corrections. 
The vector $n_\zeta$ defining the directions of the Wilson line 
in the soft factors serves as the rapidity cut-off which we indicate by the $\zeta$ dependence of the 
fragmentation function. The unsubtracted factor $D^{\mathrm{unsub}}$ is given exactly by the 
factors in figure \ref{unsubtracted} (bottom left graph in current example), defined in addition with integral over $k^-$ 
as in \eqref{pretmdepem}, and the Fourier transform from $k_\perp$ to $b_\perp$. A similar definition 
applies for the second fragmentation function associated with the region $B$. 
The final factorization formula then reads
\beq
W^{\mu\nu} \!\!\propto \frac{z_A z_B}{Q^2}H_f^{\mu\nu}(Q; \mu)\!  \int d^2 b_{\perp} e^{-iq_\perp \cdot b_\perp}
D_{H_A/f}(z_A,b_{\perp}; \zeta, \mu) \, D_{H_B/\bar{f}}(z_B,b_\perp; \zeta, \mu),
\label{tmdepem}
\eeq
where $z_{A,B}=p_{A,B}/k_{A,B}$, and  
\beq
H_f^{\mu\nu} = \mathrm{Tr} \, \slashed{\hat{k}}_AH_f^\nu \slashed{\hat{k}}_B H^{\mu \dagger}_f.
\label{hardcoeff}
\eeq
$H^\nu$ and $H^{\mu \dagger}_f$ 
stand for the hard blobs shown in figure \ref{TMDgraphs1}, defined to be irreducible in the collinear lines, 
and containing subtractions of the collinear and soft regions, just like in \eqref{subtractedfact}. 

The tensor $W^{\mu\nu}$ of course 
cannot depend on the rapidity cut-off $\zeta$, and this requirement is embedded in the Collins-Soper evolution equation of the 
fragmentation functions with respect to $\zeta$. 
In  SIDIS we instead have a convolution of one parton distribution (for the incoming target hadron) and one fragmentation function (for the final state hadron), 
\beq
W^{\mu\nu} \!\!\propto \frac{z}{Q^2}H_f^{\mu\nu}(Q; \mu)\!  \int d^2 b_{\perp} e^{-iq_\perp \cdot b_\perp}
f_{f/H_A}(x,b_{\perp}; \zeta, \mu) \, D_{H_B/\bar{f}}(z,b_\perp; \zeta, \mu)
\label{sidis}
\eeq
where $H_f^{\mu\nu}$ is given by the same expression as in \eqref{hardcoeff} (but of course the hard 
factors $H^{\mu \dagger}$ and $H^\nu$ are different in $e^+e^-$ and DIS), and $x=k_A^+/p_A^+$ 
and $z=p_B^-/k_B^-$. Thus the change is that one fragmentation function is simply 
exchanged for the parton distribution function of the target hadron.  The parton distribution $f$ is defined 
exactly as in \eqref{subtractedtmd} to include the soft factors, one simply needs to change 
$D^{\mathrm{unsub}}$ to $f^{\mathrm{unsub}}$ which means (for quarks) replacing the bottom 
left graph in figure \ref{unsubtracted} with the top left one. 

Finally in the Drell-Yan process we instead 
have two parton distributions and there is no fragmentation function since the observed final state is leptonic.
Thus
\beq
W^{\mu\nu} \!\!\propto \frac{s}{Q^2}H_f^{\mu\nu}(Q; \mu)\!  \int d^2 b_{\perp} e^{-iq_\perp \cdot b_\perp}
f_{f/H_A}(x_A,b_{\perp}; \zeta, \mu) \, f_{\bar{f}/H_B}(x_B,b_\perp; \zeta, \mu)
\label{drellyan}
\eeq
where now the hard coefficient $H^{\mu\nu}$ is the tensor for the on-shell partonic reaction $f\bar{f} \to \gamma^*$.
The extra factor $s$ in front of the integral arises from the definition of the hadronic tensor for the Drell-Yan
process which reads
\beq
W^{\mu\nu} = s \int d^4x \, e^{iq\cdot x} \langle p_A,p_B| J^\mu(x) J^\nu(0) |p_A,p_B\rangle.
\eeq

In order to obtain a reliable estimate of $H^{\mu\nu}$ it is optimal to let $\mu \sim Q$ so as to avoid 
large logarithms. The higher order corrections are then subleading 
in factors of $\alpha_s(\mu\sim Q) \ll 1$ without any logarithmic enhancements, and thus fixed order perturbative calculations
are reliable. Notice again that in all formulas above, the hard tensor $H^{\mu\nu}$ is always outside the transverse 
momentum (or coordinate) integral and the lines entering it are on-shell. 


Thus we see that the TMD parton distributions or fragmentation functions, compared to the basic parton model 
definitions, depend additionally on the variables $\zeta$ and $\mu$. They consequently satisfy evolution equations 
with respect to both these variables. The evolution in $\mu$ is given by the standard DGLAP equations while the evolution with respect to the rapidity variable $\zeta$ is given by the (Collins-Soper) CS evolution equation \cite{qcdbook}. 
The CS kernel controlling the rapidity evolution is the same for all the above reactions because it is determined by the 
soft factor which is the same in all the above examples.  

We have above outlined the fundamentals of factorization in QCD, in processes where 
a hard scale $Q$ is present, and where the collinear directions scale with $Q$.  In the small-$x$ 
region there may or may not be present a hard scale. The traditional process to study is small angle two-particle
elastic scattering where the momentum transfer $t$ is much smaller than the cms energy $s$, and 
where the collinear momenta scale with $\sqrt{s}$.  In this case the hard region, if present, has a scale 
$Q$ which is fixed, and is therefore not proportional to the asymptotic variable $\sqrt{s}$. The leading regions 
are therefore somewhat different than in the hard scattering factorization. We will outline the relevant 
regions for the small-$x$ case in section \ref{sec:diffcases} where we examine single inclusive particle 
production. We now go through the main formulations of $k_\perp$-factorization 
in the BFKL and CGC formalisms, and compare these to the hard scattering case just discussed.

\subsection{Factorization in BFKL}
\label{sec:bfklfact}

``Factorization" in the BFKL formalism refers to the Regge factorization in which a given 
$2\to n$ scattering amplitude is, in the asymptotic limit $s\to \infty$, written as a factorized product of effective vertices and 
couplings of ``reggeized gluons". This is known as the ``multi-Regge form". 
The arguments for the factorized form of the $2\to n$ 
amplitudes  go back to the pre-QCD days of Regge theory, and the so-called ``multi-peripheral" models 
\cite{Campbell:1970wy, Weis:1972ir, Bartels:1974tj, Bartels:1974tk}.  

\begin{figure}[t]
\begin{center}
\includegraphics[angle=0, scale=0.55]{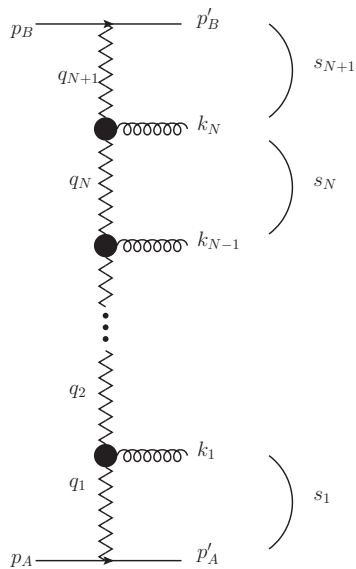}
\end{center}
\caption{\label{multiregge} The multi-Regge-factorized form of the scattering amplitude in BFKL. } 
\end{figure}

We illustrate the multi-Regge form in figure \ref{multiregge}. Here the zig-zag lines denote the 
Reggeons, and each black circle denotes the Reggeon-Reggeon-gluon vertex. Figure \ref{multiregge} 
is in the Regge theory valid when $s_i \to \infty$ for all $i$ \cite{Bartels:1974tj, Bartels:1974tk}. 

\begin{figure}[t]
\begin{center}
\includegraphics[angle=0, scale=0.55]{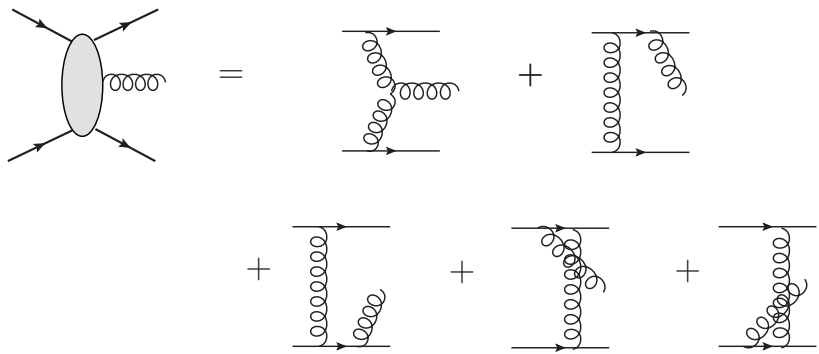}
\end{center}
\caption{\label{lipvert1} Graphs contributing to Lipatov vertex. } 
\end{figure}

In BFKL, the vertical zig-zag lines in figure \ref{multiregge} are  given by gluons whose propagators (in Feynman 
gauge) are obtained by 
\beq
P_{\mu\nu}(q_i) = \frac{-ig_{\mu\nu}}{q_i^2+i\epsilon} \to P_{\mu\nu}(q_i, s_i) =  \frac{-ig_{\mu\nu}}{q_i^2+i\epsilon} 
\left ( \frac{s_i}{q_{\perp,i}^2} \right )^{\omega(q_{\perp, i}^2)-1}
\label{reggegluon}
\eeq
where 
\beq
\omega(q_{\perp, i}^2)-1 = \alpha_s N_c \int \frac{d^{2+2\epsilon}\kappa_\perp}{(2\pi)^{2+2\epsilon}}  \frac{-q_{\perp, i}^2}{\kappa^2_\perp(\kappa_\perp-q_{\perp, i})^2},
\label{Reggefactor}
\eeq
The function $\omega$ is called the ``gluon Regge trajectory".  The vertices in figure \ref{multiregge} 
are  given by the Lipatov vertex which is an effective three-gluon 
vertex. The Lipatov vertex is  derived from the tree-level graphs of the $2\to 3$ partonic amplitude
shown in figure \ref{lipvert1}. The external partons 
may be quarks or gluons, the use of the eikonal approximations implies that the vertex is 
independent of the flavor of these particles (at least when the external particles are individual quarks or gluons). 

The fundamental assertion of the BFKL formalism is that the multi-Regge form shown in figure \ref{multiregge} 
is valid for all $2 \to n$ amplitudes. It has been argued in reference \cite{Fadin:2006bj} that the multi-Regge
result can be shown to be correct to all orders, 
once it has been shown to be correct to one-loop order for \emph{all} $2\to n$ amplitudes, by essentially using the same techniques 
($s$-channel unitarity relations) developed in Regge theory in \cite{Bartels:1974tj, Bartels:1974tk}.  We are, however, not aware of any explicit higher order calculations in 
QCD of the $2 \to n$ amplitudes for $n > 3$. For the $2\to 3$ amplitude, the multi-Regge form has been derived 
in reference  \cite{Fadin:1993wh} up to one-loop corrections to the graphs in figure \ref{lipvert1}.

As we saw in the previous section, factorization has to be shown to hold for all orders. In section 
\ref{sec:tmdgluon} we shall show some examples of higher order corrections where TMD factorization 
is know to be violated. Since the multi-Regge formula leads to a $k_\perp$-factorized form (see further \cite{ourpaper}) it is of relevance to consider such higher order graphs. As we will see, the breakdown of factorization might be hidden until 
higher order corrections. For example, figures \ref{2to2gentmd}, \ref{2to2nogentmd} and  \ref{2to2nogentmd2} 
show that
factorization breakdown is not  visible until 4 and 5 gluon exchange in the $2\to 2$ amplitudes. 
If we consider only one side of the cut $2\to 2$ amplitude, then factorization breaking graphs appear 
in 2 or 3 loop corrections to the $2\to 4$ amplitude. Similar factorization breaking terms might be present 
in the $2\to 3$ gluon amplitudes at 2 loop corrections as well. 
It may therefore very well be that one-loop corrections do not exhibit any  TMD factorization breaking.

\subsection{Factorization in the CGC}
\label{sec:cgcfact}

The Color Glass Condensate (CGC) \cite{JalilianMarian:1997jx, JalilianMarian:1997gr, JalilianMarian:1997dw, Iancu:2000hn, 
Iancu:2001ad, Ferreiro:2001qy} is a semi-classical approach developed to deal with the 
QCD physics of ``large" objects such as heavy ions. 

The set-up of the CGC formalism is rather different than the hard scattering factorization. The main assertion 
here is that the color degrees of freedom of a given hadron, such as a large nucleus,  can be
described by classical fields generated by 
a distribution of random color sources, $\rho_a$ ($a$ being the color index), which arise due to the ``fast'' moving partons, 
i.e., those partons which are in the collinear region.   These then act as sources for the 
softer gluons whose dynamics depend on the classical sources.

\subsubsection{Basics of CGC}
\label{sec:cgcbasic}

The classical fields generated by these sources are determined by the solutions to the classical equations
of motion 
\begin{eqnarray}
D_\nu F_a^{\mu\nu}(x) = J^\mu_a,
\label{YMeqs}
\end{eqnarray}
with $D_\nu$ the usual covariant derivative. 
The generic solutions to \eqref{YMeqs} give classical fields $A^a_{cl}$ that are highly non-linear in the sources $\rho_a$.
In the classical McLerran-Venugopalan (MV) model \cite{McLerran:1993ni, McLerran:1993ka}, the sources are assumed to 
originate from the valence quarks of the nucleons which are randomly distributed according to some 
weight functional, $W[\rho]$.  This is the distribution we encountered earlier in equations \eqref{cgcaverage} 
and \eqref{cgcunity}. 

In the case of a single particle traveling in the plus direction, the classical current is taken as 
\beq
J_a^\mu(x) = \delta^{\mu +} g_s\, \rho_a(x^-\!\!,x_\perp)
\label{onecurrent}
\eeq
where the classical source $\rho(x^-,x_\perp)$ has a very narrow support in $x^-$. In the case of 
two particle scattering, with the incoming hadrons traveling along the opposite light-cones, one takes 
instead
\beq
J^\mu_a(x) = \delta^{\mu +}g_s\, \rho_{1,a}(x^-\!\!,x_\perp) + \delta^{\mu -}g_s\,  \rho_{2,a}(x^+\!\!,x_\perp).
\label{twocurrent}
\eeq

The model is defined at some scale $\Lambda^{\hat{\mu}}$ which sets the applicability of the classical description. 
Here $\hat{\mu} = +$ or $\hat{\mu} = -$. 
For a hadron with large momentum $P^\mu$ along the direction $\hat{\mu}$, this means that all 
fields with $k^{\hat{\mu}} > \Lambda^{\hat{\mu}}$ are taken to be described by the classical sources $\rho$.  
The distribution $W[\rho]$ is therefore specified at the scale $\Lambda^{\hat{\mu}}$.
Physical quantities of interest in the model are calculated by functional averages using the classical 
distribution $W[\rho]$ as in \eqref{cgcaverage} for a single hadron, and 
\beq
\langle \mcal{O} \rangle = \int D\rho_1 \,D\rho_2 \, W_{\Lambda_1^+}[\rho_1] \,  W_{\Lambda_2^-}[\rho_2] 
\,\mcal{O}[\rho_1,\rho_2],
\label{twohadronav}
\eeq
in two hadron scattering. Of course, \eqref{twohadronav} is already in a factorized form. 

\subsubsection{Power counting and ``dilute" and ``dense" systems}
\label{sec:cgcpowercount}

The treatment of two particle processes is then based on a power counting argument of the 
classical sources $g_s\, \rho$. A ``dilute" particle in this power counting is defined to be 
one described by a source such that $|g_s \, \rho | \ll 1$ . For such a particle then, in 
the calculations
only the first order dependence $(g_s\, \rho)^1$ is kept. Given a functional
$\mcal{O}[\rho_1,\rho_2]$ which depends on both $\rho_1$ and $\rho_2$, expand it as a polynomial
\beq
\mcal{O}[\rho_1,\rho_2] = \sum_{n=1}^\infty \sum_{m=1}^\infty \mcal{O}_{n m} \,
(g_s \, \rho_1)^n (g_s \, \rho_2)^m.
\eeq
The definition of  particle 1 being dilute  then means that 
\beq
\mcal{O}[\rho_1,\rho_2]  \to\biggl . \mcal{O}[\rho_1,\rho_2] \biggr \vert_{\mathrm{1,dilute}} =  \sum_{m=1}^\infty \mcal{O}_{1 m} \,
(g_s \, \rho_1) (g_s \, \rho_2)^m. 
\eeq

Conversely a particle is defined to be ``dense" if it is described by a source satisfying $|g_s \, \rho | \sim 1$. 
In that case, the dependence on $g_s\, \rho$ is retained to all orders. As for real particles, a proton 
or a deuteron is defined as being ``dilute", while heavy ions such as gold or lead nuclei are defined 
to be ``dense". 
Thus ``dilute-dilute" scattering refers essentially to $pp$ or $p\bar{p}$ scattering, while ``dilute-dense" 
scattering refers to $pA$ or deuteron-Nucleus ($dA$) collisions, and finally ``dense-dense" scattering refers 
to $AA$ collisions (lead-lead or gold-gold).  Of course
a proton in the CGC becomes ``dense" at sufficiently high energies since the classical sources grow 
as a function of energy.  

In this setting, the quantum evolution is based on the logic of the leading logarithmic approximation (LLA)
where the coupling $g_s$ is fixed and small, $g_s \ll 1$. Therefore for a ``dilute" object we have 
$\rho \lesssim 1$, while for a ``dense" object we have $\rho \sim 1/g_s \gg 1$.  These assumptions lead to the formulation of factorization 
in the CGC approach \cite{Gelis:2003vh, Blaizot:2004wu, Gelis:2006yv, Gelis:2006dv, Gelis:2008rw, Gelis:2008sz}. 

\begin{figure}[t]
\begin{center}
\includegraphics[angle=0, scale=0.55]{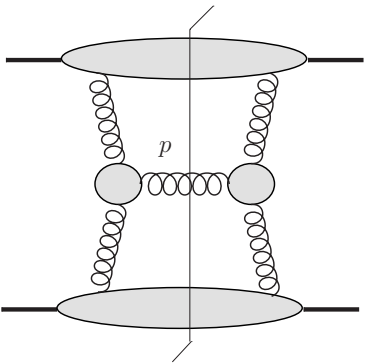}
\end{center}
\caption{\label{dilutefigure} Diagrammatic representation of  particle production in ``dilute-dilute"
scattering in the language of CGC. } 
\end{figure}

We immediately notice that this power counting is rather different in logic than the power 
counting  described in section \ref{sec:powercount}.  Here the emphasis is put on the 
classical source $\rho(x)$ specified in space-time coordinates.  Any correction beyond the classical approximation 
is calculated to order $g_s^2$ which amounts to a one-loop calculation.  For processes involving protons 
then, calculations are kept at linear order in $g_s \rho$ for each proton which in a diagrammatic analogy means that 
at most two gluon couplings are considered. In figure \ref{dilutefigure} we show an example of single inclusive gluon production in ``dilute-dilute" scattering. Thus in the dilute limit factorization is essentially identical to that in the parton model 
we considered in section \ref{sec:partonmod}. 

In general, however,  the extra gluon emissions 
between the different PSSs considered in section \ref{sec:powercount} all have small virtualities and 
they therefore couple strongly. In particular the soft gluons have all their momentum components small, 
and the QCD coupling of these gluons is therefore strong. That is, we do not have a situation where $g_s \ll 1$. 
Even in the case of a weak coupling at all relevant scales, however, such as in  QED, is the formalism outlined
in \ref{sec:powercount} and the factorization theorems rather useful for controlling the higher order corrections 
which still might be enhanced by kinematical factors. 

In the CGC higher order corrections are needed because the classical sources $\rho$ 
can have large values, $|\rho| \gg 1$, but $g_s$ itself is always small. Pure perturbative calculations 
are thus performed when $|\rho| \lesssim 1$, which happens in the case of ``dilute" particles. In the general 
treatment of factorization in QCD, or in generic field theories, however,  large contributions arise from 
surfaces in the multi-dimensional space of momentum integrals where the integration contours 
are forced to go close to the singularities of the propagators, the pinch singular surfaces. In QCD 
the momentum lines in the PSSs have large couplings.  This is the reason why factorization must be 
proven to all orders, and it is then convenient to employ the power counting analysis of the PSSs. 
Corrections are then guaranteed to be power suppressed in the large scale $Q$. In the case of small-$x$ 
therefore, ideally we would want to formulate factorization (``$k_\perp$-factorization") up to power 
suppressed corrections in $\sqrt{s}$. 

Of course the treatment of factorization cannot be purely perturbative for the reasons just explained. 
It is important to emphasize that the power counting methods of section \ref{sec:powercount} rely 
generically on dimensional analysis and Lorentz invariance, and thus not exclusively on perturbation theory. The 
explicit calculations are of course performed using Feynman graphs, but the structures 
obtained have a meaning beyond strict perturbation theory. One can therefore apply the same 
methods to the small-$x$ region where any hard scale might be absent. 

\subsubsection{The LLA and basic logic of factorization}
\label{sec:cgclla}

As the LLA is important for the formulation of factorization in the CGC, we shortly outline the logic
behind it.  An all order result can be obtained by 
calculating the one-loop graphs using the eikonal approximation, and then exponentiating the result. 
If the one loop result for a certain process is $\Gamma_1$, and 
the tree level result is $\Gamma_0$, then usually one finds 
\beq
\Gamma_1 = g_s^2 \int^Y_0 dy \, K_s(y) \, \cdot \Gamma_0,
\label{gammaoneloop}
\eeq
where $dy =\frac{dk^+}{k^+}$ and the limits on $y$ are determined by the kinematics of the given process. 
The kernel $K_s$ is found by applying the approximations appropriate for a soft term. 
We can then write the complete result up to one loop as
\beq
\Gamma_0 + \Gamma_1 = \left ( 1 +g_s^2 \int^Y_0 \!dy \, K_s(y) \right ) \Gamma_0.
\eeq
For infinitesimal change in the scale we can write this as
\beq
\Gamma_{dY} =  \left ( 1 +g_s^2 dY \, K_s(dY) \right )\Gamma_0,
\label{gammady}
\eeq
so that 
\beq
\frac{\Gamma_{dY}-\Gamma_{0}}{dY} = g_s^2\, K_s \,\Gamma_0.
\eeq
This gives the all order LLA result 
\beq
\Gamma_Y^{LLA} = \exp \left ( g_s^2 \int^Y_0 \!dy \, K_s(y) \right ) \Gamma_0.
\label{gammalla}
\eeq

A similar construction is used in the CGC
 \cite{Gelis:2003vh, Blaizot:2004wu, Gelis:2006yv, Gelis:2006dv, Gelis:2008rw, Gelis:2008sz}.  
The idea is to start with a formula at the classical level, where the correlator of the classical 
fields is calculated using \eqref{twohadronav}, and then to perform a one loop calculation as in \eqref{gammaoneloop}
and show that at this level the classical structure \eqref{twohadronav} still holds. The resulting 
one loop formula can then be resummed as in \eqref{gammady} and \eqref{gammalla} to obtain 
a final formula in the LLA. 

\begin{figure}[t]
\begin{center}
\includegraphics[angle=0, scale=0.55]{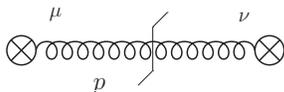}
\end{center}
\caption{\label{CGCparticleprod} Particle production by classical sources in the CGC. The crosses denote the classical field 
insertions. }
\end{figure}

Take for example the single inclusive particle production in the scattering of two hadrons, described by sources 
$\rho_1$ and $\rho_2$, which is studied in \cite{Gelis:2003vh, Blaizot:2004wu, Gelis:2006yv, Gelis:2006dv, Gelis:2008rw, Gelis:2008sz}.  The basic classical formula which is equivalent to a tree level 
calculation is given by   
\beq
\lan E_p \frac{dN}{d^3p} \ran = \frac{1}{2 (2\pi)^3} \sum_\lambda \lan |\mcal{M}_\lambda(p)|^2 \ran 
\label{classicalglueprod}
\eeq
where 
\beq
\mcal{M}_\lambda(p) =p^2 A^{cl}_\mu(p) \epsilon^{*\mu}_{(\lambda)} (p).
\eeq
We illustrate this in figure \ref{CGCparticleprod} where the crosses denote the insertions of the 
classical fields $A^{cl}(p)$.  Note that $A^{cl}(p)$ 
is a function of both $\rho_1$ and $\rho_2$ so it contains the effects of both hadrons.  At the pure classical 
level, one evaluates \eqref{classicalglueprod} using \eqref{twohadronav}. This gives 
\beq
\langle A_\nu A_\mu \rangle_0 = \int D \rho_1 \, D \rho_2\, W_{\Lambda^+}[\rho_1] \, W_{\Lambda^-}[\rho_2] \, 
(A^{cl}_\nu A^{cl}_\mu)[\rho_1,\rho_2]
\eeq
where the subscript on the left hand side is to denote that this corresponds to the tree level calculation. 
The weight functionals can at this level be fully parametrized using the MV model from which an explicit result 
can be obtained for \eqref{classicalglueprod}.  

The one-loop correction to the tree level result is then found to be \cite{Gelis:2008rw, Gelis:2008sz}
\beq
\langle A_\nu A_\mu \rangle_1 \!=\! \int D \rho_1 \, D \rho_2 \, W_{\Lambda^+}[\rho_1] \, W_{\Lambda^-}[\rho_2] 
\left [ \ln\frac{\Lambda^+}{p^+} H_1 + \ln\frac{\Lambda^-}{p^-} H_2 \right ] (A^{cl}_\nu A^{cl}_\mu)[\rho_1,\rho_2]
\label{cgconeloop}
\eeq
where each $H_i$ corresponds to the ``JIMWLK Hamiltonian". Each  $H_i$ 
is a Hermitian differential kernel \cite{Iancu:2002xk} (in the sense of functional differentiation) that acts on the classical fields $A^{cl}_\nu A^{cl}_\mu$ in \eqref{cgconeloop}. 
We see that this result is analogous to 
\eqref{gammaoneloop}. To understand the logarithmic factors in \eqref{cgconeloop}, note that if $K_s(y)$ 
is independent of $y$ (which it nearly always is), then the integral in \eqref{gammaoneloop} simply gives 
$Y\cdot K_s$. The rapidity $Y$ exactly corresponds to the logarithmic factors in \eqref{cgconeloop} and 
we see that $K_s$ corresponds\footnote{The JIMWLK Hamiltonian is of order $g_s^2$ in the quantum fluctuations, but it contains the classical sources
$g_s \, \rho$ to all orders. } to $H_i$.   Using that the $H_i$ are Hermitian one can then rewrite 
the complete one-loop result \eqref{cgconeloop} as \cite{Gelis:2008rw, Gelis:2008sz}
\beq
\langle A_\nu A_\mu \rangle_1 \! +\! \langle A_\nu A_\mu \rangle_0  \!\!&=& \!\!
\int D \rho_1 \, D \rho_2 \left ( 1 + \ln\frac{\Lambda^+}{p^+} H_1 \right) W_{\Lambda^+} [\rho_1] 
\left (1 + \ln\frac{\Lambda^-}{p^-} H_2 \right) W_{\Lambda^-}[\rho_2]  A^{cl}_\nu A^{cl}_\mu \nonumber \\
&\equiv& \!\! \int D \rho_1 \, D \rho_2\,  W_{p^+} [\rho_1] W_{p^-} [\rho_2] (A^{cl}_\nu A^{cl}_\mu)[\rho_1,\rho_2]
\label{cgcfact}
\eeq
In this rewriting one uses that formally the term containing the product $H_1H_2$ in \eqref{cgcfact}
is of higher order (it is not of LLA) and thus neglected. 
One then gets exactly as in \eqref{gammalla} the LLA result
\beq
W_{dY}=(1+dY\, H)W_0 \to W^{LLA}_Y = \exp \left ( \int^Y_0 dy \, H(y) \right ) W_0.
\eeq
Equation \eqref{cgcfact} is referred to as the ``high energy factorization", or ``JIMWLK factorization", formula \cite{Gelis:2003vh, Blaizot:2004wu, Gelis:2006yv, Gelis:2006dv, Gelis:2008rw, Gelis:2008sz}. 

\subsubsection{Comparison to TMD factorization}
\label{sec:cgctmd}

In its derivation, \eqref{cgcfact} is rather different than the TMD factorization described in section \ref{sec:TMD}. 
For example, in \eqref{cgcfact} there is a factorized product of the classical weight functionals $W[\rho_i]$ 
rather than a product of parton distributions and/or fragmentation functions.

Equation \eqref{cgcfact} is in the literature implied to be a generalization of ordinary TMD factorization. 
In section 5 of reference \cite{Gelis:2008rw} we can for example read that \\
``\emph{JIMWLK factorization proven here is far more general and robust in 
comparison to the $k_\perp$-factorization often discussed in the literature.}" 

The statement on the wider generality of the CGC formula is motivated by the observation that 
 one can for ``dilute" systems obtain from \eqref{cgcfact} a formula which looks like a $k_\perp$-factorized formula. 
Since this ``dilute" limit involves a simplified approximation within the CGC formalism, it is therefore said 
that \eqref{cgcfact} is more general.  For example, 
for the single inclusive gluon production using \eqref{classicalglueprod} and \eqref{cgcfact} one gets
in the ``dilute" limit a formula that looks like equation \eqref{GLRfact} below which is the $k_\perp$-factorization formula canonically 
used in the small-$x$ region.  Moreover,
within the CGC, the TMD gluon distribution can be calculated explicitly if $W[\rho]$ is given. For example, the 
WW gluon distribution  can be calculated from \eqref{WWdistrbadj} once $W[\rho]$ is specified. The converse 
statement on the other hand is not true: It is not enough to have an explicit formula for \eqref{WWdistrbadj} 
in order to extract $W[\rho]$ uniquely. 

In this sense, it can indeed be said that  \eqref{cgcfact} is more general than the TMD factorization. However, 
from a different perspective we find that this statement is misleading and not correct.  Moreover, as we shall explain now,
the factorization explained in section \ref{sec:hardscatfact} is actually more robust. 

Equation \eqref{cgcfact} is namely only derived at one loop order using the logic of the 
LLA while the TMD factorization is much more general and accurate than that.  
The LLA result for example
gives no hint at all as to what the higher order corrections might look like. 
There are even instances where it gives the wrong result, 
even qualitatively, an example being the Drell-Yan cross section at zero transverse momentum where 
the LLA gives a vanishing result while the true result that can be obtained from the factorization approach
is non-zero \cite{Collins:1981va}. 
Contrary to the LLA, in the factorization 
approach the higher order corrections are well controlled, and even if the explicit calculations of the higher order 
corrections can be difficult in practice, one can nevertheless make reliable estimates of their importance \cite{qcdbook}. 
It is therefore not correct to say that the ``JIMWLK factorization" is more robust than the TMD factorization. 
In fact the opposite is clearly true with regards to the accuracy of the derivation. 

Moreover, when in the CGC the dilute limit is taken, the TMD gluon distribution that appears in the 
factorization formula is given by \cite{Iancu:2002xk, Blaizot:2004wu, Gelis:2008rw}
\beq
\bigl . f(x,k_\perp; \zeta) \bigr \vert_{\mathrm{dilute}} \!\!\!&=& \! \frac{1}{k_\perp^2}\langle \rho(k)\rho (-k) \rangle_{W_{\zeta P^+}} \!\!= 
\langle F^{+i}(k)F^{+i}(-k) \rangle_{W_{\zeta P^+}} \nonumber \\
&=& \!\!\! \int d^3x \, d^3y \, e^{ixP^+ (x^-\!\! -y^-) - ik_\perp (x_\perp-y_\perp)}
\langle F_a^{+i}(x)F_a^{+i}(y)\rangle_{W_{\zeta P^+}}.
\label{dilutegluon}
\eeq
The subscripts on the correlators imply that the averages using $W[\rho]$ are performed at the scale $\zeta P^+$.
Acting with the dilute limit of the JIMWLK Hamiltonian on the classical sources in \eqref{dilutegluon} one then recovers 
the BFKL equation  for the object $f(x,k_\perp; \zeta)$ (for a simple demonstration of this, see \cite{Iancu:2002xk}). 
Thus the BFKL formalism can be identified with the 
dilute limit of the JIMWLK formalism. Since for example the CCH formalism \cite{Catani:1990xk, Catani:1990eg}   is based upon BFKL it is 
indeed correct that \eqref{cgcfact} presents a generalization of the work in \cite{Catani:1990xk, Catani:1990eg}. 
Moreover, as the work in \cite{Catani:1990xk, Catani:1990eg} is frequently referred to as the ``$k_\perp$-factorization" formula, 
in this sense (\emph{i.e.} if `$k_\perp$-factorization" is understood to refer to  \cite{Catani:1990xk, Catani:1990eg})  
\eqref{cgcfact} is more general than ``$k_\perp$-factorization". The CCH formalism is, however, also based on the 
LLA, and neglected terms are therefore not power-suppressed.

The argument for factorization in \cite{Catani:1990xk, Catani:1990eg} is  based on the use of the 
light-cone gauge (in DIS) or axial gauge (in hadron-hadron collisions).
The
final expression in \eqref{dilutegluon} actually equals the earlier light-cone gauge expression in \eqref{numberdens}. 
A similar definition also appears in the factorization approach as we discussed in reference to figure 
\ref{nongaugeDIS}.  It is, however, important to realize that \eqref{dilutegluon} is supposed to hold in the dilute limit for \emph{any} gauge, even a covariant gauge. 
This is in fact in line with the power counting we discussed in section \ref{sec:cgcpowercount}
above, where the definition of the dilute limit is that $g_s \, \rho \ll 1$. This is of course why equation \eqref{dilutegluon}
is second order only in $\rho$ (the first order term $\langle \rho \rangle$ vanishes when, as usual, the distribution 
$W[\rho]$ is taken to be a Gaussian). 

It is then important, however, to realize that the distribution thus obtained in \eqref{dilutegluon} is \emph{not} the 
TMD gluon distribution in the TMD factorization approach.  One cannot in the TMD factorization in covariant 
gauge simply drop the Wilson lines because as mentioned above, the soft gluons exchanged between different 
regions have strong coupling. The TMD factorization therefore does not correspond to the dilute limit of the 
CGC. The factorization \eqref{cgcfact} does indeed represent a different 
structure than the TMD factorization, but it cannot be said to be more general since it contains only a one-loop 
calculation while the TMD factorization is valid to leading power, rather than to leading logarithm. 

We want to emphasize that this point is important and not merely a technical detail.  The reason is that 
if we wish to establish factorization for a given process, then a possible breakdown of factorization 
may not show up until higher order corrections are considered, beyond the dilute limit. 
In section \ref{sec:tmdgluon} below we shall discuss this point in the context of the small-$x$ single 
inclusive gluon production formula. As we explain there for example, the factorization breaking graphs
studied in  \cite{Collins:2007jp, Rogers:2010dm} do not show up until one considers two gluon corrections 
to the parton model graphs, see figures \ref{2to2gentmd} and \ref{2to2nogentmd}. In terms of Feynman 
diagrams, the parton model graphs themselves are already at two loop order, so the factorization breaking does
not appear until 4 loop graphs. In the dilute limit considered above, or in the logic of the LLA, however, this would have been completely missed. 

It is therefore difficult to  discuss the validity of factorization at one loop order, or in a 
``dilute" approximation in the sense described in section \ref{sec:cgcpowercount}. In that case for 
example proton-proton collisions become rather trivial but the real situation is
far more complicated than that, as should be obvious from our discussion in section \ref{sec:hardscatfact}.

\subsubsection{Causality and factorization}

\begin{figure}[t]
\begin{center}
\includegraphics[angle=0, scale=0.45]{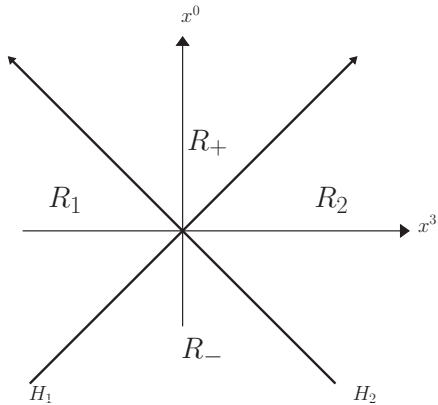}
\end{center}
\caption{\label{spacetime}  Space-time illustration of the scattering of two hadrons $H_1$ and $H_2$. In the 
classical solutions of CGC, one has independent solutions in the non-casually-connected regions 
$R_1$ and $R_2$. The solutions in the forward light-cone region $R_+$ are however non trivial and 
give rise to the so-called Glasma \protect \cite{Lappi:2006fp}.} 
\end{figure}

An argument given for the validity of \eqref{twohadronav} is based partly on causality (see 
for example \cite{Gelis:2008rw}), namely that two fast moving hadrons as shown in figure \ref{spacetime}
cannot interact with each other prior to the collision. This by itself, however, does not imply that there 
must be a factorized structure for the observable under study. In covariant gauge, it is true that the hadrons 
cannot interact prior to the collision, and they are therefore causally disconnected before the collision. To write a factorization 
formula, however, one must be able to factorize the soft emissions which can occur at late times after 
the collision. Even though the hadrons are casually disconnected prior to the scattering, the scattering 
might produce color entangled states which break factorization (see section \ref{sec:tmdgluon} below).

Moreover, the causality argument does not hold in ``physical gauges", such as the Coulomb gauge 
or the axial gauge, where manifest Lorentz invariance is broken and faster-than-light propagation is 
possible in individual graphs. The causality violating contributions should cancel in the final, physical 
results, but the proofs can be very non-trivial. It was in fact early reported  that \cite{Bodwin:1981fv, Bodwin:1988fs} 
the faster-than-light interactions in the physical gauges would correlate the hadrons prior to the collision and 
break factorization in hadron-hadron collisions such as in the Drell-Yan process.

Factorization, both collinear and TMD, in fact holds in Drell-Yan \cite{qcdbook}. The problematic gluons are 
precisely the Glauber (Coulomb) gluons which complicate the proofs. However, in covariant gauge one can 
consistently deform the integration contours away from the Glauber region and restore factorization. Whether 
this can be done for more complicated interactions is of course the real question. We discuss this more in 
section \ref{sec:gluonprod} below. What is clear, however, is that the proof of factorization is much more intricate 
than what general causality arguments would suggest.

\subsection{Hybrid formalisms}
\label{sec:hybrid}

Some of the applications of the CGC model falls into a category that we shall call the ``hybrid formalisms", since they combine the CGC 
treatment above with that of collinear hard scattering factorization (see e.g. \cite{Gelis:2002ki, Gelis:2002fw, Gelis:2002nn, Dumitru:2005gt, Dumitru:2005kb, Gelis:2006hy}). 
These formalisms are used especially in proton-nucleus ($pA$) collisions. 
Typical examples include photon production, Drell-Yan, and soft particle production in 
the forward region (all in $pA$ collisions). As we shall show here, however, these formalisms do not address 
the question whether there is factorization for the given process, and the validity of the proposal 
to mix collinear factorization with the CGC treatment is not at all clear to us.

\begin{figure}[t]
\begin{center}
\includegraphics[angle=0, scale=0.6]{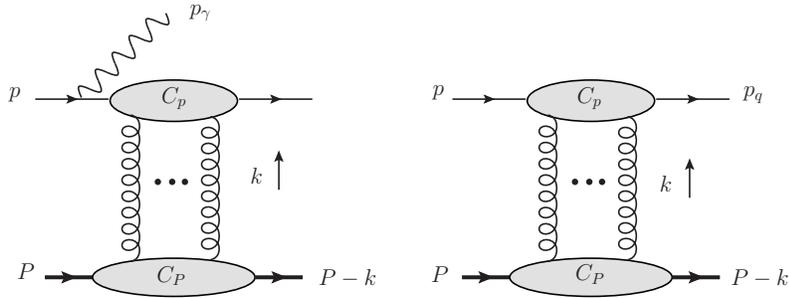}
\end{center}
\caption{\label{promptphoton} Examples of processes considered in the hybrid formalisms. Left: Photon 
production in quark-Nucleus scattering. Right: Quark production in quark-Nucleus scattering. } 
\end{figure}

We illustrate in figure \ref{promptphoton} two examples of the processes considered in this framework.  
The upper incoming line refers to a quark of momentum $p$ while the lower thick line with momentum $P$
refers to the nucleus. The proton is therefore not treated explicitly. Only interactions between the active 
quark and the nucleus are considered as indicated in the figure. The gluon attachments between the 
lower and the upper blobs is described by a Wilson line exactly as in \eqref{Wilsonfund}.  

Consider the quark production case. The incoming quark is here on-shell and has zero transverse momentum 
\cite{Gelis:2002nn, Dumitru:2005gt}. Thus the transverse momentum of the ``observed" final state quark is determined by the 
momentum transferred from the nucleus. This dependence is then given directly by the Fourier transform 
of the Wilson line  \eqref{Wilsonfund}, 
\beq
\hat{W}(k_\perp) = \int d^2 x_\perp e^{-ik_\perp \cdot x_\perp} W(x_\perp).
\label{wilsonkt}
\eeq

There are two possibilities, that the observed particle has low transverse momentum,
of order of the typical intrinsic transverse momentum, \emph{i.e.} $k_\perp \sim m$, or that 
it has large transverse momentum, of the order of a hard scale $Q$. 
The cases in figure \ref{promptphoton} suggest that the particle is produced at low 
transverse momentum, since the $k_\perp$ dependence is  directly
determined by \eqref{wilsonkt}. In that case, however,  there is no 
reason to neglect the transverse momentum  from the proton side,  as this could completely change the kinematics of the observed 
final state particle. One must therefore formulate a TMD factorization formula, with the TMD parton distribution 
and fragmentation function
of the proton taken into account.
If on the other hand the produced particle has large transverse momentum, then a hard region 
must properly be included in the process. This, however, is not the case in figure \ref{promptphoton}.

The central idea of the hybrid formalisms is based on what is called the ``factorization of mass 
singularities".  Here an emphasis is put on the mass divergences that appear in massless on-shell 
partonic reactions  \cite{pinkbook}.  This procedure is in fact widely found in the literature when dealing with 
collinear factorization.  Despite its wide use, however, it is a physically misleading procedure. 
It is in fact a rather different approach than the factorization explained in section \ref{sec:hardscatfact}
above. 
In this approach it is first \emph{asserted} that a hadronic cross section $\sigma_h$, or a structure
function $W_{h}$, is  a convolution of the corresponding partonic cross section $\sigma_p$, or structure function 
$W_p$, and a so-called ``bare parton density'', $f^{\rm bare}$: 
\beq
W_{h} (q,P) = W_p(q,\xi P) \otimes  f^{\rm bare}(\xi).
\label{massfact1}
\eeq
The convolution in the variable $\xi$ is here the same as in equations \eqref{simplequarkfact} and \eqref{simplegluonfact}. 
In the appendix of \cite{Dumitru:2005gt} (see also \cite{Chirilli:2011km}) it is for example asserted for single 
inclusive hadron production that the differential cross section is given by 
\beq
d \sigma_h(p,p_h;P) =  f^{\rm bare}(\xi) \otimes D^{\rm bare}(z) \otimes d \sigma_p(z,\xi; P)
\label{hybridfact}
\eeq
where $p$, $p_h$ and $P$ are the momenta of the incoming proton, the produced hadron and the 
incoming nucleus respectively. For the forward particle production shown in figure \ref{promptphoton} 
(right graph), the incoming quark has momentum $\xi \, p$ while the outgoing quark has momentum 
$p_h/z$ and it subsequently fragments to produce the observed hadron $p_h$.  

In both cases, the calculations are then 
performed with massless partons and with the parton entering the scattering taken to be on-shell with zero transverse momentum. With these assumptions, collinear divergences appear in the partonic cross sections. 
 It has been shown in the case of \eqref{massfact1}
that the result for $W_p$ can be written as a convolution of a divergent factor, $D$ (not to be confused 
with the fragmentation function), 
and a finite cross section $\hat{\sigma}$ \cite{Ellis:1978ty, Curci:1980uw}. 
Using the associativity of the convolution operation $\otimes$, one can then write 
\beq
W_h  =  (\hat{\sigma} \otimes D) \otimes f^{\rm bare} = \hat{\sigma} \otimes (
D \otimes f^{\rm bare}) = \hat{\sigma} \otimes f^{\rm ren}
\label{massfact2}
\eeq
where the ``renormalized'' parton distribution is given by $f^{\rm ren} =
D \otimes f^{\rm bare}$.  This final result is actually just like that in \eqref{pdfrenorm}. 
The just outlined procedure is, however, problematic for several reasons. 

To begin with, there is no proof for the assertion \eqref{massfact1} or \eqref{hybridfact}, which 
actually \emph{is} the statement of factorization. In the hybrid 
formalisms, it is simply stated that the proton side can be treated by integrated distributions. It is also in this case
not exactly  clear what the ``bare parton density" is. According to the set up of the formalism, 
it is supposed to represent a distribution
of on-shell and massless partons in the proton. This, however, is physically an ill-defined concept since 
quarks and gluons never exist as on-shell particles inside real hadrons. Moreover, if quark masses are retained 
in the calculations, there are no collinear
divergences. It is therefore dangerous to emphasize the importance of the mass divergences 
since they appear only due to the approximation of using massless on-shell partons, and are therefore of 
a spurious nature.  The 
``regularization" procedure just above is therefore conceptually different than \eqref{pdfrenorm}, and crucially, it 
is not in any way related to factorization even if this might seem to be implied. 

In the analysis of section \ref{sec:hardscatfact} what factorization means is that a given cross section 
or structure function can be written in a factorized form where each factor is associated with a 
given momentum region. For example, 
in the case of DIS it means that we can factorize the hadronic tensor as
\beq
W \sim \int_x^1 \frac{dz}{z} C^{(0)}_j(Q/\mu, z/x, \epsilon) \, f_j^{(0)}(z;\mu,\epsilon)
\label{baredisfact}
\eeq
up to power-suppressed corrections. We can also write this simply as 
\beq
W \sim C^{(0)}_j \otimes f_j^{(0)}.
\eeq
The meaning of the bare parton distribution is then that it is the gauge invariant 
integrated or TMD parton distribution constructed out of the bare fields of the Lagrangian. An example 
is the light-cone gauge definition of the bare integrated gluon distribution in \eqref{intgluonpdf}. In fact
any gauge invariant definition of a parton distribution involving suitable Wilson lines, as for example in the WW 
distribution \eqref{WWdistrbadj} or the dipole distribution \eqref{dipgluedistrb}, must refer to the bare distribution, because
the gauge transformation properties  are obeyed by the gauge links constructed out of the bare fields. So 
strictly speaking we should have denoted all those distributions as in \eqref{intgluonpdf} and 
\eqref{baredisfact}, \emph{i.e} by a 
superscript $f^{(0)}$.  It is important, however, to realize that this bare distribution, constructed out of the bare 
fields, cannot be the same as the undefined quantities in \eqref{massfact1} and \eqref{hybridfact}.  
For it is clear that it does not represent any distribution of on-shell, massless partons as is implied by 
\eqref{massfact1} and \eqref{hybridfact}. 
Once factorization has been proved as in \eqref{baredisfact} (or in \eqref{tmdepem}, \eqref{sidis} 
and \eqref{drellyan}), which itself is a very non-trivial statement, then renormalization is a matter of removing UV divergences by a suitable redefinition of the parameters of the 
Lagrangian. Order by order in perturbation theory this means adding the necessary counter terms from 
the Lagrangian, for example in the $\overline{\mathrm{MS}}$ scheme. One then finds the renormalized parton 
distribution via a formula as in \eqref{pdfrenorm}.  For \eqref{baredisfact} we find that 
\beq
W \sim C^{(0)}_j \otimes f_j^{(0)} &=& C^{(0)}_{j'} \otimes \delta_{j'j} \otimes  f_j^{(0)} \nonumber \\ 
&=& C^{(0)}_{j'} \otimes (Z^{-1} \otimes Z)_{j'j}\otimes  f_j^{(0)} \nonumber \\
&=& (C^{(0)}_{j'} \otimes Z^{-1}_{j'j''}) \otimes (Z_{j'' j} \otimes  f_j^{(0)} ) \nonumber \\ 
&=& C_{j} \otimes f_j
\label{renormdisfact}
\eeq
where $f_j$ is the renormalized distribution given by \eqref{pdfrenorm}, and the Kronecker delta in the 
first line also includes delta functions with respect to the momentum convolutions.  This procedure still 
applies if the quark masses are retained in which case there are no collinear divergences at all.

Now, in the factorization approach, one can indeed approximate the momentum entering the hard scattering 
factor as massless and on-shell.  It is crucial, however, that the hard scattering factor, $C$ in \eqref{renormdisfact},  
is defined 
with suitable subtractions (as we indicated in \eqref{subtractedfact} and showed in figures \ref{dissubtracted} 
and \ref{dissubtracted2}) so that it genuinely describes a wide angle scattering with scale $Q$ (we also note that the UV divergences of the subtraction terms are regulated by $Z^{-1}$ in \eqref{renormdisfact}). In the TMD 
factorization in section \ref{sec:TMD} for example, the errors in neglecting the transverse momenta, $q_\perp$, in the hard factor goes as $q_\perp/Q$ which indeed is small in the validity region of the formalism. 
In \eqref{massfact1} and \eqref{hybridfact}, however, this is no longer the case (in particular in \eqref{hybridfact}
the partonic part still contains the scattering off the nucleus). 
Moreover for particle production at low transverse momentum, the 
 neglected transverse momentum, from the proton side, is of the same order as the transverse momentum of the final state particle which means that the error is substantial. 
 
What is also non-trivial is that TMD factorization is mixed into the formalism of the factorization of mass 
singularities. 
If in fact we want to treat the given problem using TMD distributions, then in the small-$x$ case where 
the produced particle is typically soft, one must consider off-shell matrix elements, precisely because 
of the reason just explained above. The off-shell matrix elements must then carefully be specified, to ensure gauge 
invariance (or rather gauge-independence), and one cannot use on-shell incoming partons. 
For the lowest order contributions, gauge independent off-shell scattering coefficients have been calculated 
in the CCH approach  \cite{Catani:1990xk, Catani:1990eg}, and an explicit all order definition in the case 
of BFKL is given in  \cite{Collins:1991ty}. See also \cite{Hautmann:2008vd, Deak:2009xt, Deak:2011gj, Ermolaev:2011aa}
for more recent considerations.

To summarize this section, the hybrid formalisms do not really address the question of factorization. 
Factorization is in a sense  assumed from the start, via equation \eqref{massfact1} or \eqref{hybridfact}. 
In fact the real problem is to show a factorization like in \eqref{baredisfact} to start with. 
Moreover, the procedure which is referred to as the renormalization of the parton densities is conceptually 
very different from what is the case in the hard scattering factorization. It is moreover physically a misleading 
procedure since the basic structures are not well-defined. 
Additionally we have seen that for particles produced at low transverse momentum, 
TMD distributions must be used also from the proton side, but then of course one must first formulate a valid 
TMD factorization formula first, which might not be possible. We will in the coming sections analyze single particle production in the small-$x$ region.

\section{The fundamentals of single inclusive particle production}
\label{sec:gluonprod}

We will now  give a comprehensive analysis of single inclusive particle production in high energy 
QCD, explaining many details which are usually overlooked. We will start by going through the 
basics of particle production, giving an overview of the leading regions in different kinematical 
situations. We then 
go on to analyze single inclusive gluon production in hadron-hadron scattering 
which is a process that has been widely studied (see e.g. \cite{Gribov:1981kg,  Gribov:1983fc, Gribov:1984tu, Kharzeev:2003wz, Gelis:2003vh, Blaizot:2004wu, Marquet:2004xa, Kharzeev:2004if, Gelis:2006yv, Gelis:2006dv, Gelis:2008rw, Gelis:2008sz, Levin:2010dw, Levin:2010zy, Albacete:2010bs,  Levin:2011hr, ALbacete:2010ad, Levin:2010br}
and references therein) in the small-$x$ region.  
We will first go through the process using the axial gauge which is essentially the gauge on which the 
arguments for factorization are based, for example in \cite{Gribov:1981kg,  Gribov:1983fc, Gribov:1984tu, Kovchegov:1998bi}.
We will in detail explain the technical difficulties of the axial gauge, and why after all it is not  convenient for proving factorization.  We will then discuss hadron production from a more complete point of view, by building upon the analysis of the leading regions for the different kinematical cases.  Finally we shall address the exact form of the TMD gluon 
distribution associated with this process, finishing with a discussion of the validity of factorization.

\subsection{The different cases of particle production}
\label{sec:diffcases}

\begin{figure}[t]
\begin{center}
\includegraphics[angle=0, scale=0.55]{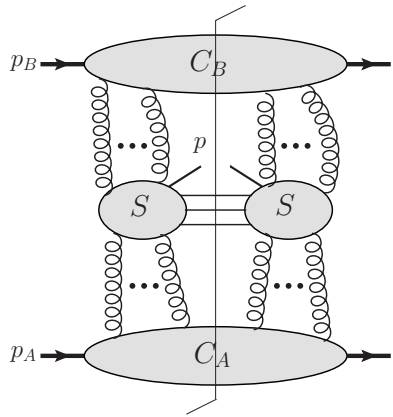}
\end{center}
\caption{\label{softhadronprod} Production of soft hadrons in the small-$x$ limit. The observed hadron $p$ 
is associated with the soft region.  } 
\end{figure}

\begin{figure}[t]
\begin{center}
\includegraphics[angle=0, scale=0.55]{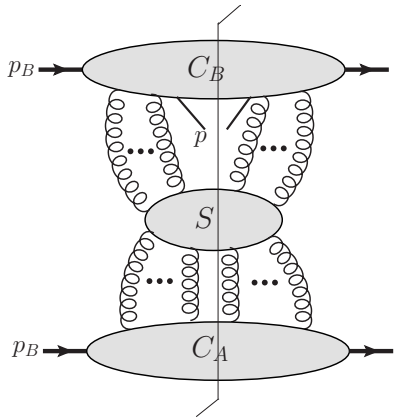}
\end{center}
\caption{\label{collhadronprod} Production of hadrons in the fragmentation region of particle $B$ in the small-$x$ limit. The observed hadron $p$ has rapidity close to that of $B$. A similar graph exists for production in the opposite 
direction close to $A$.  These cases require the use of fracture functions rather than ordinary parton distributions 
and fragmentation functions. } 
\end{figure}

\begin{figure}[t]
\begin{center}
\includegraphics[angle=0, scale=0.55]{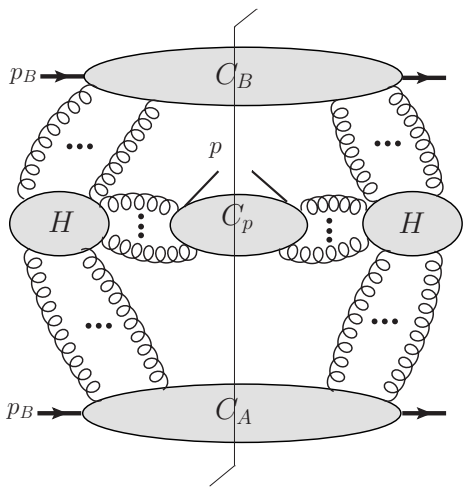}
\end{center}
\caption{\label{hardhadronprod} Production of hadron in the presence of a hard factor.  The soft region 
coupling the collinear regions is also present, and additional collinear factors emerging from the hard 
scattering may be present as well, but for simplicity we do not show these here. } 
\end{figure}

In figures \ref{softhadronprod}, \ref{collhadronprod} and \ref{hardhadronprod} we list the possible 
scenarios for single inclusive particle production at small-$x$.  In this section we explain the physics 
of the different cases. 

Figure \ref{softhadronprod} represents a typical scenario of particle production in the Regge region, 
namely that of a soft particle produced at a typical small angle scattering event. In this case there is 
no hard region. All virtualities are of the typical soft scale $m^2$. The momentum $p$ of the produced particle therefore 
typically scales as $|p^\mu| \sim m$.  This case is relevant for soft particle production at mid-rapidity. 
The inclusive charged particle spectrum at mid-rapidity, 
\beq
\left . \frac{dN_{ch}}{d\eta} \right \vert_{\eta=0},
\eeq
has been measured by the different experimental groups at the LHC; ATLAS \cite{:2010ir}, CMS \cite{Khachatryan:2010nk}
and ALICE  \cite{Aamodt:2010ft, Aamodt:2010pp}. This also happens to be the mostly studied case in the applications of small-$x$ physics \cite{Kharzeev:2003wz,  Kharzeev:2004if,  Levin:2010dw, Levin:2010zy, Albacete:2010bs,  Levin:2011hr, ALbacete:2010ad, Levin:2010br, Grinyuk:2012mc}. 

Next, in figure \ref{collhadronprod} we show particle production in the case where the produced particle 
is close in rapidity to one of the hadron beams. This case therefore covers the forward production of particles. 
At the LHC, the CMS detector can detect particles in the pseudorapidity range $|\eta| < 5$ thanks to the
hadronic forward calorimeters. Since the particles traveling in the forward region have enormous 
longitudinal momentum, they must of course have high $p_\perp$ as well, since otherwise they would 
have too large rapidity and escape detection via the beam pipes.  
In CMS for example \cite{Chatrchyan:2012gw} forward \emph{jets} (not hadrons) in the rapidity range 
$3.2 < |\eta| < 4.7$ have $p_\perp \geq 35$ GeV.  One can also arrange for events where a hard di-jet 
is produced at central rapidity, to accompany the forward jet. The correlations between the forward 
jet and the central jets then offer important insight into the parton kinematics, see e.g. \cite{Deak:2010gk, Deak:2011gj,
Deak:2011ga}.  Actually if the momentum of the produced hadron belongs to either $C_A$ or $C_B$, then one has to use so-called fracture functions 
rather than ordinary fragmentation functions or parton distributions.

Finally in figure \ref{hardhadronprod} we show the case where the hadron is produced with large 
rapidity separation to both beams (for example in the central region) and where a hard region is present.
This could for example be the case where the components $p^\mu$ are typically of order $Q \gg m$ 
or where we are looking at an event where a hard collision is present, that is jets with large $p_\perp$ 
are produced in addition to the particle we tag (we do not show these additional jets in figure \ref{hardhadronprod}). 
The region decomposition here needs some explanation. 

In section \ref{sec:powercount} we classified momenta according to different possible scalings. 
The external scales in that case were set by $Q$ which also happens to be the hard momentum scale 
in the process. Therefore such a classification is appropriate when the components of the hard momenta scale with the 
longitudinal momenta of the external particles. The decomposition is thus appropriate when $x$ is not too small. 
In that case we noticed that the only real possibilities for a pinch of a given momentum $k^\mu$ 
were as follows:
\begin{itemize}
\item{ None of $k^\mu$ scales with $Q$. Then we can characterize $k^\mu$ by the typical soft scale $m$, 
\emph{i.e.} $k^\mu \sim m$.  Then $k \in S$.}

\item{A longitudinal component, say $k^+$ or $k^-$ scales with $Q$. Then we have 
$k^+ \sim Q$, $k^-\sim m^2/Q$, $k_\perp \sim m$ and vice versa. In this case 
$k \in C_A$ (or $k \in C_B$ in opposite case).}

\item{$k_\perp \sim Q$ in which case also $k^+k^- \sim Q^2$. Thus $k^\mu \sim Q$ and in that 
case $k\in H$.}

\end{itemize}

Using this classification we then saw that a power counting analysis gives that at leading 
power, $C_A$ and $C_B$ can be connected to $S$ via arbitrarily many soft longitudinally 
polarized gluons, while again arbitrarily many collinear gluons can be exchanged between 
$H$ and the respective collinear region. 

In the small-$x$ case we have a different situation. In this case the large components of the 
external particles scale with $\sqrt{s}$ but the momentum transfer remains fixed as $\sqrt{s} \to \infty$.
Thus in this case there is no region in which all momentum components scale with the asymptotic 
parameter $\sqrt{s}$. In the soft production case one has the possibilities that
\begin{itemize}
\item{ None of $k^\mu$ scales with $\sqrt{s}$. Then generally $k \in S$.}

\item{$k^+$ or $k^-$ scales with $\sqrt{s}$. In this case $k \in C_A$ or $k \in C_B$ respectively.}

\end{itemize}

There may, however, also be present hard collisions which give rise to jets or hadrons 
of several tens of GeV. Thus we may very well have regions where $k^\mu \sim Q$. We then propose the following classification
\begin{itemize}
\item{If $k^+$ or $k^-$ scales with $\sqrt{s}$, then just as above we let $k \in C_A$ or $k \in C_B$ 
respectively. }
\item{ Let $|k^\mu|/\sqrt{s} \to 0$ as $\sqrt{s} \to \infty$, but such that for example $|k^+|/|k^-| \gg 1$
and $|k^+|/|k^i| \gg 1$. Then even though $k^+ \ll \sqrt{s}$, we shall let $k \in C_A$. In the opposite 
case we of course let $k \in C_B$. To characterize such cases we shall let $k^{+} \sim Q \ll \sqrt{s}$
(or  $k^{-} \sim Q \ll \sqrt{s}$) where $Q \gg m$.}

\item{ We define the region where $k^+k^- \sim Q^2$ to be the hard region. Thus in figure \ref{hardhadronprod}
there is  momentum $k^- \sim Q$ flowing into $H$ from $C_B$, and momentum $k^+ \sim Q$ flowing 
in from $C_A$. The momenta going out from $H$ to the final state is then characterized by the scale $Q$. } 

\item { Momenta such that $|k^\mu| \sim m \ll Q$ are as before classified as soft. In figure  \ref{hardhadronprod}
we do not explicitly draw the soft subgraph to keep the notation simple. }

\end{itemize}

With this classification we can then understand figure \ref{hardhadronprod}.  Notice that the momentum 
lines whose large components scale with $\sqrt{s}$, and therefore belong to one of the collinear regions,
cannot join the collinear region to the hard region $H$, since in that case a large momentum $\sqrt{s}$ 
would be transferred to $H$, and we would no longer be in the small-$x$ region. Thus in figure \ref{hardhadronprod}
the lines joining $C_{A,B}$ to $H$ belong to the second class above.  This is a different situation then 
in section \ref{sec:powercount} where any line in $C_{A,B}$ can join that region to $H$. 

We shall now argue that the 
power counting is essentially the same as in section \ref{sec:powercount}, despite the somewhat different 
kinematics. 
\begin{figure}[t]
\begin{center}
\includegraphics[angle=0, scale=0.55]{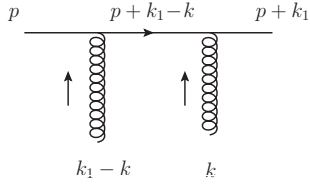}
\end{center}
\caption{\label{2gluontohard} Coupling of two gluons from $C_A$ to $H$.} 
\end{figure}
In figure \ref{2gluontohard} we show an example where two gluons from $C_A$ couple to $H$ as defined above. 
These two gluons have $k^+ \sim Q$, and in the lower end (not shown in the figure) they couple to collinear-to-$A$
gluons which may have momenta scaling as $\sqrt{s}$ in the plus direction.  The leading contribution is then 
given by  
\beq
\int d^4k_1 \int d^4 k \frac{\slashed{p}+\slashed{k}_1}{(p+k_1)^2}
\gamma^-\frac{\slashed{p}+\slashed{k}_1-\slashed{k}}{(p+k_1-k)^2}\gamma^-\frac{\slashed{p}}{p^2}\,
\frac{1}{k_1^2}\, \frac{1}{k^2} \, A^{++}(k_1,k,p_A)
\eeq
where $p\in H$. We then write this expression as 
\beq
\int d^4k_1 \int d^4 k \frac{\slashed{p}}{2p^-k_1^+}\gamma^-\frac{\slashed{p}}{2p^-(k_1^+-k^+)}
\gamma^-\frac{\slashed{p}}{p^2}\frac{1}{k_1^2}\, \frac{1}{k^2} \, A^{++}(k_1,k,p_A).
\eeq
Now, as in section \ref{sec:powercount} we characterize the momentum coupling $C_A$ to $H$ 
by a scale $\lambda_A$, such that $k^- \sim \lambda_A^2/Q$ and $k_\perp \sim \lambda_A$. When 
the momentum $k$ in figure \ref{2gluontohard} couples to $A^{++}(k_1,k,p_A)$, there will be a 
typical contribution of 
\beq
\frac{\sqrt{s}}{(p_A+k)^2} \sim \frac{\sqrt{s}}{p_A^+k^-} \sim \frac{\sqrt{s}}{\sqrt{s}\lambda_A^2/Q}
= \frac{Q}{\lambda_A^2}.
\label{Reggecount}
\eeq
The factor $\sqrt{s}$ in the numerator comes from the large boost of $A$ in the + direction. Remember 
that in the case covered in section \ref{sec:powercount} we have 
\beq
\frac{Q}{(p_A+k)^2} \sim \frac{Q}{Q\lambda_A^2/Q} = \frac{Q}{\lambda_A^2}.
\label{hardcount}
\eeq
As we see \eqref{Reggecount} agrees with \eqref{hardcount}. We therefore essentially have the same situation 
as before, that is arbitrarily many longitudinally polarized gluons of the second type in the classification above 
can connect the collinear regions to $H$ in figure \ref{hardhadronprod}.  Indeed the contribution from figure \ref{2gluontohard}
gives 
\beq
\left [\int d^4k_1 \frac{\slashed{p}+\slashed{k}_1}{(p+k_1)^2}\gamma^-\frac{\slashed{p}}{p^2}\frac{1}{k_1^2} A^+\right ]
\int d\lambda_A \, \lambda_A^3 \, \frac{1}{Q}\,\frac{1}{\lambda_A^2}\,  \frac{Q}{\lambda_A^2}.
\eeq
The term in the brackets corresponds to the contribution from gluon $k_1$ only. The factor outside 
therefore gives the contribution from attaching the additional gluon $k$ and we see that it 
gives a logarithmic contribution 
\beq
\int \frac{d\lambda_A}{\lambda_A}
\eeq
so that there is no power suppression for coupling the extra gluon $k$ to $H$. 

To ensure the validity of all these arguments it is again important that one can perform contour deformations
out of the Glauber region. We will in the next sections give a careful analysis of the 
factorization arguments that are based on the use of axial gauge, and we will show the difficulties associated 
with such arguments. We will continue the general discussion of single particle production in section \ref{sec:singlehadron}
below. Before that, however, we want in the coming sections to concentrate on the small-$x$ single inclusive gluon production
cross section that has been widely used for phenomenological applications. 

\subsection{The small-$x$ formula for gluon production}
\label{sec:smallxgluon}

\begin{figure}[t]
\begin{center}
\includegraphics[angle=0, scale=0.55]{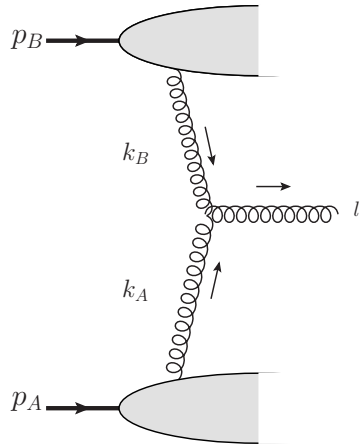}
\end{center}
\caption{\label{glrgluonprod} Single inclusive gluon production in hadron-hadron scattering 
according to equation \protect \eqref{GLRfact}. } 
\end{figure}

The most basic process for gluon production  is depicted in figure \ref{glrgluonprod}
where the idea is that two gluons, $k_A$ and $k_B$, each belonging to one of the incoming hadrons, fuse to produce 
a gluon of momentum $l$ which then emerges in the final state.  The argument for the validity of figure \ref{glrgluonprod} 
is based on the use of axial gauge. The situation is similar to that in figure \ref{nongaugeDIS} where the use of the light-cone gauge 
eliminates all higher order gluon exchanges to leading power. 

The factorization formula being used  is given by \cite{Gribov:1984tu, Kovchegov:1998bi, Kharzeev:2003wz, Kharzeev:2004if,  Levin:2010dw}
\beq
\frac{d\sigma}{d^2l_\perp dy} = \frac{2\alpha_s}{C_F \, l_\perp^2} \int d^2k_\perp \,
f_A(y,k_{A,\perp})\, f_B(Y\!-\!y, k_{B,\perp})
\label{GLRfact}
\eeq
where 
\beq
k_A \equiv k, \,\,\,\, k_B \equiv l-k,
\eeq
$y$ is the rapidity of the produced gluon with respect to the right moving hadron $p_A$. The functions 
$f_A$ and $f_B$ represent the respective TMD gluon distributions, and we shortly write 
down the definitions used. 
The origin of equation \eqref{GLRfact} goes back to the GLR papers \cite{Gribov:1981kg,  Gribov:1983fc, Gribov:1984tu}
where the function $f$ is ``defined'' as the derivative of the integrated gluon distribution (which is called the ``gluon structure function'' in \cite{Gribov:1981kg,  Gribov:1983fc, Gribov:1984tu}) 
\beq
f(y,k_\perp) = \frac{\partial xG(x,k_\perp)}{\partial k_\perp^2}, \,\,\,\,  y = \ln 1/x   .
\label{intvsunintglr}
\eeq
We note that this relation (or rather the inverted integral version of it) is a direct application of the parton model result \eqref{intvsunintpm}, although in the parton model the integral over the unintegrated distribution is over all $k_\perp$. 
There are several good reasons for why one should be very cautious with the naive application of the parton model result. 
We will discuss this more in \cite{ourpaper}, and see also the comments just after equation \eqref{firstorderKTgluon} below. 

As for the validity the factorization formula \eqref{GLRfact}, it is in the literature common to cite the works \cite{Kovchegov:1998bi, Kovchegov:2001sc}.
Reference \cite{Kovchegov:2001sc} makes use of the dipole formalism in studying the deep inelastic scattering 
on a large nucleus, where the nucleus is taken to be described by the classical MV model. In this case
the ``unintegrated gluon distribution'' is taken to be  
\beq
f(k_\perp;y) = \frac{N_c}{(2\pi)^4\alpha_s}\int d^2r_\perp \int d^2 b_\perp e^{-ir_\perp \cdot k_\perp} \nabla_r^2\, 
\mcN_G(r_\perp, b_\perp;y),
\label{KTgluon}
\eeq 
where $\mcal{N}_G$ has the same meaning as $\mcal{N}$ in \eqref{dipN} but is instead written in the 
adjoint representation as
\beq
\mcN_G(r_\perp, b_\perp;y) \equiv 1 - \frac{1}{N_c^2-1}
\lan \mathrm{Tr} \{ \tilde{W}(b_\perp \!\!+\! r_\perp/2)\tilde{W}^\dagger (b_\perp\!\!-\!r_\perp/2)
\}\ran_y. 
\label{dipadj}
\eeq
The Wilson line $\tilde{W}$ has the same form as in  \eqref{Wilsonfund}, but with 
the replacement $t_F^a \to T^a$ where  $T^a$ is the adjoint color matrix. 
As can be seen we have indicated the dependence on 
the rapidity variable $y$ a bit differently in \eqref{KTgluon} than in \eqref{GLRfact}. We have 
in fact done this in purpose and it should later on be clear why we have done so. Notice 
for now that \eqref{KTgluon} is essentially the dipole distribution \eqref{dipgluedistrb}, with 
the difference that it is here written using Wilson lines in the adjoint representation. 
It is important to note, however, that \eqref{KTgluon} is not directly derived from the 
formalism in \cite{Kovchegov:2001sc}. Its form is rather asserted by the \emph{assumption} that the dipole formalism 
used in \cite{Kovchegov:2001sc} is equivalent to the factorization formula \eqref{GLRfact}.  

The results of \cite{Kovchegov:2001sc} are in turn partly based on \cite{Kovchegov:1998bi} where the light-cone gauge is employed and 
it is argued that the leading regions have the structure shown in the figure \ref{glrgluonprod}. We also note that 
a similar factorized formula is found in the classical DDT paper \cite{Dokshitzer:1978hw} from the early days of QCD.

We will therefore now go through the light-cone gauge calculation. First, however, we need to specify the kinematics more carefully. 

\subsubsection{The kinematics}

We denote as usual the incoming momenta by $p_A$ and $p_B$. In the cms frame in the limit of very high energy 
and neglecting the masses one has 
\beq
p_A &=& (\sqrt{s/2},0,0_\perp) \\
p_B &=& (0, \sqrt{s/2}, 0_\perp) 
\eeq
so that $s=2p_A \cdot p_B=2p_A^+p_B^-$. 

\begin{figure}[t]
\begin{center}
\includegraphics[angle=0, scale=0.65]{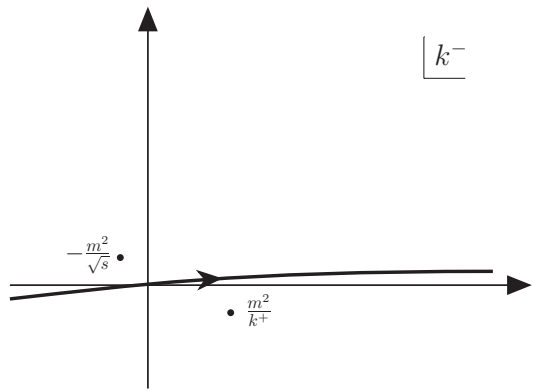}
\end{center}
\caption{\label{kmcontour} Poles in the plane of $k^-$ and possible integration contour. } 
\end{figure}

We now ask which of the cases in section \ref{sec:diffcases} above that is relevant here for figure \ref{glrgluonprod}.
From figure \ref{glrgluonprod} we see that there will be a typical contribution of the type
\beq
\frac{\mathrm{Numerator}}{(k_A^2+i\epsilon)(k_B^2+i\epsilon)((p_A-k_A)^2+i\epsilon)((p_B-k_B)^2+i\epsilon)}
\times (\mathrm{Rest}\,\, \mathrm{of} \,\, \mathrm{graph} ).
\eeq

Let us now consider the $k_A$ part, and the integral over $k_A^-$. We note that if $k_A^+ < 0$ or $k_A^+ > p_A^+$ 
then the poles in the $k_A^-$ plane are either both in the upper or in the lower half plane respectively. In those 
cases we can deform away from the poles simultaneously and we get a power suppressed contribution. Thus we have 
$0 < k_A^+ < p_A^+$. In this case the pole from the $k_A$ propagator is in the lower half plane, while the pole 
from the lower blob is in the upper half plane and the integration contour is therefore trapped. We show the 
pole structure in figure \ref{kmcontour}. 
We here simply denote the order of the magnitude of the poles, setting 
$k_\perp \sim m$. 
If we denote the two poles in $k_A^-$ by $k^-_1$ and $k^-_2$ we see that the distance 
between them satisfies 
\beq
|k_1^--k_2^-| &=& \frac{k_{A,\perp}^2}{2}\left ( \frac{1}{k_A^+}+\frac{1}{p_A^+-k_A^+} \right ) \nonumber \\
&\sim & \frac{k_\perp^2}{2k_A^+}.
\eeq
Thus when $k_\perp\to 0$ (and all masses in the theory are neglected) we get an exact pinch. As 
$k_A^+ \to \sqrt{s}$ we also see that the poles are increasingly pinched and there is potentially a large 
contribution (from the collinear PSS). This, however, corresponds to the non-Regge region and is 
therefore not relevant for us. 
 Now, we can let
the integration contour pass near the $p_A-k_A$ pole in which case $|k_A^-| \sim m^2/\sqrt{s}$ (if actually 
the lower blob consists of a single spectator line then this pole becomes exact because there will be 
a delta function setting the spectator line on-shell).  

We might, however, also ask what happens if there is a hard region as in figure \ref{hardhadronprod}.
Assume for example that $l_\perp \sim Q$. 
As described in section \ref{sec:diffcases}, we must then have $k^+ \sim Q$ and $(l-k)^- \sim Q$
(we now use that $k_A=k$ and $k_B=l-k$).  Then 
\beq
k^+ \sim Q, \,\, k^- \sim Q^2/\sqrt{s},
\eeq
and thus 
\beq
k^+k^- \sim \frac{Q}{\sqrt{s}}Q^2 \ll Q^2 \sim k_\perp^2.
\eeq
The last estimate comes from $k_\perp \sim |l_\perp-k_\perp| \sim l_\perp \sim Q$. This, however 
implies $k^2 \sim Q^2$, which means that $k$ is actually not in the collinear-to-$A$ PSS. It instead belongs to $H$ 
and one can see that this case is suppressed. To get a leading contribution we want $k^2 \sim m^2$, 
and similarly $(l-k)^2 \sim m^2$, but then it is easy to see that we cannot have $l_\perp^2 \sim Q^2$. 
Thus for the graph shown in figure \ref{glrgluonprod} we do not have the situation in figure \ref{hardhadronprod}. 
To have a situation with a hard region like in figure \ref{hardhadronprod} we must instead consider an 
additional collinear, unobserved, jet that emerges from $H$. This, however, makes the situation rather 
complicated and changes the physics involved quite a bit. We shall briefly come back to this case in 
the discussions in sections \ref{sec:singlehadron} and \ref{sec:tmdgluon} below.

For analyzing the small-$x$ formula \eqref{GLRfact} we consider the situation where $k^+ \sim m$. 
That is we essentially have the soft (or perhaps semi-hard) case in figure \ref{softhadronprod}. 
A similar analysis as in above for the $l-k$ line implies that in this case
\beq
|k^-| \!\! \sim m^2/\sqrt{s}, \,\,\,\,\, |l^+\!\!\!-\!k^+| \sim m^2/\sqrt{s}.
\label{gluonprodkin1}
\eeq
Therefore
\beq
|k| \sim (m, m^2/\sqrt{s},m),
\label{ksmallx}
\eeq
so that 
\beq
k^+ = l^+ + \mathcal{O}(m^2/\sqrt{s}), \,\,\,\,\, l^- \gg |k^-|.
\label{kandlrelation}
\eeq
Thus 
\beq
k^+ \!\!\sim k^i, \,\,\,\,\, (l^-\!\!\!-\!k^-)\sim l^i-k^i,
\eeq
and
\beq 
|k^+k^-| &\sim& m^3/\sqrt{s} \ll m^2 \sim k_\perp^2  \\ 
|(l^+ \!\!\!- \!k^+)(l^-\!\!\!-\!k^-)| &\sim& m^3/\sqrt{s} \ll m^2 \sim (l_\perp\!\!-\!k_\perp)^2.
\label{gluonprodkin2}
\eeq 
Both gluons $k$ and $l-k$ are therefore in the Glauber region where the transverse momentum 
components dominate.  In light of what we have said earlier it would seem that we better avoid 
the Glauber region. Note, however, that there is no Glauber pinch here so we can deform out of the 
Glauber region if necessary. 

\subsection{The use of the light-cone gauge}
\label{sec:lcgauge}

The main argument for the validity of \eqref{GLRfact} given in 
\cite{Kovchegov:1998bi} is based on the use of the light-cone gauge.  Since an axial gauge is also used 
in \cite{Gribov:1984tu} to argue for the validity of \eqref{GLRfact}, we now go in through the derivation in these 
gauges. We shall start with the light-cone gauge in this section and then in the next section give an account 
based on the non-light-like axial gauge. We notice that axial or light-cone gauge is also used in establishing 
the factorization formulas in the CCH \cite{Catani:1990eg} and CCFM \cite{Catani:1989sg} formalisms. 

There is in fact problem with the kinematical arguments given above in the light-cone gauge.  
If we choose the gauge $A^+=0$ then 
the treatment of the $A$ part is as we just described. However, we do get a problem of the treatment 
of the $B$ side. Similarly we do get a problem of the treatment of the $A$ side if we work in $A^-=0$ 
gauge. In fact the latter is the gauge on which the arguments in \cite{Kovchegov:1998bi}
are based.  What we want to demonstrate in this section is that the 
light-cone gauge is clearly improper for the treatment of hadron-hadron collisions (be it proton-proton, 
proton-nucleus or nucleus-nucleus collisions). We will offer 
several reasons for this, and we return to the just mentioned issue at the end of this section. We 
will now simply push forward with the light-cone gauge and then see that it leads to severe problems. 

Let us now denote the gluon propagators by
\beq
P^{\mu\nu}(k) = \frac{-iN^{\mu\nu}(k)}{k^2+i\epsilon}. 
\eeq
Then in the light-cone gauge $n\cdot A=0$ we have 
\beq
N^{\mu\nu}(k) = g^{\mu\nu} - \frac{n^\mu k^\nu}{k\cdot n} -  \frac{n^\nu k^\mu}{k\cdot n}.
\label{LCprop}
\eeq
We shall write $N^{\mu\nu}(k)$ as 
\beq
N^{\mu\nu}(k) = \overrightharp{G}^{\mu\nu}(k) - \overleftharp{K}^{\mu\nu}(k),
\label{KMKG}
\eeq
where
\beq
\overrightharp{G}^{\mu\nu}(k)  &\equiv& g^{\mu\nu} - \frac{n^\mu k^\nu}{k\cdot n } \\
 \overleftharp{K}^{\mu\nu}(k) &\equiv& \frac{ k^\mu n^\nu}{k\cdot n}.
\eeq
Our notation here is inspired by the so-called $K$-$G$ decomposition introduced by Grammer and Yennie \cite{Grammer:1973db}.
The directions of the harpoons indicate whether it is the left or the right Lorentz index that is carried by
the momentum $k$; $\overrightharp{G}^{\mu\nu}(k)$ (and $\overrightharp{K}^{\mu\nu}$) 
contains $k^\nu$, while $ \overleftharp{K}^{\mu\nu}(k)$ (and $\overleftharp{G}^{\mu\nu}$)
contains $k^\mu$.  Notice that the standard Grammer-Yennie decomposition which is applied to the 
Feynman gauge propagators is in this notation given by 
\beq
N_{Feyn}^{\mu\nu} = g^{\mu\nu} =  \overleftharp{G}^{\mu\nu}(k) +  \overleftharp{K}^{\mu\nu}(k) 
 = \left ( g^{\mu\nu} - \frac{k^\mu n^\nu}{k\cdot n } \right ) + \frac{k^\mu n^\nu}{k\cdot n }.
 \label{kgdecomp}
\eeq
The $K$-$G$ decomposition is important in proving factorization in the hard 
scattering domain since Ward identities can be applied to the $K$ terms which are the dominant 
contributions.  Remember from the 
analysis in section \ref{sec:hardscatfact} that there can be arbitrarily many longitudinally polarized 
gluons exchanged between the hard and collinear regions, $H$ and $C$, and between the collinear 
and the soft regions, $C$ and $S$.  These gluons precisely correspond to the $K$ terms. 
If we choose $n$ such that $n\cdot A = A^+$, then for the $G$ terms we have
\beq
\overrightharp{G}^{-+}(k) = g^{-+} \!\!-  \frac{k^+}{k^+} = 0, \,\,\, \overleftharp{G}^{+-}(k) = g^{+-} \!\!-  \frac{k^+}{k^+} = 0,
\label{Gplusminus}
\eeq
while for the $K$ terms
\beq
\overrightharp{K}^{-+}(k) =  \frac{k^+}{k^+} = 1, \,\,\, \overleftharp{K}^{+-}(k)=  \frac{k^+}{k^+} = 1.
\label{Kplusminus}
\eeq
For the dominant polarization $N^{-+}$ we therefore see that only the $K$ terms contribute. 
The key step to proving factorization is then to repeatedly apply the Ward identities on the $K$ terms. 

If, however, 
$k$ is dominated by its transverse component, then one can no longer neglect the transverse 
$G$ contributions to which the Ward identities do not apply.  If for example we have momentum which scales 
as $l-k$ in the above example, then 
\beq
|\overrightharp{G}^{-i}(l-k)| = \left \vert \frac{(l-k)^i}{(l-k)^+} \right \vert \gg 1, \,\,\, |\overleftharp{K}^{i-}(l-k)|=
\left \vert \frac{(l-k)^i}{(l-k)^+}\right \vert \gg 1.
\eeq
This means that the transverse components cannot be neglected in favor of the $+-$ components. 
Moreover, even for the $K$ terms, 
the application of the Ward identities leave non-factorizing remainder terms which are complicated. 
These can be neglected in the collinear limit but not in the Glauber region. Therefore in all the higher order
corrections to figure \ref{glrgluonprod} we must be able make all necessary contour deformations so as 
to power suppress these contributions.

\begin{figure}[t]
\begin{center}
\includegraphics[angle=0, scale=0.55]{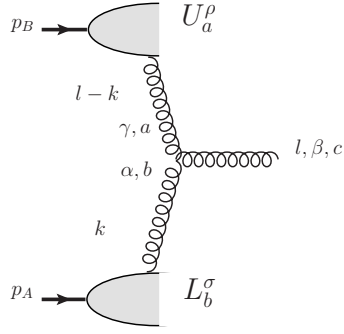}
\end{center}
\caption{\label{glrgluonprod2} The graphical representation of formula \protect \eqref{GLRfact}. } 
\end{figure}

In the axial gauge, the singular propagators must be regularized. A canonical regularization is obtained 
by treating the singularities as principal values. 
Now, in  \cite{Kovchegov:1998bi} the regularization is instead performed by choosing
\beq
N^{\mu\nu}(k) = g^{\mu\nu} - \frac{n^\mu k^\nu}{k\cdot n - i\epsilon} -  \frac{n^\nu k^\mu}{k\cdot n +i\epsilon}.
\label{KMprop}
\eeq
Here the momentum flows from $\mu$ towards $\nu$. The vector $n$ is now chosen so that $n \cdot A = A^-$. 
There is, however, a fundamental problem with this gauge, and it shows up already for the lowest
order contribution in figure \ref{glrgluonprod2}. It is related to the fact that the light-cone gauge does not 
treat the hadrons symmetrically. We now demonstrate this problem by calculating the contribution in 
figure \ref{glrgluonprod2}. 

The polarization vector of the produced gluon $l$ is chosen in  \cite{Kovchegov:1998bi} to satisfy $\epsilon^+(l) = 0$. 
Since $l \cdot \epsilon(l) = 0$ one has $\epsilon^-(l) = l^i\epsilon^i/l^+$. 
The contribution from the process depicted in figure \ref{glrgluonprod2} is given by 
\beq
-g_s\epsilon^*_\beta(l)U_a^\rho(p_B,l-k)  (\overrightharp{G}_{\rho\gamma}(l-k) - \overleftharp{K}_{\rho\gamma}(l-k)
) V_{abc}^{\gamma\alpha\beta}(\overrightharp{G}_{\sigma\alpha}(k) - \overleftharp{K}_{\sigma\alpha}(k))L_b^\sigma(p_A,k) 
\nonumber \\
\label{graphone}
\eeq 
where $V$ is the three-gluon vertex.  The dominant component of the lower part is $L^+ \propto \sqrt{s}$, 
while the dominant component of the upper part is $U^- \propto \sqrt{s}$.  We notice that in the above expression, 
\beq
U_a^\rho(p_B,l-k) \overleftharp{K}_{\rho\gamma}(l-k) = L_b^\sigma(p_A,k)\overleftharp{K}_{\sigma\alpha}(k) 
= 0
\eeq
by the use of the Ward identity.
One is then left with 
\beq
-g_s\epsilon^*_\beta(l)U_a^\rho(p_B,l-k) \,  \overrightharp{G}_{\rho\gamma}(l-k) \, V_{abc}^{\gamma\alpha\beta}\, \overrightharp{G}_{\sigma\alpha}(k)\, L_b^\sigma(p_A,k).
\label{graphtwo}
\eeq
It is easily seen that the leading contributions are 
\beq
\overrightharp{G}^{\sigma\alpha}(k) \, L_{b,\sigma}(p_A,k) \approx \overrightharp{G}^{-\alpha}(k) \, L_{b}^+(p_A,k) =
g^{-\alpha}L_b^+(p_A,k)
\label{Lleading}
\eeq
and 
\beq
U_{a,\rho}(p_B,l-k)\,  \overrightharp{G}^{\rho\gamma}(l-k)  \approx U_a^-(p_B,l-k)\,  \overrightharp{G}^{+\gamma}(l-k).
\eeq
For the $\gamma$ index, the $\gamma=-$ component gives zero because of the gauge $A^-=0$, while 
\beq
 |\overrightharp{G}^{++}(l-k)| =  \frac{|l^+-k^+|}{|l^--k^-|} \ll \frac{|l^i-k^i|}{|l^--k^-|} = |\overrightharp{G}^{+i}(l-k)| 
\eeq
so the leading term comes from $\gamma=i$.  From \eqref{Lleading}, taking also into account the contribution 
from the complex conjugate amplitude,  we see that we have for the lower part (we neglect the color indices for the moment)
\beq
L^{+ \dagger}(p_A,k) L^+(p_A,k)\! &=&\! \sum_X \int d^4x \, e^{ik\cdot x} \langle p_A | A^+(x)
|X,\mathrm{out}\rangle \langle X, \mathrm{out}| A^+(0) |p_A \rangle \nonumber \\
\! &=&\!  \sum_X \frac{1}{(k^-)^2}\int d^4x \, e^{ik\cdot x} \langle p_A | (k\cdot A(x)+k^iA^i(x)) |X,\mathrm{out}\rangle
\nonumber \\
&& \hspace{35mm} \langle X, \mathrm{out}| (k\cdot A(0)+k^iA^i(0)) |p_A \rangle 
\nonumber \\
\! &=&\!  \sum_X \frac{1}{(k^-)^2}\int d^4x \, e^{ik\cdot x} \langle p_A | k^iA^i(x)|X,\mathrm{out}\rangle \langle X, \mathrm{out}|  k^iA^i(0) |p_A\rangle \nonumber \\
\label{Lpdfdef}
\eeq
where in the second equality we used the fact that $A^-=0$, while in the last equality we used the Ward identity.  
For the upper part we instead have for the leading term
\beq
 \left (U^-\,  \overrightharp{G}^{+i}\right )^\dagger \left (U^-\,  \overrightharp{G}^{+i}\right )= \int d^4x \, e^{i(l-k)\cdot x} 
\langle p_B | A^i(x) A^i(0) |p_B \rangle. 
\label{Updfdef}
\eeq

In the gauge $A^-=0$, the canonical definition of the TMD gluon distribution is (which directly corresponds 
to the parton model definition \eqref{eq:pdf.lf.def})
\beq
&k^-& \!\!\!\!\int \frac{dk^+}{(2\pi)^4} \int d^4x \, e^{ik\cdot x} \langle p_B| A^{i}(x)A^{i}(0)|p_B\rangle \nonumber \\
&=& \int \frac{dk^+}{(2\pi)^4}\frac{1}{k^-} \int d^4x \, e^{ik\cdot x} \langle p_B| F^{-i}(x)F^{-i}(0)|p_B\rangle.
\label{pdfcanonical}
\eeq
This is for example also 
the case for the Weizsacker-Williams distribution in the CGC, with the only trivial difference being that in 
that case the pre-factor in the first line above is taken to be $(k^-)^2/p_B^-=\bar{x}k^-$ instead of $k^-$ 
(with $\bar{x} = k^-/p^-_B$). 
We notice, however, that the so-called dipole gluon distribution cannot be really fully consistent 
with  \eqref{pdfcanonical}.  The reason is that  for the dipole 
gluon distribution, in the corresponding derivation one must actually set $k^-$ to 0
(this is why the Wilson lines \eqref{Wilsonfund} are integrated from $-\infty$ to $+\infty$ in the 
longitudinal direction).
One can therefore not multiply the definition with $k^-$ as above,
in order to obtain the canonical form \eqref{pdfcanonical}. In that case one may instead 
multiply the integral by $p_B^-$. 

While it is straightforward to put \eqref{Updfdef} into the proper form, this is not so with the 
lower component \eqref{Lpdfdef}. Going now back to the evaluation of 
the graph in figure \ref{glrgluonprod2} we thus have
\beq
&&g_s\epsilon^{*\beta}(l) U_a^-(l-k)L_b^+(k) \frac{(l-k)_\perp^\gamma}{l^-\!\!-k^-}g^{-\alpha}V_{\gamma\alpha\beta}^{abc} \nonumber \\
&=& -g_s U_a^-(l-k)L_b^+(k)f^{abc}\frac{1}{l^-\!\!-k^-} [ -\epsilon^{*-}(l_\perp^2\!-\!k_\perp^2)-\epsilon^{*i} (l^i\!-\!k^i)
(k^-\!\!\!-\!2l^-) ] \nonumber \\
&\approx & - g_sU_a^-(l-k)L_b^+(k)f^{abc}\left [- \frac{\epsilon^{*i}l^i}{l^-l^+}(l_\perp^2\!-\!k_\perp^2)+ 2\epsilon^{*i} (l^i\!-\!k^i) \right ]
\eeq
where $k^-$ has been neglected with respect to $l^-$.  Using $2l^+l^-=l_\perp^2$
one then gets
\beq
g_s\frac{2U_a^-(l-k)L_b^+(k)f^{abc}}{l_\perp^2} ( \epsilon^{*i}l^i (l_\perp^2\!-\!k_\perp^2) - \epsilon^{*i} (l^i\!-\!k^i)l_\perp^2).
\eeq
Squaring and summing over polarization and color indices, and integrating over $k$, we have 
\beq
\frac{g_s^2}{N_c^2-1}\!\!\sum_{aa'bb'c} \int \frac{d^4k}{(2\pi)^4} 4(U_a^-U_{a'}^{-\dagger})(L_b^+L_{b'}^{+\dagger})f^{abc}f^{a'b'c} \,
\frac{k_\perp^2(l_\perp-k_\perp)^2}{l_\perp^2}. 
\label{glueprodres}
\eeq
Now, to write a factorization formula for this result we have to untangle the color flow and at the same
time make the appropriate kinematical approximations. Using \eqref{kandlrelation}, we now neglect $k^-$ with respect
to $l^-$ in the $U$ factors, and we set $k^+=l^+$ in the $L$ factors. The $k^+$ ($k^-$) integral then acts only on the 
$U$ ($L$) factors.

For obtaining the differential single inclusive cross section we project the diagonal 
color components in $U$ and $L$, and we find that the result can be written as
\beq
\frac{d\sigma}{dy \, d^2l_\perp}=\frac{1}{2s} \frac{1}{2(2\pi)^3}\frac{4\, g_s^2\, N_c}{N_c^2-1} \frac{(2\pi)^4}{l_\perp^2}\int d^2k_\perp  \left [ \int \frac{dk^+}{(2\pi)^4} \sum_a U_a^-U_{a}^{-\dagger} \,
(l_\perp\!\!-\!k_\perp)^2 \right ]_{k^-\!=0} \nonumber \\
\times  \left [  \int \frac{dk^-}{(2\pi)^4} \sum_b L_b^+L_{b}^{+\dagger} \, k_\perp^2 \right]_{k^+\!=l^+}. 
\label{glueprodres2}
\eeq
We notice that up to this point the arguments have followed very closely those in section \ref{sec:simplefact}
that lead to equation \eqref{simplegluonfact}.  However, as we discussed after equation \eqref{simplegluonfact}, 
a more careful treatment is needed since the integration over the momentum will include contributions which 
are not strictly in the region where the above kinematics holds.  What we saw in equation \eqref{subtractedfact}
was that this could be treated by including subtractions in the hard factor. In this case, we instead need 
subtractions in the last factor of \eqref{glueprodres}. In fact one must correctly treat the gluon production 
factor, the analog of the hard region, to all orders and make sure it is gauge independent. This, however, 
does not affect the definition of the gluon distribution. 

Now, for the first bracket 
containing the upper blobs in \eqref{glueprodres2} we have from \eqref{Updfdef} (we keep the summation over the color indices
implicit)
\beq
\frac{1}{p_B^-}\left [ \int \frac{dk^+}{(2\pi)^4}  U_a^-U_{a}^{-\dagger} \, (l_\perp\!\!-\!k_\perp)^2 \right ]_{k^-\!\!=0} \!\!\!\!\!
=  \frac{(l^-)^2}{p_B^-} \int \frac{dk^+}{(2\pi)^4} \int d^4x \, e^{i(l-k)\cdot x} \langle p_B | A_a^i(x) A_a^i(0) |p_B \rangle \nonumber \\
= \int \frac{dx^+d^2x_\perp}{ (2\pi)^3\, p_B^-}\, e^{il^- x^+\! -(l_\perp\!-k_\perp)\cdot x_\perp} \langle p_B | F_a^{-i}(x^+\!\!,0^-\!\!,x_\perp) F_a^{-i}(0) |p_B \rangle \nonumber \\
\label{Updfdef2}
\eeq
where we have chosen to include the factor $1/p_B^-$ into the definition.  For the lower blobs, however, we cannot get the standard formula because of the asymmetric 
gauge choice $A^-=0$.  Using \eqref{Lpdfdef} we have (again keeping summation over color indices 
implicit)
\beq
&&\frac{1}{p_A^+}\left [  \int \frac{dk^-}{(2\pi)^4}  L_b^+L_{b}^{+\dagger}\, k_\perp^2 \right]_{k^+=l^+}  
\nonumber \\
&=&\sum_X \frac{1}{p_A^+}\int \frac{dk^-}{(2\pi)^4} \frac{k_\perp^2}{(k^-)^2}\int d^4x \, e^{i k x}
\langle p_A | k^iA_b^i(x)|X,\mathrm{out}\rangle \langle X, \mathrm{out}|  k^iA_b^i(0) |p_A\rangle\biggl . \biggr \vert_{k^+=l^+}.
\nonumber \\
\eeq
This expression is clearly different than \eqref{pdfcanonical} or \eqref{Updfdef2}, and does not correspond to 
any know distribution.  One therefore does not obtain formula \eqref{GLRfact}.

Let us now explain the other difficulty with the light-cone gauge that we mentioned just above equation 
\eqref{graphone}. As we have seen, in $A^-=0$ gauge we have a problem with the definition of the 
parton distribution for particle $A$ which moves in the + direction. Similarly if we chose $A^+=0$ gauge, 
then we will have a problem with the definition for particle $B$. Let $k_{A,B}$
denote momenta attached between the collinear regions $A, B$ and any other region such as $S$
or $H$. Where $k_A$ attaches to $A$, the collinear lines of $A$ will force $k_A^-$ to  generally be 
small as in figure \ref{kmcontour}. If we now work in the $A^-=0$ gauge it means we 
additionally have the $1/k^-_A$ pole at the origin, and the combined poles from the propagator and 
the collinear lines of $A$ will then generally pinch $k_A^-$ at the origin. This, however, means that the higher order terms cannot be deformed out to $k^-_A \sim Q$
to power suppress them (terms for example such as $\overleftharp{G}^{i+}$ will be large). The gauge $A^-=0$ therefore fails for the gluons attaching to $A$. A similar
argument for $B$ shows that the $A^+=0$ gauge similarly is not useful.

\subsection{Non-light-like symmetric axial gauge}
\label{sec:axialgauge}

To get a formula that looks like \eqref{GLRfact} one must instead choose a gauge that treats the two 
hadrons symmetrically, this can for example be done by choosing the non light-like axial gauge $A^++A^-=0$, 
\emph{i.e.} the temporal gauge $A^0=0$.  Using this gauge, one can again eliminate the extra gluon 
couplings to the collinear regions.
We will here use this gauge to 
derive \eqref{GLRfact} and at the same time we will see what the definition of the TMD gluon distribution is.
However,  in 
section \ref{sec:singlehadron} we will explain the general case, and  demonstrate the problem that is inherent 
in this axial gauge treatment as well. 

In the gauge $A^++A^-=0$, the numerator of the gluon propagator is given by
\beq
N^{\mu\nu}(k) = g^{\mu\nu} - \frac{n^\mu k^\nu+n^\nu k^\mu}{n\cdot k} + \frac{k^\mu k^\nu n^2}{(n\cdot k)^2}
\label{temporalprop}
\eeq
where $n \cdot k = k^++k^-$ for any $k$.  The contribution in  figure \ref{glrgluonprod2} gives 
\beq
-g_s\epsilon^*_\beta(l)U_a^\rho(p_B,l-k)  N_{\rho\gamma}(l-k)
 V_{abc}^{\gamma\alpha\beta}N_{\sigma\alpha}(k)L_b^\sigma(p_A,k).
\eeq
The last term proportional to $n^2$ in \eqref{temporalprop} then cancels in both propagators above 
when the Ward identity is applied on $U$ and $L$. One is then left with the same expression as in \eqref{graphone}
which again reduces to \eqref{graphtwo} when applying the Ward identity. As before we have that 
\beq
\overrightharp{G}^{\sigma\alpha}(k) \, L_{b,\sigma}(p_A,k) \approx \overrightharp{G}^{-\alpha}(k) \, L_{b}^+(p_A,k) 
\eeq
and 
\beq
U_{a,\rho}(p_B,l-k)\,  \overrightharp{G}^{\rho\gamma}(l-k)  \approx U_a^-(p_B,l-k)\,  \overrightharp{G}^{+\gamma}(l-k),
\eeq
but in this case the leading $G$ terms are different. We have 
\beq
&&\left \vert \overrightharp{G}^{++}(l-k) \right \vert = \left \vert \frac{l^+-k^+}{l^+-k^++l^--k^-} \right \vert 
\sim \left \vert \frac{l^+-k^+}{l^-} \right \vert \sim \frac{m}{\sqrt{s}} \ll 1 \\
&& \left \vert \overrightharp{G}^{+-}(l-k) \right \vert = \left \vert 1 - \frac{l^--k^-}{l^+-k^++l^--k^-} \right \vert 
\sim \left \vert \frac{k^-}{l^-} \right \vert \sim \frac{m}{\sqrt{s}}\ll 1 \\
&&\left \vert \overrightharp{G}^{+i}(l-k) \right \vert = \left \vert \frac{l^i-k^i}{l^+-k^++l^--k^-} \right \vert 
\sim \left \vert \frac{l^i-k^i}{l^-} \right \vert \sim \frac{m}{m} = 1, 
\eeq
and 
\beq
&&\left \vert \overrightharp{G}^{--}(k) \right \vert = \left \vert \frac{k^-}{k^++k^-} \right \vert 
\sim \left \vert \frac{k^-}{k^+} \right \vert \sim \frac{m}{\sqrt{s}}\ll 1 \\
&& \left \vert \overrightharp{G}^{-+}(k) \right \vert = \left \vert 1 - \frac{k^+}{k^++k^-} \right \vert 
\sim \left \vert \frac{k^-}{k^+} \right \vert \sim \frac{m}{\sqrt{s}}\ll 1 \\
&&\left \vert \overrightharp{G}^{-i}(k) \right \vert = \left \vert \frac{k^i}{k^++k^-} \right \vert 
\sim \left \vert \frac{k^i}{k^+} \right \vert \sim  \frac{m}{m} =  1.
\eeq
The leading contributions are therefore the transverse components in both sides. 
Squaring the contribution from figure \ref{glrgluonprod2} and summing over gluon polarizations 
one is then left with (we neglect for simplicity the color factors since they are exactly the same 
as in the light-cone gauge calculation above) 
\beq
&&g_s^2 (U^-U^{-\dagger})(L^+L^{+\dagger})\frac{1}{(l^+-k^++l^--k^-)^2}\frac{1}{(k^++k^-)^2} \times 
 \nonumber \\
&&\sum_\lambda \left [ -2(k^+\epsilon_\lambda^-+k^-\epsilon_\lambda^+)(k_\perp^2-l_\perp\cdot k_\perp) 
+\epsilon^i_{\lambda} k^i l_\perp^2 -\epsilon^i_{\lambda} l^i (-k_\perp^2
+2l_\perp \cdot k_\perp)  \right ]^2.
\eeq

We shall next choose the external polarization vector to satisfy $\epsilon^-=0$, which means 
that $\epsilon^+ = \epsilon^i l^i/l^-$. Then the first term in the sum above gives
\beq
-2\frac{k^-}{l^-}\epsilon^i l^i(k_\perp^2-l_\perp\cdot k_\perp) 
\eeq
which is of the order of a transverse component multiplied by $k^-/l^- \sim m/\sqrt{s} \ll 1$ and can therefore be neglected
compared to the other transverse terms. One then gets
\beq
g_s^2 (U^-U^{-\dagger})(L^+L^{+\dagger})\frac{1}{(l^+-k^++l^--k^-)^2}\frac{1}{(k^++k^-)^2} 
l_\perp^2 k_\perp^2 (l_\perp-k_\perp)^2.
\eeq
Inserting now all pre-factors and color indices, we get for the gluon production cross section
\beq
\frac{d\sigma}{dy \, d^2l_\perp}=\frac{1}{4} \frac{1}{2(2\pi)^3}\!\!\!\!\!&&\!\!\!\!\!\frac{(2\pi)^4\, g_s^2\, N_c}{N_c^2-1} 
\label{sigmatemp1}\\
&\times& \!\!\!\int d^4 k  \left [  \frac{U_a^-U_{a}^{-\dagger}(l_\perp \!\!-\!k_\perp)^2}{p_A^- (2\pi)^4} \right ]
\left [  \frac{L_b^+L_{b}^{+\dagger}k_\perp^2}{p_B^+ (2\pi)^4} \right ] \frac{l_\perp^2}{(l^+\!\!-k^+\!\!+l^-\!\!-k^-)^2
(k^+\!\!+k^-)^2}. \nonumber 
\eeq
To define the TMD gluon distribution we now notice that 
\beq
U_a^-U_{a}^{-\dagger}(l_\perp \!\!-\!k_\perp)^2 &=& (l^+\!\!-k^+\!\!+l^-\!\!-k^-)^2
\int d^4x e^{i(l-k)\cdot x} \langle p_A| A^i_a(x)A^i_a(0)|p_A\rangle \nonumber \\
&=& 2\int d^4x e^{i(l-k)\cdot x} \langle p_B|F_a^{0i}(x)F_a^{0i}(0)|p_B\rangle, 
\label{nonlighttmdup}
\eeq
where $F^{0i}=(1/\sqrt{2})(F^{+i}+F^{-i})$.  Similarly
\beq
L_b^+L_{b}^{+\dagger}k_\perp^2 = 2\int d^4x e^{ik\cdot x} \langle p_A|F_b^{0i}(x)F_b^{0i}(0)|p_A\rangle.
\label{nonlighttmddown}
\eeq
To obtain the canonical forms of the two gluon distributions, we notice that we can drop the $F^{-i}$ contribution 
in \eqref{nonlighttmddown}, since it gives rise to the contributions $k^-L^i$, $k^iL^-$ and $L^{i-}$ which 
are all power-suppressed. Therefore we might as well replace $F^{0i}$ by $F^{+i}/\sqrt{2}$. Similarly for 
the expression in \eqref{nonlighttmdup} we can replace $F^{0i}$ by $F^{-i}/\sqrt{2}$. 
To get the factorization formula, one further needs to approximate $k^-=0$ in the upper part, 
and $k^+=l^+$ lower part. Furthermore we applied the approximations from the kinematics in 
\eqref{gluonprodkin1}-\eqref{gluonprodkin2} in the last factor in \eqref{sigmatemp1}
which can then be written as (up to power-suppressed corrections)
\beq
\frac{l_\perp^2}{(l^+\!\!-k^+\!\!+l^-\!\!-k^-)^2(k^+\!\!+k^-)^2} \sim \frac{l_\perp^2}{(l^-)^2(l^+)^2} = 
\frac{4}{l_\perp^2}. 
\eeq 
Thus we find
\beq
\frac{d\sigma}{dy \, d^2l_\perp}&=& \frac{2\pi^2 \, \alpha_s}{C_F \, l_\perp^2} \int d^2k_\perp 
 \left [ \int \frac{dk^+}{(2\pi)^4}\frac{1}{p_B^-} U_a^-U_{a}^{-\dagger}(l_\perp \!\!-\!k_\perp)^2\right ] 
 \left [  \int \frac{dk^-}{(2\pi)^4}\frac{1}{p_A^+} L_b^+L_{b}^{+\dagger} k_\perp^2\right ]  \nonumber \\
 &=& \frac{2\pi^2 \, \alpha_s}{C_F \, l_\perp^2} \int d^2k_\perp f_B(x_B,l_\perp-k_\perp) f_A(x_A,k_\perp),
 \label{particleprodfact}
\eeq
with 
\beq
 f_A(x_A, k_\perp) = \int \frac{dx^-d^2x_\perp}{(2\pi)^3\, p_A^+}e^{ix_Ap_A^+ x^- - i k_\perp x_\perp} 
 \langle p_A|F_a^{+i}(0^+\!\!,x^-\!\!,x_\perp)F_a^{+i}(0)|p_A\rangle,
 \label{temporalpdf1}
\eeq
and
\beq
f_B(x_B,l_\perp-k_\perp) = \int \frac{dx^+d^2x_\perp}{(2\pi)^3\, p_B^-}e^{ix_Bp_B^-x^+ - i(l-k)_\perp x_\perp} 
 \langle p_B|F_a^{-i}(x^+\!\!,0^-\!\!,x_\perp)F_a^{-i}(0)|p_B\rangle,
 \label{temporalpdf2}
\eeq
where $x_A=l^+/p_A^+$ and $x_B=l^-/p_B^-$.  

\subsubsection{The coefficient of the formula}

As for the coefficient in front of formula \eqref{particleprodfact}, we note that different
values appear in the literature. Let us denote the coefficient in \eqref{particleprodfact} by 
\beq
C= \frac{2\pi^2 \, \alpha_s}{C_F}=  \frac{4\pi^2 \, N_c \, \alpha_s}{N_c^2-1}.
\label{Ccoeff}
\eeq
In the papers \cite{Kovchegov:2001sc, Kharzeev:2004if, Levin:2010dw, Levin:2010zy, Levin:2011hr, Levin:2010br} 
we instead find the formula (this is the value we used in writing \eqref{GLRfact})
\beq
C = \frac{2\alpha_s}{C_F} = \frac{4\, N_c \, \alpha_s}{N_c^2-1},
\label{ckt}
\eeq
while in \cite{Gribov:1984tu} we find,
\beq
C = \frac{N_c \, \alpha_s}{(2\pi)^6}, 
\label{cglr}
\eeq
and in \cite{Gribov:1983fc}
\beq
C = 2\pi N_c \alpha_s.
\label{cglr2}
\eeq
Similarly we find in \cite{Albacete:2010bs}
\beq
C = \frac{(2\pi)^8\, C_F\, \alpha_s^3}{\pi N_c^2},
\label{cam}
\eeq
and in \cite{ALbacete:2010ad}
\beq
C = \frac{2 \pi^2 \, K\,C_F\,  \alpha_s}{N_c^2} = \frac{\pi^2 \, K\,(N_c^2-1) \alpha_s}{N_c^3}
\label{cad}
\eeq
where $K$ is a fit parameter which is quoted to be of the numerical value 1.5-2. We see that the coefficients in \eqref{ckt}, \eqref{cglr}, \eqref{cglr2},
\eqref{cam} and \eqref{cad} are all different from each other.  It appears also that none agrees with the result above, 
equation \eqref{Ccoeff}. Our result \eqref{Ccoeff} on the other hand agrees with the result in \cite{Schafer:2012yx}
where it was indeed observed that an extra factor $\pi$ for each TMD distribution must be included to agree with 
\eqref{ckt} above.

The numerical differences between the pre factors used in different papers are clearly rather 
important. It should also further be noted that in the papers \cite{Kharzeev:2003wz, ALbacete:2010ad}
the $k_\perp$ integration  is performed only up to $l_\perp$ while such a bound does not appear 
in the other papers.  Moreover in most of the phenomenological applications the coupling $\alpha_s$ is taken 
to run with some scale which also differs from paper to paper.

\subsection{Higher order terms in axial gauge, and more complete view}
\label{sec:singlehadron}

From the contribution in figure \ref{glrgluonprod2} we have thus seen that we can in the non-light-like
axial gauge, $A^++A^-=0$, obtain the formula \eqref{GLRfact} where the TMD distributions are given by 
\eqref{temporalpdf1} and \eqref{temporalpdf2}. We notice that exactly the same gauge is used in the 
CCH formalism \cite{Catani:1990eg} and in the GLR paper \cite{Gribov:1984tu}. 
 
 The question is of course what happens when we include higher order corrections to figure \ref{glrgluonprod2}. 
 We will now in this section first prove that the axial gauge does indeed eliminate to leading 
 power the couplings to the collinear regions, and at the same time we will see  what kinematics is necessary 
 for this result to hold. We will show that the kinematics is actually opposed to the usual small-$x$ kinematics. 
 Thus for the higher order corrections to be generally negligible we will need contour deformations to ensure 
 the desired kinematics. We shall then give an argument for why the needed contour deformations generally fail
 in  the axial gauge.  
 
 Assume now that we have a collinear region $C$ which carries momentum lines that are large in some direction $w_C$. For example this could be region $C_A$ which has large momentum in the $+$ direction.  Let $\tilde{w}_C$ 
 be the conjugate direction to $w_C$, such that $w_C \cdot \tilde{w}_C=1$.  The large component of $C^\mu$ is 
 then given by $\tilde{w}_C \cdot C$.  We now choose the axial gauge $n\cdot A = 0$ where $n$ is not necessarily 
 light-like.  Let $V$ be any vector. We then have 
 \beq
 V\cdot C = V\cdot w_C \,\, \tilde{w}_C\cdot C + \mathrm{p.s.c.}
 \eeq
 where ``p.s.c." as before stands for ``power suppressed corrections".  Now we let $V=n$, and using that 
 we are in $n\cdot A=0$ gauge, we obtain
 \beq
 0 = n\cdot C = n\cdot w_C \,\, \tilde{w}_C\cdot C + \mathrm{p.s.c.}
 \eeq
 Assuming now that $n \cdot w_C \neq 0$, we can separately scale the gauge vector 
 \beq
 n \to \frac{n}{n\cdot w_C}
 \eeq
 for each collinear region in the graph to get
 \beq
 0 = n\cdot C = \tilde{w}_C\cdot C + \mathrm{p.s.c.}
 \eeq
 Thus we conclude that the leading term vanishes in the axial gauge, and only power-suppressed 
 contributions remain. Notice that if $n \cdot w_C=0$ then we cannot necessarily conclude that the 
 leading contribution is eliminated. It might also be that, depending on the exact kinematics, several 
 directions of $C^\mu$ simultaneously become important. In that case the advantage of the axial 
 gauge vanishes. Let us illustrate these points with some examples. 
 
 Consider now  a gluon $k$ coupling to region $C_A$, and 
 denote $\tilde{C}_A^\mu=N^{\mu\nu}(k)C_{A,\nu}$. It is actually then $\tilde{C}_A$ that corresponds to $C$ 
 above (since $n\cdot \tilde{C}_A = 0$ but $n \cdot C_A \neq 0$).  
 Assume we are in the $A^-=0$ gauge.Then 
 \beq
 \tilde{C}_A^+ &\sim& N^{+-}C_A^+ \!=\left (1 - \frac{k^-}{k^-} \right ) C_A^+ = 0, \\
  \tilde{C}_A^i &\sim&  N^{i-}C_A^+ \!= 0, \\
  \tilde{C}_A^- &\sim& N^{--} \!\!= 0.
 \eeq
 Therefore only power suppressed contributions from $A$ will remain (we could have also immediately seen this 
 from the fact that $n\cdot C_A = 0 + \mathrm{p.s.c}$). On the other hand if we choose the gauge 
 $A^+=0$ then 
  \beq
 \tilde{C}_A^+ &\sim& N^{+-}C_A^+\!=\left (1 - \frac{k^+}{k^+} \right ) C_A^+ = 0, \\
  \tilde{C}_A^i &\sim&  N^{i-}C_A^+ \!= -\frac{k^i}{k^+}C_A^+, \\
  \tilde{C}_A^- &\sim& N^{--}C_A^+ \!= -\frac{k^-}{k^+}C_A^+.
 \eeq
 Here we see that $ \tilde{C}_A^i$ and $\tilde{C}_A^- $ are suppressed only if $k^+$ is the dominant 
 component of $k$. If not, then in the higher order terms all contributions can be important and the 
 situation obviously gets complicated. The gauge $A^+=0$ is useful in DIS where  the 
 target hadron has large $P^+$.  In hadron--hadron collisions, however, as we have seen, the
 light-cone gauge cannot be used. There is  moreover the problem with rapidity divergences 
 which appear in TMD distributions via integrals like \eqref{rapdiv} (the light-cone distribution \eqref{Updfdef}
 for example leads to divergences and is therefore ill-defined). These divergences become visible starting 
 from one loop calculations.  Now assume we are instead in $A^++A^-=0$ gauge.  Then 
 \beq
 \tilde{C}_A^+ &\sim& N^{+-}C_A^+ \!=\left (1 - \frac{k^+\!\!+k^-}{k^+\!\!+k^-} +\frac{k^+k^-n^2}{(k^+\!\!+k^-)^2}\right ) C_A^+
 = \frac{k^+k^-n^2}{(k^+\!\!+k^-)^2}C_A^+ , \\
  \tilde{C}_A^i &\sim&  N^{i-}C_A^+ \!=\left (- \frac{k^i}{k^+\!\!+k^-} +\frac{k^ik^-n^2}{(k^+\!\!+k^-)^2}\right ) C_A^+ , \\
  \tilde{C}_A^- &\sim& N^{--}C_A^+ \!= \left ( - \frac{2k^-}{k^+\!\!+k^-} +\frac{(k^-)^2n^2}{(k^+\!\!+k^-)^2}\right ) C_A^+.
 \eeq
 If for example $k$ is collinear to $C_A$, then indeed the contributions are power suppressed. 
  
 Thus for the axial gauge to be useful, the momenta emerging from $C_A$ ($C_B$) should be collinear to $C_A$ ($C_B$).
 Actually none of the momentum components need to scale with $\sqrt{s}$, but the dominant component 
 should be $k^+$ (or $k^-$ for $C_B$).  Remember indeed from our classification scheme in section \ref{sec:diffcases}
 that momenta which have no components scaling with $\sqrt{s}$ but whose components along $C_A$ 
 dominates are still classified as belonging to $C_A$. 
 If, however, we are in a region where for example $k_\perp$ dominates, then we see that we have a large 
 contribution from the transverse components.  In that case we cannot neglect the higher 
 order corrections. This is why we must be able to always deform the contour into the region where $k^+$
 (or $k^-$ for $C_B$) is  the large component.      

\begin{figure}[t]
\begin{center}
\includegraphics[angle=0, scale=0.55]{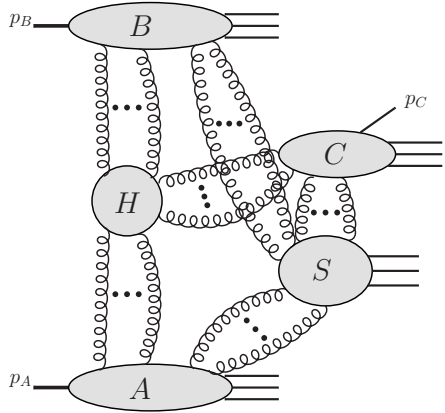}
\end{center}
\caption{\label{singlehadronprod} Leading regions for single inclusive hadron 
production via gluon initiated jet in hadron-hadron collisions. There is an additional collinear region associated with the produced hadron $p_C$. There will generally also be additional collinear regions associated with unobserved 
jets, these are not shown here for simplicity.} 
\end{figure}
 
\begin{figure}[t]
\begin{center}
\includegraphics[angle=0, scale=0.55]{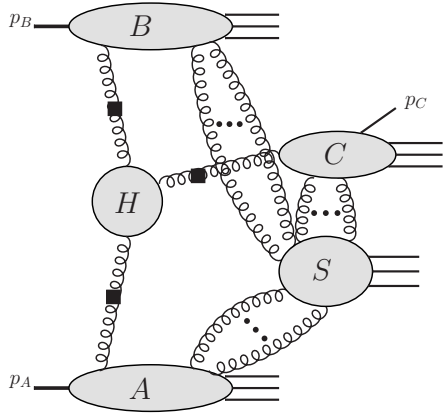}
\end{center}
\caption{\label{singlehadronprodLC} Single hadron production in axial gauge where the extra collinear-to-hard 
can be eliminated. The collinear regions then couple to the hard region via a single transversely polarized gluon, 
indicated by the black squares, each. } 
\end{figure}

 The analysis above and in  section \ref{sec:diffcases} suggests a general 
 picture like in figure \ref{singlehadronprod}. We consider the case where the observed hadron, $p_C$, has 
 some component scaling with $Q$, the reason being that the scale $Q$ is needed to suppress the higher 
 order corrections as seen above. 
 The regions in figure  \ref{singlehadronprod} are to be 
 understood in the classification presented in section \ref{sec:diffcases}. The momentum $Q$ is fixed 
 and $Q/\sqrt{s} \to 0$ asymptotically. There are actually further lines going out from the hard region 
 which give undetected collinear regions but we do not show them in figure \ref{singlehadronprod}  for simplicity.  
 According to what we have just seen above, in axial gauge we generally 
 expect the contributions in figure \ref{singlehadronprod} to be reduced to that of figure \ref{singlehadronprodLC}. 
 Here the extra collinear-to-hard gluons are missing, and the remaining gluons coupling to $H$ are 
 transversely polarized (indicated by black squares).

Note from figure \ref{singlehadronprodLC}
that the soft region still remains. Indeed the analysis above does not directly apply to the soft region since 
we needed a scale $Q$ to suppress the higher order terms. To simplify the expression completely then, one 
must be able to show that the soft region can be eliminated or neglected. 

\begin{figure}[t]
\begin{center}
\includegraphics[angle=0, scale=0.55]{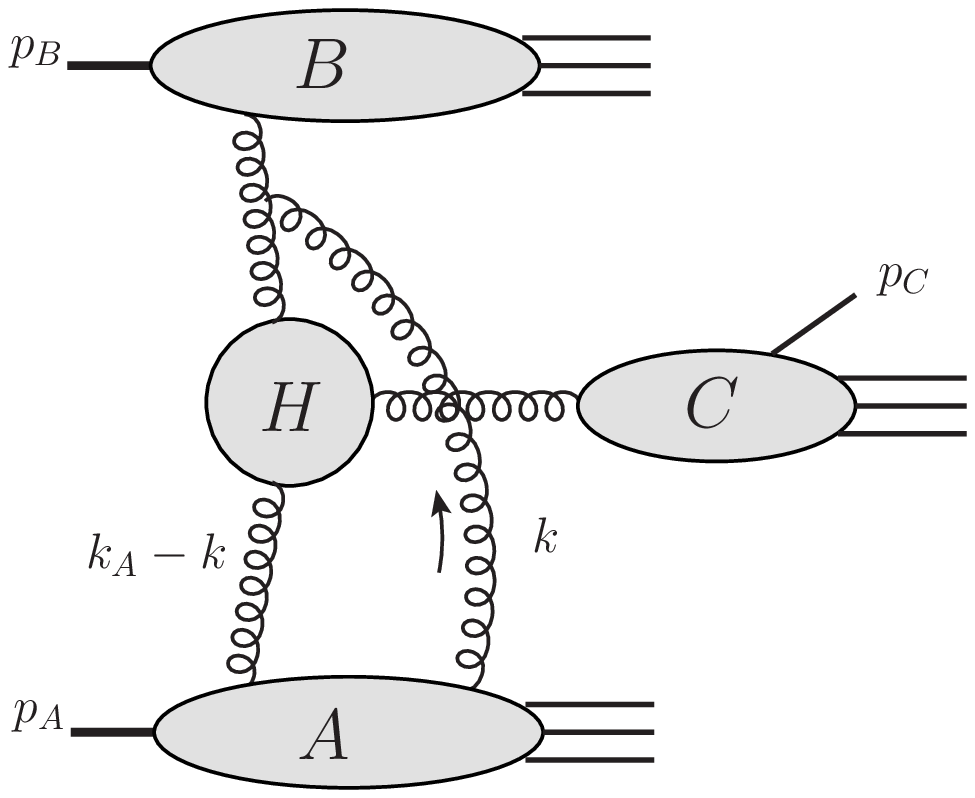}
\includegraphics[angle=0, scale=0.55]{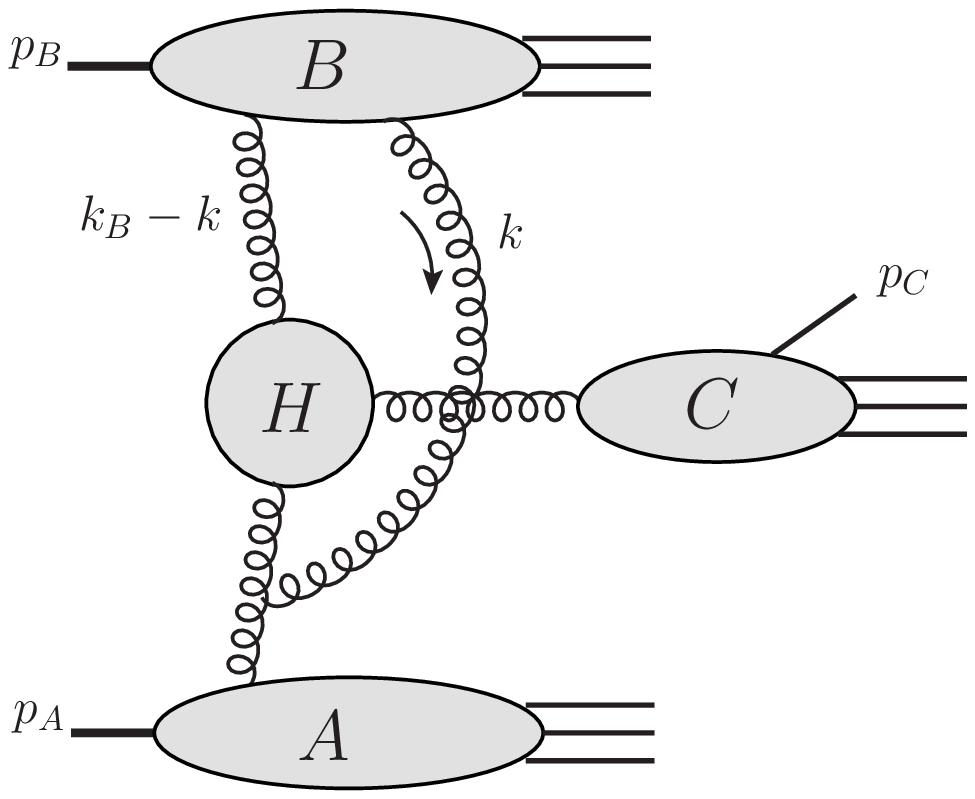}
\includegraphics[angle=0, scale=0.55]{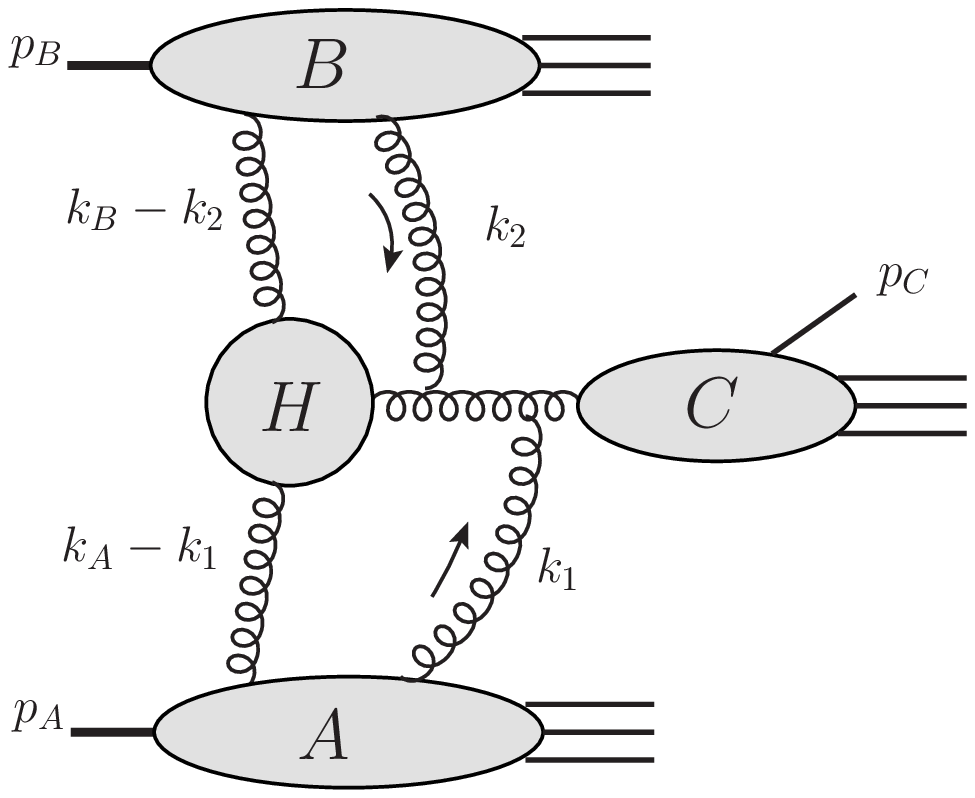}
\end{center}
\caption{\label{singlehadronsoft} Examples of graphs in axial gauge where soft gluons are exchanged between the 
collinear regions. Each of these type of emissions require contour deformations in different directions to stay 
out of the Glauber region. } 
\end{figure}

In figure \ref{singlehadronsoft} we show examples of soft gluons exchanged between the different regions. 
In the first graph (top left) the gluon $k$ attaches to the collinear-to-$B$ gluon that goes into the hard 
scattering.  The momentum $k$ then runs in a loop from top to down, counterclockwise, via $H$ into $A$
and back again.   The line $k_A-k$ then gives a pole (taking all $k_\perp$ to be of order $m$) 
\beq
k^- \sim \frac{m^2}{k_A^+} - i\epsilon \sim \frac{m^2}{Q} - i\epsilon.
\eeq
Inside the lower blob $A$, $k$ will run along the large momentum $p_A$, and so there 
will be a typical pole of the type 
\beq
k^- \sim \frac{m^2}{p_A^+} +i \epsilon \sim \frac{m^2}{\sqrt{s}} +i \epsilon.
\eeq
We thus see that these poles pinch the integration contour of $k^-$.
It might also be that $k$ in the lower blob attaches to a line with plus component only 
of order $Q$ instead of $\sqrt{s}$, but in any case we see that $k^-$ is at least forced to be 
small as $m^2/Q$.  One can still save the power counting arguments if $k^+$ can be deformed 
far out so that $k^+k^- \sim k_\perp^2$.  

We must now, however, exactly specify how to treat the singularities of the axial gauge propagator
\eqref{temporalprop}.  The canonical regularization of these singularities is given by the
principal value prescription. 
The canonical regularization is useful because the corresponding 
generalized functions then obey elementary relations, such as ordinary differentiation, that are obeyed by the 
corresponding regular functions
\cite{gelfand}. 
The use of principal value, however, also implies that one cannot deform the contours. The variable $k\cdot n$ 
must therefore remain on the real axis. 

As we have seen above, for the contributions in figure \ref{singlehadronsoft} we must deform 
in the first graph (top left) $k^+$ but not $k^-$, in the second graph (top right) $k^-$ but not $k^+$ while 
in the last graph (bottom) we must simultaneously deform $k_1^+$ and $k_2^-$ while keeping 
$k_1^-$ and $k_2^+$ fixed. We then, however, see that these requirements are in contradiction 
with the fact that we cannot deform on $k\cdot n$. For example, deforming in the first case $k^+$, 
\emph{i.e.} letting $k^+ \to k^+ + iC$ for some large $C \sim Q$, but keeping $k^-$ fixed implies 
that 
\beq
k\cdot n = k^++k^- \to (k^++iC)+k^- =k\cdot n +iC
\eeq
which is not allowed. The required contour deformations therefore fail.  We thus conclude 
that the treatment in axial gauge is not complete. 

One may also consider the possibility of using the so-called ``planar gauge" introduced 
in \cite{Dokshitzer:1978hw}. In this gauge, the gauge vector $n$ is non-light-like, so that 
$n^2 \neq 0$, but the last term in the axial gauge propagator \eqref{temporalprop} is eliminated 
(by a clever choice of the gauge fixing term in the Lagrangian). Moreover, as shown in 
\cite{Dokshitzer:1978hw}, Faddeev-Popov ghosts are still absent, just like in axial gauge. This 
gauge has thus all the advantages of the axial gauge, and in addition is free from the double 
pole in the propagator. It is therefore certainly much better behaved. However, the unphysical 
singularity $1/k\cdot n$ still remains and must be treated via the principal value. Therefore the 
above arguments still apply to this gauge. In \cite{Dokshitzer:1978hw} the authors argue that,
since the propagator poles are unphysical and have to all cancel at the end of the day, one might
as well treat $1/k\cdot n$ as a regular function, excluding this pole from loop integrals. The problem, 
however, is that one still needs to perform the contour deformations to prove factorization, and 
in doing this the term $1/k\cdot n$ cannot be neglected in the intermediate steps, even if the 
final result should be free from unphysical poles.

It is of course possible that one chooses a regularization which is not principal value. For example, 
we saw above that the choice in  \cite{Kovchegov:1998bi}  for the light-cone gauge is given by \eqref{KMprop}.
In any case, however, it is very hard to see how exactly a systematic procedure is developed that 
is capable of treating graphs of arbitrarily high order, as is required for the full proof of factorization. 
As far as we aware of, this has never been done. We leave the possibility open that 
a treatment in axial gauge might work out, but it is  difficult to see how this would be 
achieved.

\subsection{The gluon distribution function}
\label{sec:tmdgluon}

We have systematically gone through single inclusive particle production at high energies, 
and we have concentrated especially on the small-$x$ factorization formula \eqref{GLRfact}. 
In this section we examine more closely  
the exact definition of the TMD gluon distribution. We will moreover at the end of the section make some final 
comments on the validity of factorization. 


According to \eqref{KTgluon}, the gluon distribution is a (modified) Fourier transform of the dipole 
scattering amplitude in the adjoint representation. The expression \eqref{KTgluon} is appropriate in 
a covariant gauge, and not in an axial gauge. In the canonical definition of the parton distributions, 
the direction of the Wilson lines in \eqref{dipadj} are 
taken opposite to the hadron, \emph{i.e} for a hadron moving with momentum $p_A$ ($p_B$), the direction 
is taken as $n_B$ ($n_A$), which is parallel to $p_B$ ($p_A$).  To leading power we can also take the directions to be  $n_A+n_B$ for both hadrons, and 
the axial gauge $(n_A+n_B)\cdot A=0$ then sets the Wilson lines to unity. At first sight, however, this does not 
seem to be strictly
correct because if in \eqref{dipadj} we set the Wilson lines to be unity then we find that \eqref{dipadj}
vanishes, $\mcal{N}=0$, which obviously cannot be true. Part of the answer is that a fully gauge invariant definition 
of \eqref{dipadj} requires that we also insert transverse gauge links at infinity, and these are non-zero 
in any axial gauge. However, to match the axial gauge expressions, \eqref{temporalpdf1} and \eqref{temporalpdf2},
one must also express the distribution \eqref{dipadj} using the field tensors $F^{+i}$ and $F^{-i}$.  
Let us now see how this can be done. 

It is in fact a fundamental property of all gluon distributions  that the field tensors $F^{\mu\nu}$ appear in the definitions.
The underlying reason for this comes from the elementary parton model definition \eqref{eq:pdf.lf.def}.  
As the QCD definitions are appropriate modifications and generalizations of the parton model result, it 
is then natural that the field tensors appear in the definitions of the integrated and TMD gluon distributions
\cite{qcdbook}. This is also the case in the construction scheme for the generalized TMD distributions 
given in \cite{Bomhof:2004aw, Bomhof:2006dp}.  It should therefore also be possible to write the dipole distribution \eqref{KTgluon} using the field tensors, if it indeed is a TMD gluon distribution as claimed.

Consider the lowest order contribution from \eqref{dipadj} where we insert a set of outgoing states 
$|X,\mathrm{out}\rangle$ between the Wilson lines and then expand each Wilson line to first order in $g_s$. We will 
\emph{assume} that the averaging in \eqref{dipadj} is given by an ordinary expectation value between 
momentum eigenstates of the hadron, but we are not actually sure whether this is consistent with the 
formalism from which \eqref{dipadj} is supposed to arise. Nevertheless, without this assumption we cannot 
make any real comparison.  We also neglect for the moment the regulator $y$ in \eqref{KTgluon} and 
\eqref{dipadj}.
The first order expansion of \eqref{dipadj} in \eqref{KTgluon} for a hadron with momentum $p_A$
gives
\beq
&&f_A^{(1)}(k_\perp) = \frac{N_c}{(2\pi)^4\alpha_s}k_\perp^2 \int d^2x_\perp\!\! \int d^2 y_\perp e^{-ik_\perp \cdot (x_\perp
-y_\perp)}  \nonumber \\
&&\sum_X\frac{g_s^2N_c}{N_c^2-1}  \int dx^-\! \int dy^- \frac{\langle p_A | A_a^+(x^-,x_\perp) |X,\mathrm{out} \rangle 
\langle X,\mathrm{out}| A_a^+(y^-,y_\perp) |p_A\rangle}
{\langle p_A|p_A\rangle}.
\label{f1formula}
\eeq

The argument to convert $k^i A_a^+$ into $F^{+i}_a$ can now be made as follows. In the power counting 
of the contributions from the region collinear to $p_A$, the largest contribution arises from the + component as we 
have seen in sections \ref{sec:powercount} and \ref{sec:diffcases}.  In the $N$ gluon exchange term, the biggest contribution 
therefore arises from the terms where we pick up the contribution $A^{+\cdots +}$ for all the $N$ collinear-to-$p_A$ gluons.
For every contribution where we change one of the gluon polarizations from the longitudinal index $+$ to a transverse index $i$, we lose one power of $\sqrt{s}$. 
Thus one can let
\beq
k^i A_a^+ \to k^i A_a^+ - k^+ A_a^i
\label{fieldtotensor}
\eeq
since the correction produces a power suppressed term.  It is important to notice that this exchange 
is not permissible in the hard scattering factorization. From the power counting in section \ref{sec:powercount}
we actually see that $k^iA^+ \sim m\,Q$ and $k^+A^i \sim Q\, m$ for a collinear-to-$A$ gluon $k$. In the 
small-$x$ case, however, $k^+ \ll \sqrt{s}$, so that $k^+A^i \ll \sqrt{s} \, m \sim k^iA^+$.

For the lowest order term in \eqref{f1formula}
this is enough to convert each $k^i A_a^+$ into $F^{+i}_a$ since the commutator in $F^{+i}$ contributes 
at higher order. 
Removing the sum over the states $X$, one can then rewrite \eqref{f1formula} as 
\beq
f_A^{(1)}(k_\perp) &=& \frac{N_c}{2\pi \alpha_s}\frac{g_s^2N_c}{N_c^2-1}\int \frac{dx^-d^2x_\perp}{(2\pi)^3 2p_A^+} 
e^{-ik_\perp \cdot x_\perp} \langle p_A | F_a^{+i}(0^+\!,x^-\!,x_\perp)  F_a^{+i}(0) |p_A\rangle \nonumber \\
&=& \frac{N_c^2}{N_c^2-1} \int \frac{dx^-d^2x_\perp}{(2\pi)^3 p_A^+} 
e^{-ik_\perp \cdot x_\perp} \langle p_A | F_a^{+i}(0^+\!,x^-\!,x_\perp)  F_a^{+i}(0) |p_A\rangle.
\label{firstorderKTgluon}
\eeq
In the dipole model from which \eqref{dipadj} arises, the large $N_c$ limit is 
employed which means that the coefficient $N_c^2/(N_c^2-1)$ is set to unity.  The result \eqref{firstorderKTgluon}
then very strongly resembles \eqref{temporalpdf1}.  

We note, however, that in \eqref{firstorderKTgluon}, there is no $x$ dependence as in  \eqref{temporalpdf1}. 
This is a characteristics of the dipole formalism where the longitudinal component of the total momentum 
coupling to the collinear region is neglected. The rapidity dependence of the dipole distribution therefore 
purely arises from the rapidity cut-off. 
In \eqref{temporalpdf1}, the rapidity cut-off is 
not yet included, and the $x_A$ variable which is the longitudinal momentum fraction of the gluon $k$ 
in figure \ref{glrgluonprod2} clearly does not play the role of a rapidity cut-off.  This is also one of the reasons 
why the dipole distribution \eqref{KTgluon} or \eqref{dipgluedistrb} cannot be directly related to the integrated
distribution as in \eqref{intvsunintglr}, since the meanings of the longitudinal variables in \eqref{intvsunintglr} 
are completely different on the right and the left hand sides. Despite this, however, the relation
\eqref{intvsunintglr} is still widely advocated in the small-$x$ literature. 

When all the gluons coupling to the collinear region contribute with their longitudinal polarizations, 
however, there must be certain cancellations due to the Ward identities. In Feynman gauge the easiest 
way to see this is to use the $K$-$G$ decomposition \eqref{kgdecomp}.  Ward identities apply on the 
$K$ terms, and these correspond to the longitudinally polarized gluons. For the region collinear to $p_A$, 
we choose the vector $n$ in the $K$-$G$ decomposition \eqref{kgdecomp} to be in the opposite 
direction to $p_A$, \emph{i.e.} $n=n_B$ (and the other way around for the $B$ terms). Then 
as we saw in \eqref{Gplusminus} and \eqref{Kplusminus}, the longitudinal components vanish
for the $G$ terms while for the $K$ terms we get unity. The largest contribution therefore arises 
from the terms where we only pick up the $K$ terms. Ward identities, however, imply that part of this 
largest contribution cancel, leaving behind a reminder term which is of the same order as the contributions 
where one gluon contributes as $G^{i-}$, while all the other terms contribute via the
$K^{+-}$ terms \cite{qcdbook, Collins:2008sg}. It is then the combination of the $G^{i-}$ 
term and the remainder term from the Ward identity cancellations that give rise to the field tensor 
term $F^{+i}$ (including the commutator term) 
while the sum over all the $K^{+-}$ terms give the Wilson lines. We explain this in the context of the small-$x$ calculations 
in \cite{ourpaper} where we derive the TMD gluon distribution that looks like \eqref{WWdistrbadj}. That is, 
a gluon distribution including the $F^{+i}$ factors is naturally constructed. 

Let us now extend the above analysis to all orders. 
In \cite{Bomhof:2004aw, Bomhof:2006dp} a construction scheme of TMD parton distributions was proposed.
The proposed scheme is a method of converting the collinear-to-hard gluons to Wilson lines, 
thus giving the ``unsubtracted" TMD parton distributions.  We now apply the scheme to the present 
process. 

\begin{figure}[t]
\begin{center}
\includegraphics[angle=0, scale=0.55]{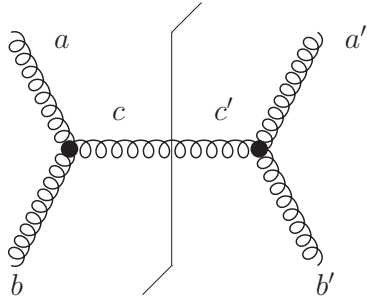}
\end{center}
\caption{\label{2to1gluon}  The elementary graph for the gluon production.} 
\end{figure}

The scheme starts from studying the elementary 
``hard" graph for the process under consideration, that is figure \ref{2to1gluon}. 
Of course  here this graph does not 
involve any hard momenta, but that does not really affect the structure of the Wilson lines which parametrize 
the non-perturbative structure. According to \cite{Bomhof:2004aw, Bomhof:2006dp} then, the contribution from the process in 
figure \ref{2to1gluon} to the TMD gluon distribution of the lower particle (with momentum $p_A$) is
\beq
F_{b'}(x)\, F_b(0) \, if^{abc}\, if^{a'b'c'} (W_B^{(+)})_{cc'}(W_B^{(-)})_{aa'}
\label{gaugelinks1}
\eeq
where 
\beq
W_B^{(\pm)}= W_B(0;\pm \infty^-\!\!,0_\perp)W_T(\pm \infty^-\!\!, 0_\perp;\pm \infty^-\!\!, x_\perp)W_B(\pm \infty^-\!\!,x_\perp;x^-\!\!, x_\perp),
\label{wilsonpm}
\eeq
and 
\beq
\label{wilsonlong}
W_B(x;y) = P \exp \left ( -ig_s \int_x^y d z\,  n_B\cdot A_a (z) T^a \right ) , \\
W_T(x,y) = P \exp \left ( -ig_s \int_x^y d z_\perp \cdot A_{\perp,a} (z) T^a \right ).
\label{wilsontrans}
\eeq
If we instead consider the TMD distribution of the upper hadron with momentum $p_B$ then 
the longitudinal direction in \eqref{wilsonpm} should be $+$ instead of $-$, and in \eqref{wilsonlong} 
$n_B \to n_A$. Notice that in \eqref{gaugelinks1} the Wilson lines are in the adjoint representation 
as is clear from the color subscripts.  We now use $T^b_{ac}=if^{abc}$ for the adjoint representation 
to rewrite \eqref{gaugelinks1} as
\beq
F_{b'}(x)\, F_b(0) \, T^b_{ac}\, T^{b'}_{a'c'}\,  (W_B^{(+)})_{cc'}(W_B^{(-)})_{aa'} \nonumber \\
= F_{a'c'}(x)F_{ac}(0) \,  (W_B^{(+)})_{cc'}(W_B^{(-)})_{aa'}\label{gaugelinks2}
\eeq
where we have defined 
\beq
F_{ac} \equiv F_b \, T^b_{ac}.
\label{adjtensor}
\eeq
From equation \eqref{gaugelinks2} one then finds the following contribution to the 
correlator in the gluon distribution 
\beq
\langle p_A| \left( F(x)W_B^{(+)\,\dagger} \right )_{a'c}\left ( F(0)W_B^{(-)} \right )_{ca'} |p_A\rangle \nonumber \\
= \mathrm{Tr} \langle p_A|  F(x) W_B^{(+)\,\dagger} F(0)W_B^{(-)}|p_A\rangle. 
\eeq
The trace is taken with respect to the adjoint representation with the field tensor 
defined as in \eqref{adjtensor}. The (unsubtracted) gluon distribution function is then given by 
\beq
f_A(x_A,k_\perp) = \! \int \frac{dx^-d^2x_\perp}{(2\pi)^3\,p_A^+}e^{ix_Ap_A^+x^-\!-ik_\perp\cdot x_\perp}
 \mathrm{Tr} \langle p_A|  F^{+i}(0^+\!\!,x^-\!\!,x_\perp) W_B^{(+)\,\dagger} F^{+i}(0)W_B^{(-)}|p_A\rangle. \nonumber \\
\label{gaugelinkpdf}
\eeq
Actually, note that in the canonical definitions \eqref{eq:pdf.lf.def} and \eqref{pdfcanonical} we would 
instead of $1/p_A^+$ insert the factor $1/k_A^+=1/(x_Ap_A^+)$. The reason we choose $1/p_A^+$ here
is that we will connect the above distribution with that of the dipole result \eqref{KTgluon} and remember 
from above that the dipole result cannot be obtained if we have the factor $1/k_A^+$ (see also remarks 
just below). 

Strictly speaking \eqref{gaugelinkpdf} involves only the bare fields. Remember from section \ref{sec:simplefact}
that the gluon distribution has to be renormalized as in equation \eqref{pdfrenorm}.  The soft region 
must also properly be subtracted to cancel the rapidity divergences in \eqref{gaugelinkpdf}.  
A similar definition is easily obtained for the gluon distribution associated with $p_B$
\beq
f_B(x_B,k_\perp)= \! \int \frac{dx^+d^2x_\perp}{(2\pi)^3\,p_B^-}e^{ix_Bp_B^-x^+\!-ik_\perp\cdot x_\perp}
 \mathrm{Tr} \langle p_B|  F^{-i}(x^+\!\!,0^-\!\!,x_\perp) W_A^{(+)\,\dagger} F^{-i}(0)W_A^{(-)}|p_B\rangle. \nonumber \\
\label{gaugelinkpdfb}
\eeq

Exchanging to leading order the Wilson line directions to $n_A+n_B$ in both cases and applying 
the axial gauge $(n_A+n_B)\cdot A=0$ we then obtain \eqref{temporalpdf1} and \eqref{temporalpdf2} 
respectively. There is an additional factor $N_c$ arising from the color traces in \eqref{gaugelinkpdf} 
and \eqref{gaugelinkpdfb} (exactly as in \eqref{firstorderKTgluon}). Thus we can see \eqref{gaugelinkpdf} 
and \eqref{gaugelinkpdfb} as possible generalizations of \eqref{temporalpdf1} and \eqref{temporalpdf2} 
to arbitrary gauge. 

The connection to the dipole formula \eqref{KTgluon} and \eqref{dipadj} can now be made as 
follows. We consider the transverse derivatives in \eqref{KTgluon} acting on the Wilson lines in \eqref{dipadj}.
The effect of the derivative can be written as (for the hadron $p_A$)
\beq
\partial_x^i \tilde{W}(x_\perp) = -i g_s\int dx^- W_B(x;\infty^-\!\!,x_\perp)\partial_x^i A^+_a(x)T^aW_B(-\infty^-\!\!,x_\perp;x)
\label{Wilsondiff}
\eeq
where as we recall $\tilde{W}$  is given by taking \eqref{Wilsonfund} with the adjoint color matrices 
while $W_B(x;\infty^-\!\!,x_\perp)$ and $W_B(-\infty^-\!\!,x_\perp;x)$ are given by \eqref{wilsonlong}. 
We can again use \eqref{fieldtotensor} since the correction is power suppressed. One can also 
argue that the commutator of the field tensor is subleading since at given order in $g_s$ it 
contains one factor $A^i$ which replaces a factor $A^+$ from the Wilson line. In that case we could replace $ -i\partial_x^i A^+_a(x)T^a \to F^{+i}_aT^a=F^{+i}$ 
in \eqref{Wilsondiff}.  This would imply that \eqref{KTgluon} contains the same structure as 
in \eqref{gaugelinkpdf}, once we also set $x=0$ in \eqref{gaugelinkpdf} which as we remember from 
above is the standard approximation in the dipole formalism. 

Thus as we have seen, in a sense the formula \eqref{KTgluon} together with \eqref{dipadj} 
contains the contributions from the gluon field tensors as in \eqref{gaugelinkpdf}.  We motivated this 
by the power counting arguments, but a word of caution is in order here. We have mentioned above
that the $K$ terms in the $K$-$G$ decomposition are subject to certain cancellations from the Ward identities. This implies actually that terms 
containing one factor of $A^i$ at each side of the cut become leading. As explained above, these arise 
from the $G^{i-}$ terms. Thus the transverse components
in $F^{+i}$, including the commutator, may not be automatically dropped. 
The expression in  
\eqref{gaugelinkpdf} is therefore more correct than \eqref{KTgluon}, assuming of course that 
factorization holds. If not, then neither expression needs to be correct. Let us therefore now finish 
our analysis with a discussion on the validity of factorization. 

What we have thus seen is that \eqref{KTgluon} and \eqref{dipadj} can be related to the 
distribution, \eqref{gaugelinkpdf} or \eqref{gaugelinkpdfb}, constructed using the scheme of \cite{Bomhof:2004aw, Bomhof:2006dp}. However,  the scheme in \cite{Bomhof:2004aw, Bomhof:2006dp} by itself 
does not prove whether factorization holds or not. When a TMD parton distribution 
associated with a given collinear region is being constructed, one considers the attachments of the collinear-to-hard gluons 
to each line of the hard graph, and replace each set of connections by a Wilson line that correctly carries 
the color of the hard line. Since TMD factorization is used for two particle production in the almost back-to-back 
region, as in the examples of $e^+e^-$ annihilation and Drell-Yan production in section \ref{sec:TMD}, 
the relevant hard graphs are usually $2 \to 2$ partonic graphs, and one can then use these basic graphs 
to construct the possible gauge links for a given collinear region. An extensive list of possible gauge links 
is given in \cite{Bomhof:2006dp}.  

For proving factorization, however, one must consider all gluon attachments from the collinear regions 
to the hard graph \emph{simultaneously}, as well as all possible soft attachments between the collinear regions. 
For example, in \eqref{gaugelinks1}, following \cite{Bomhof:2004aw, Bomhof:2006dp}, the attachments from the 
collinear regions $C_A$ and $C_B$ in figures \ref{softhadronprod}, \ref{hardhadronprod} or \ref{singlehadronprod},
are considered separately, and each is summed into the Wilson lines in \eqref{gaugelinks1}. Considering all possible attachments, however, as for example in the graphs in figure 
\ref{singlehadronsoft},  it may very well be that the resulting structure is more complicated than 
in \eqref{gaugelinks1} or that it is not even possible to identify any gauge link contributions to the TMD distributions. 
At the same time, one must be able show that deformations out of the Glauber region are possible, or that 
the poles producing the Glauber pinch cancel. Cancellation of the Glauber region has been demonstrated 
explicitly in the case of Drell-Yan (Ch 14, \cite{qcdbook}), but difficulties may easily arise for the 
more complicated processes studied in \cite{Bomhof:2004aw, Bomhof:2006dp}. 

\begin{figure}[t]
\begin{center}
\includegraphics[angle=0, scale=0.6]{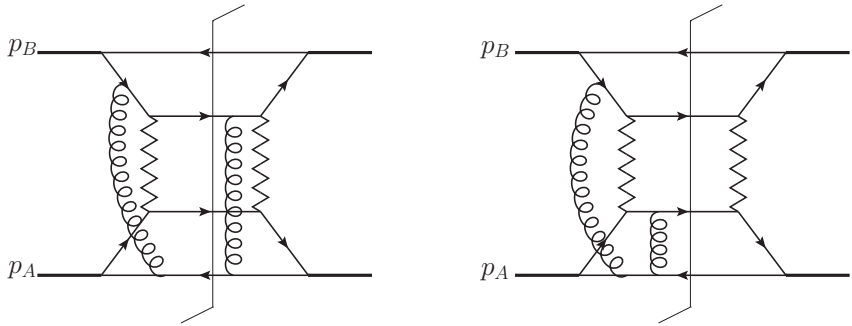}
\end{center}
\caption{\label{2to2gentmd}  Production of two hadrons in an elementary model considered in \protect \cite{Collins:2007jp}. We indicate the hard scattering by the exchange of the zig-zag lines. The additional gluon contributions correspond to breakdown 
of ordinary factorization.  } 
\end{figure}

\begin{figure}[t]
\begin{center}
\includegraphics[angle=0, scale=0.55]{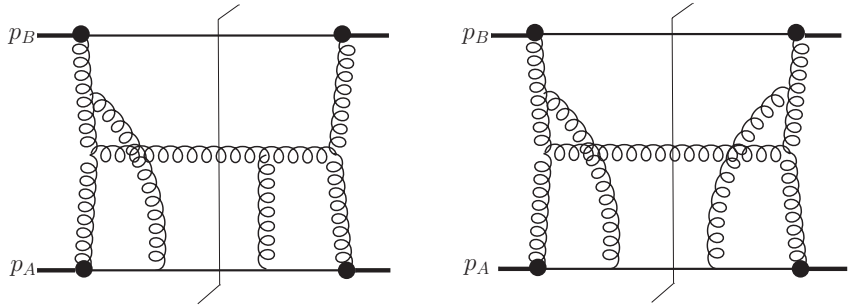}
\end{center}
\caption{\label{4glueexch} Examples of the type of graphs that are taken into account in the construction scheme of 
equation \protect \eqref{gaugelinks1}. The solid lines indicate the spectator parts of each hadron. } 
\end{figure}

In reference \cite{Collins:2007jp}, the breakdown of ordinary TMD factorization (\emph{i.e.} the TMD 
factorization that is relevant for the processes in section \ref{sec:TMD}) was explicitly demonstrated in di-hadron 
production in hadron-hadron collisions at the level of 2 gluon exchange between the hard part and the 
collinear part. We illustrate in figure \ref{2to2gentmd} two examples of the type of graphs considered in \cite{Collins:2007jp}. 
To distinguish the hard scattering we draw the hard gluons by zig-zag lines, while the collinear-to-hard
gluons are illustrated by curly lines. In the elementary model considered in \cite{Collins:2007jp}, the gluons are massive Abelian gluons, and the active lines that enter the hard scattering are scalar ``di-quarks" while the spectator lines are fermions.
The breakdown of ordinary factorization is then understood as being due to the attachments of the collinear gluons 
from the lower hadron lines to the upper active ``quark" line which is of course color connected to the upper hadron. 
The collinear-to-$p_A$ gluons in figure \ref{2to2gentmd} which couple to the upper active lines of the hard part 
are precisely the gluons  that  in the scheme of \cite{Bomhof:2004aw, Bomhof:2006dp} give rise to the gauge 
links of the generalized TMD distributions. The construction in \eqref{gaugelinks1} therefore contains these 
contributions. We illustrate these in the single gluon production case in figure \ref{4glueexch}. 

\begin{figure}[t]
\begin{center}
\includegraphics[angle=0, scale=0.6]{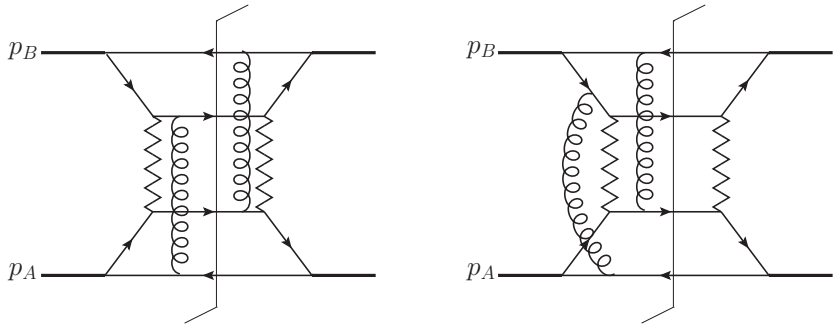}
\end{center}
\caption{\label{2to2nogentmd}  Examples of the class of graphs  considered 
in \protect \cite{Rogers:2010dm} that lead to the breakdown of TMD factorization for DSA. We indicate the hard scattering by the exchange of the zig-zag lines.} 
\end{figure}

\begin{figure}[t]
\begin{center}
\includegraphics[angle=0, scale=0.6]{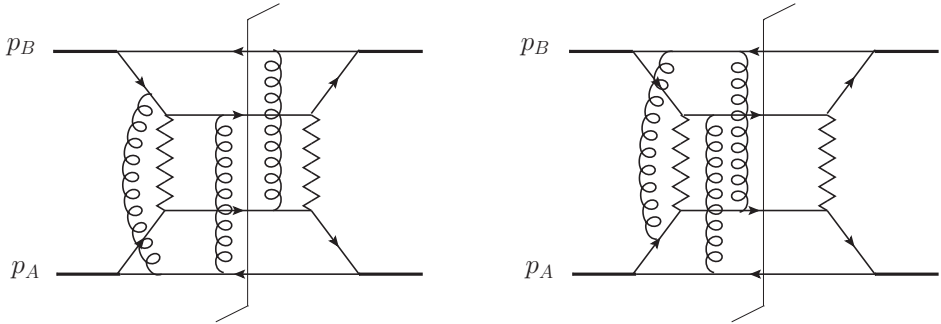}
\end{center}
\caption{\label{2to2nogentmd2}  Examples of graphs where TMD factorization is broken for the unpolarized 
cross section. We indicate the hard scattering by the exchange of the zig-zag lines.} 
\end{figure}

As discussed above, however, for a complete proof of factorization one must also consider the simultaneous gluon couplings 
between the upper hadron and the hard part.  This was 
considered in reference  \cite{Rogers:2010dm} which calculated in a slightly different model than \cite{Collins:2007jp} 
the type of graphs shown in figure \ref{2to2nogentmd} (the zig-zag lines for example correspond to a massive 
color singlet scalar boson).  These graphs have an entangled  color structure 
which makes it impossible to factorize the color flows even in the scheme of \cite{Bomhof:2004aw, Bomhof:2006dp}. 
The examples shown in figure \ref{2to2nogentmd} then break factorization for the Double Spin Asymmetry (DSA), 
while in the specific model considered the contributions from figure \ref{2to2nogentmd} to the unpolarized 
cross section cancel.  Breakdown of factorization for the unpolarized cross section instead appears 
for graphs where three additional gluons are exchanged, with at least one gluon coupling to each 
hadron. We illustrate this in figure \ref{2to2nogentmd2}.

\begin{figure}[t]
\begin{center}
\includegraphics[angle=0, scale=0.55]{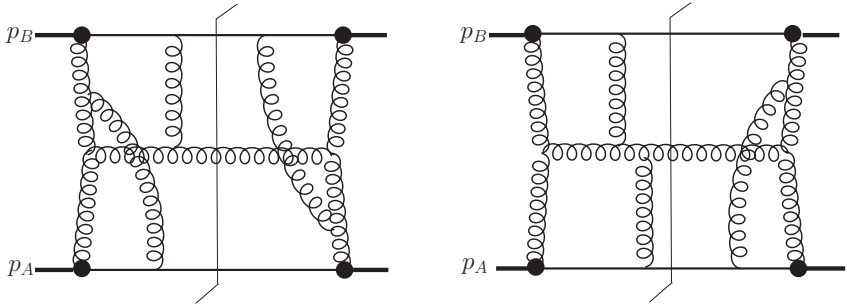}
\end{center}
\caption{\label{5glueexch} Examples of the type of graphs that may go beyond the construction scheme of 
equation \protect \eqref{gaugelinks1} in QCD. The solid lines indicate the spectator parts of each hadron. } 
\end{figure}

What this shows to us in the case of gluon production at small-$x$ is that to answer the question 
of factorization one needs to consider graphs like in figure \ref{5glueexch}. These graphs have 
non-trivial color flows that do not seemingly factorize into color singlet factor associated 
with each collinear region. In that case one must demonstrate explicitly that such contributions cancel. Given, 
however that they do not even cancel in the simple models considered in \cite{Collins:2007jp, Rogers:2010dm} 
it seems rather difficult to see how they would in full QCD. Indeed we note that the results from \cite{Rogers:2010dm} have 
further been systematized in \cite{Buffing:2011mj} where simultaneous couplings to different 
parts are considered, generalizing the scheme in \cite{Bomhof:2004aw, Bomhof:2006dp}. 
The difficulties with the color entangled contributions are there clearly demonstrated. 

\begin{figure}[t]
\begin{center}
\includegraphics[angle=0, scale=0.55]{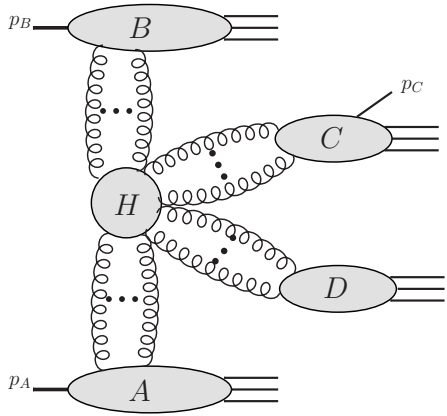}
\end{center}
\caption{\label{hardhadronprod2} Single hadron production where the second jet emerging from the 
hadron is integrated over. Arbitrarily many gluons can be exchanged between each collinear 
region and the hard region, as indicated by the dots. We do not show the soft region. } 
\end{figure}

We mentioned earlier that the gluon production in figures \ref{glrgluonprod}, \ref{glrgluonprod2}, 
\ref{2to1gluon}, \ref{4glueexch} and \ref{5glueexch} corresponds to the case of soft particle production, 
illustrated in figure \ref{softhadronprod}. To instead consider hard gluon (or rather hadron) production 
with large transverse momentum, so that a scale $Q$ is present which can be used to suppress 
transverse polarizations, we need to 
take into account that the hard part contains additional jets. It can be shown that the case where 
more than two jets emerge from the hard region is suppressed in the almost back-to-back region \cite{qcdbook}.
 We then consider the case where two gluon jets emerge from the hard region, and where only one of them 
 contains the detected hadron. We illustrate this case in figure \ref{hardhadronprod2}.

The case in figure \ref{hardhadronprod2} equals to taking di-hadron production and then 
integrating over one of the hadron momenta.  The $2\to 2$ hard scattering is now more intricate,
and the scheme of \cite{Bomhof:2004aw, Bomhof:2006dp} becomes rather complicated 
as can be seen from table 8 in reference \cite{Bomhof:2006dp}. More importantly, however, the 
results in \cite{Collins:2007jp, Rogers:2010dm, Buffing:2011mj}  become highly relevant and show us that generally
factorization is broken in di-hadron production. Cancellation of the factorization breaking terms 
occur for the integrated distribution, but not if we merely integrate over the momentum of one of detected 
final state particles.  In fact this can be seen in \cite{Buffing:2011mj} where simplifications occur 
only when one integrates over \emph{all} momenta except for a single hadron. Even in that case, 
however, the simplification only occurs for contributions that are termed ''tree-level".
It may be of course that the color structures simplify 
in the strict large $N_c$ limit where $N_c \to \infty$. The factorization breaking graphs studied in 
\cite{Rogers:2010dm}, see figure \ref{2to2nogentmd}, are for example non-leading in $N_c$.  Their 
effect on the production cross section may still, however, be important if there is no kinematical 
suppression. 

Finally we note that in more general processes like in figure \ref{hardhadronprod2} there is also the 
soft factor which will now be more complicated than in standard TMD factorization. Assuming that factorization
holds, according to \cite{Bomhof:2006dp} the unsubtracted TMD gluon distribution is a highly complicated 
function containing many different Wilson lines. Each light-like Wilson line produces rapidity divergences 
that must be regulated. In addition to the rapidity divergences there appear divergences related to the 
self energy corrections of the Wilson lines. 
All these divergences are regulated by subtracting the soft factor from the collinear region, 
which leads to definitions like in equation \eqref{subtractedtmd}. 
In the case of the gauge link structures that appear for  figure \ref{hardhadronprod2} using the scheme 
of \cite{Bomhof:2006dp}, however, we dare not even ask how exactly all these issues would be dealt 
with. It appears to be an immensely difficult task to obtain final definitions of the highly complicated TMD distributions which 
are free from all divergences.  Yet this would be extremely helpful for precise phenomenological applications.

\section{Summary}
\label{sec:summary}

Our main aim has been to provide a coherent analysis of TMD factorization 
and the TMD gluon distribution, especially as used in the small-$x$ region, and to examine many
important points that usually are not well explained or are overlooked in the literature. 


In section \ref{sec:factorization} we have given a unified analysis of the concept of factorization in different formalisms, 
the hard scattering formalism (section \ref{sec:hardscatfact}), the BFKL formalism (section \ref{sec:bfklfact}) and the CGC formalism (section \ref{sec:cgcfact}).  We also analyzed in section \ref{sec:hybrid} what we called hybrid approaches which combine collinear factorization with the use of TMD distributions.

The main point in section \ref{sec:hardscatfact} has been to explain what exactly is meant by factorization 
in the hard scattering case, and what approximations and methods are built into the analysis. We have then compared 
these to the small-$x$ treatments which use somewhat different methods. We emphasized in section 
\ref{sec:cgctmd} the difference between 
factorization which is constructed to be valid to leading power and the leading logarithmic approximation (LLA) 
that is based on the one-loop calculation.  As we have explained the former is of much greater accuracy and 
generality which is important to understand when comparing the different treatments.  

In section \ref{sec:hybrid}
we explained the idea behind the so-called factorization of mass singularities that is built into the hybrid formalisms. 
Let us note here that it has been demonstrated in \cite{qcdbook} that for the simplest partonic reactions as relevant 
for DIS, the method gives the same results as the hard scattering factorization for the massless limit of the hard 
scattering coefficient. It is, however, not clear to us whether this still holds in the cases studied in the hybrid formalisms, 
where one includes also TMD distributions, and studies  proton-nucleus collisions. 
We also note that the CCH and CCFM formalisms essentially base 
their underlying formulas on the same approach. The use of the method in these formalisms is discussed in \cite{ourpaper}.
We have explained here why this procedure is physically misleading, and caution should be taken before 
trying to move on to more complicated reactions. 

In section \ref{sec:gluonprod} we have given an extensive analysis of single particle production in the small-$x$ 
region. We started by showing in section \ref{sec:diffcases} that one can perform a power counting analysis 
very much as in section \ref{sec:powercount} to identify the leading structure. This is crucial to understand 
when the higher order corrections can be neglected and how the asserted formulas can be 
justified. 
The main  factorization formula \eqref{GLRfact} has been extensively used in phenomenological 
applications of small-$x$ QCD, at both RHIC and the LHC. It is therefore crucial to understand the physics 
behind it and the justifications given for its validity. We noted that 
many treatments in the literature are based on the axial gauge, and we therefore examined the application of the axial 
gauge in justifying the factorization formula \eqref{GLRfact}.  We showed in section 
\ref{sec:lcgauge} why the light-cone gauge is inappropriate for the formulation while in section \ref{sec:axialgauge}
we showed how one can obtain the standard factorization formula in a symmetric axial gauge.  

Then in section \ref{sec:singlehadron} we demonstrated the technical difficulties with the use of the axial gauge
and suggested that a more complete treatment be based instead on  covariant gauge. 
In section \ref{sec:tmdgluon} we then 
discussed the gluon distribution that is associated with  \eqref{GLRfact} and how it could generally be 
constructed from Feynman graphs, and we examined the graphs that are problematic for the full proof of factorization.  

There have lately been many applications of TMD factorization in the small-$x$ region, in
$pp$, $pA$ and $AA$ collisions. 
To fully prove factorization, however, one must show that the graphs of the type we showed in section \ref{sec:tmdgluon} 
cancel. 
In the case of $pA$ 
collisions we emphasize that the gluon couplings from the proton side cannot neglected. In particular it does not 
follow that one can automatically treat the proton using integrated parton distributions and fragmentation 
functions. If the observed particle is at low $p_\perp$ then the transverse momentum of the collinear region 
of the proton and the soft region cannot be neglected outside of these regions, and as a consequence 
TMD distributions must be used everywhere. A more complete factorization formula must then be constructed, 
taking into account the difficulties outlined in sections \ref{sec:singlehadron} and \ref{sec:tmdgluon}. 

Finally, a point which did not discuss much here concerns the scattering coefficient in the gluon 
production formula, equation \eqref{GLRfact}. Note that this factor diverges as $l_\perp \to 0$. This is 
in fact a sign that the standard treatment cannot be complete.
One should provide   for the scattering factor a full definition that is valid to all orders, is gauge independent, 
and which contains 
necessary subtractions to remove all divergences. An example for the scattering factor in heavy 
$q\bar{q}$ production is given in \cite{Collins:1991ty}.


\section*{Acknowledgements}

I would like to thank John Collins, Anna Stasto and Bo-Wen Xiao for useful discussions during 
an extended period of time. This work is supported by U.S. D.O.E. grant number~DE-FG02-90-ER-40577

\bibliographystyle{JHEP}
\bibliography{refs2}

\end{document}